\documentclass{elsart}
\usepackage{setspace}
\usepackage{amssymb}
\usepackage{amsmath}
\usepackage{graphicx}
\usepackage{amsfonts}
\makeindex

\begin{document}

\begin{frontmatter}

\title{Nuclear Astrophysics with Radioactive Beams}
\author[TAMU]{C.\ A.\ Bertulani}
\ead{Carlos\_Bertulani@tamu-commerce.edu} and
\author[NSCL,MSU] {A.\ Gade}
\ead{gade@nscl.msu.edu}
\address[TAMU]{Department of Physics, Texas A \& M University, Commerce, Texas
  75429, USA}
\address[NSCL]{National Superconducting Cyclotron Laboratory, Michigan
State University, East Lansing, MI 48824, USA}
\address[MSU]{Department of Physics \& Astronomy, Michigan State University,
  East Lansing, MI 48824, USA}
\begin{abstract}
The quest to comprehend how nuclear processes influence
astrophysical phenomena is driving experimental and theoretical
research programs worldwide. One of the main goals in nuclear
astrophysics is to understand how energy is generated in stars, how
elements are synthesized in stellar events and what the nature of
neutron stars is. New experimental capabilities, the availability of
radioactive beams and increased computational power paired with new
astronomical observations have advanced the present knowledge. This
review summarizes the progress in the field of nuclear astrophysics
with a focus on the role of indirect methods and reactions involving
beams of rare isotopes.
\end{abstract}


\end{frontmatter}

\tableofcontents

\pagebreak
 
  \maketitle

\pagenumbering{roman}

\setcounter{page}{5}

\pagenumbering{arabic} \setcounter{page}{1}

\section{Introduction}
Nuclear reactions in stars and stellar explosions are responsible for the
ongoing synthesis of the
elements~\cite{Burbridge1957,Fowler1984,Kratz1993,Wallerstein1997,Kaeppeler1998,Schatz1998,Arnould1999,Langanke2001,Arnould2003}.
Nuclear physics plays an important role as it determines the signatures of
isotopic and elemental abundances found in the spectra of stars, novae,
supernovae, and X-ray burst as well as in characteristic $\gamma$-ray radiation
from nuclear decays, or in the composition of meteorites and presolar grains
(see \cite{Grawe2007} for a review on the nuclear structure input to nuclear
astrophysics). 

The rapid neutron capture process (r process) is responsible for the
existence of about half of the stable nuclei heavier than iron; yet a site that
can produce the observed elements self-consistently has not been
identified~\cite{Cowan1991,Qian2003}. Capture cross sections for most of the
nuclei involved are hard if not impossible to measure in the laboratory and
indirect experimental approaches have to be employed to gather the relevant
nuclear structure information. Nuclear masses and $\beta$-decay half-lives are
among the few direct observables that are input for calculations that model
nucleosynthesis in the r process. X-ray bursts provide a unique window into
the physics of neutron stars. They are the most frequent thermonuclear
explosions known. The brightness, frequency and opportunity to be observed with
different telescopes makes them unique laboratories for explosive nuclear
burning at extreme temperatures and densities~\cite{Champagne1992,SchatzRehm2006,Fisker2008}. The reaction sequence
during an X-ray burst proceeds through nuclei at or close to the proton drip
line mainly by $(p,\gamma)$ and $(p,\alpha)$ reactions and $\beta$ decays
($\alpha$p and rp process)~\cite{Wallace1981}. Most rp-process reaction rates
are still  based exclusively on theory. Energy in explosive hydrogen burning
events such as 
X-ray bursts is initially generated in the CNO cycle and as the temperature
increases, $\alpha$-capture on unstable oxygen and neon (\nuc{15}{O} and
\nuc{18}{Ne}) leads to a break out and an ensuing chain of proton captures that
can go as far as tin. Supernovae play a crucial role in the understanding of the
universe as they are the major source of nucleosynthesis and possibly of cosmic
rays. Core-collapse supernovae~\cite{Woosley2005} are one of the proposed sites
of the r process. Thermonuclear supernovae (type Ia) are powered by explosive
carbon and oxygen burning of a white dwarf that has reached the Chandrasekhar
mass limit. For both types of supernovae, the driving processes are not well
understood and weak interaction rates play a key
role~\cite{Langanke2003}. Temperatures and densities are so high that electron
captures on unstable nuclei become crucial.    

Most aspects in the study of nuclear physics demand beams of energetic particles
to induce nuclear reactions on the nuclei of target atoms. It was from this need
that accelerators were developed. Over the years, many ways of accelerating charged
particles to ever increasing energies have been devised. Today we
have ion beams of all elements from protons to uranium available at energies
well beyond those needed for the study of atomic nuclei. The quantities used in
nucleosynthesis calculations are reaction rates. A thermonuclear reaction rate
is a function of the density of the interacting nuclei, their relative velocity
and the reaction cross section. Extrapolation procedures are often needed to
derive cross sections in the energy or temperature region of astrophysical
relevance. While non-resonant cross sections can be rather well extrapolated to
the low-energy region, the presence of continuum, or sub-threshold resonances,
can complicate these extrapolations. We will mention some of the important
examples.

In the Sun, the reaction $^{7}$Be$\left( {\rm p},\gamma\right) ^{8}$B
plays a major role for the production of high energy neutrinos from
the $\beta$-decay of $^{8}$B. These neutrinos come directly from the
center of the Sun and are ideal probes of the Sun's structure. {\it
John Bahcall} frequently said that this was the most important
reaction in nuclear astrophysics \cite{John}. Our understanding of 
this reaction has improved considerably with the advent of rare-isotope beam
facilities. The reaction $^{12}$C$\left( \alpha,\gamma\right)
^{16}$O is extremely relevant for the fate of massive stars. It
determines if the remnant of a supernova explosion becomes a
black-hole or a neutron star \cite{Woosley}. These two reactions are
only two examples of a large number of reactions, which are not yet
known with the accuracy needed in astrophysics.

In this review, we summarize recent developments and achievements in nuclear
astrophysics with a focus on theoretical approaches and experimental techniques
that are applicable to or utilize rare-isotope beams, respectively. Section 2
will cover reactions within stars, Section 3 is devoted to nuclear reaction
models, Section 4 reviews the effect of environment electrons, Section 5
outlines approaches with indirect methods and Section 6 summarizes recent
nuclear astrophysics experiments with rare-isotope beams.  Finally, in Section 7
we present our outlook for the present and future of this field.

\section{Reactions within stars}

\subsection{Thermonuclear cross sections and reaction rates}

The nuclear cross section for a reaction between a nuclear target $j$\ and a nuclear projectile
$k$\ is defined by%
\begin{equation}
\sigma=\frac{\mathrm{{number\ of\ reactions\ target^{-1}sec^{-1}}}%
}{\mathrm{flux\ of\ incoming\ projectiles}}={\frac{{r/n_{j}}}{{n_{k}v}}%
},\label{astrophys1}%
\end{equation}
where the target number density is given by $n_{j}$, the projectile number
density is given by $n_{k}$, and $v$ is the relative velocity between target
and projectile nuclei. The number of reactions per unit volume and time
can be expressed as $r=\sigma vn_{j}n_{k}$, or, more generally, by%
\begin{equation}
{\small r}_{j,k}{\small =}\int{\small \sigma|v}_{j}{\small -v}_{k}%
{\small |d}^{3}{\small n}_{j}{\small d}^{3}{\small n}_{k}{\small .}%
\label{astrophys2}%
\end{equation}

The evaluation of this integral depends on the type of particles and
their distributions. For nuclei $j$\ and $k$\ in an astrophysical
plasma, obeying a Maxwell-Boltzmann distribution (MB),%
\begin{equation}
d^{3}n_{j}=n_{j}\left({\frac{{m_{j}}}{{2\pi kT}}}\right)^{3/2}\exp\left(-{\frac
{{m_{j}^{2}v}_{j}}{{2kT}}}\right)d^{3}{v}_{j},\label{astrophys3}%
\end{equation}
Eq. \eqref{astrophys2} simplifies to $r_{j,k}=\left<\sigma
v\right>n_{j}n_{k}$, where the reaction rate $\left<\sigma v\right>$\ is the average of $\sigma v$\
over the temperature distribution in \eqref{astrophys3}.\ More
specifically,
\begin{align}
r_{j,k}  & =\left<\sigma\mathrm{v}\right>_{j,k}n_{j}n_{k}\label{astrophys4}\\
 \left<\sigma\mathrm{v}\right>_{j,k}&=\left({\frac{8}{{m_{jk}\pi}}}\right)^{1/2}%
(kT)^{-3/2}\int_{0}^{\infty}E\sigma(E)\exp\left(-{E\over kT}\right)dE.\label{astrophys5}%
\end{align}
Here $m_{jk}$\ denotes the reduced mass of the target-projectile system.

\subsubsection{Photons}

When in Eq. \eqref{astrophys2} particle $k$\ is a photon, the
relative velocity is always $c$\ and there is no need to integrate quantities
over $d^{3}n_{j}$. Thus, one obtains
$r_{j}=\lambda_{j,\gamma}n_{j}$\ where $\lambda_{j,\gamma}$\ results
from an integration of the photodisintegration
cross section over a Planck distribution for photons of temperature $T$%
\begin{equation}
d^{3}n_{\gamma}   ={\frac{{E_{\gamma}^{2}}}{{\pi^{2}(c\hbar)^{3}}}}{1 \over {\exp(E_{\gamma}/kT)-1}}dE_{\gamma},\label{astrophys6}
\end{equation}
which leads to
\begin{equation}
r_{j}   =\lambda_{j,\gamma}(T)n_{j}=\frac{1}{{\pi^{2}(c\hbar)^{3}}}{{\int
d^{3}n_{j}}}\int_{0}^{\infty}{\frac{{c\sigma(E_{\gamma})E_{\gamma}^{2}}%
}{{\mathrm{exp}(E_{\gamma}/kT)-1}}}dE_{\gamma}.\label{astrophys7}%
\end{equation}

There is, however, no direct need to evaluate photodisintegration cross
sections, because, due to detailed balance, they can be expressed by the
capture cross sections for the inverse reaction $l+m\rightarrow j+\gamma
$\ \cite{FCZ67}
\begin{equation}
\lambda_{j,\gamma}(T)=\left({\frac{{\xi_{l}\xi_{m}}}{\xi_{j}}}\right)\left({\frac{{A_{l}A_{m}}%
}{A_{j}}}\right)^{3/2}\left({\frac{{m_{u}kT}}{{2\pi\hbar^{2}}}}\right)^{3/2}
 \left<\sigma\mathrm{v}\right>_{l,m}
\mathrm{exp}%
\left(-{Q_{lm}\over kT}\right),\label{astrophys8}%
\end{equation}
where $m_u=m_{^{12}C}/12$ is the mass unit, $Q_{lm}$ is the reaction $Q$-value,  $T$ is the temperature, $ \left<\sigma\mathrm{v}\right>_{j,k}$
is the inverse reaction rate,  $\xi (T)=\sum
_{i}(2J_{i}+1)\exp(-E_{i}/kT)$ are partition functions, and $A$ are the mass numbers  of the participating
nuclei in a thermal bath of temperature $T$.

\subsubsection{Electron, positron and neutrino capture}

The electron is about 2000 times less
massive than a nucleon. Thus, the velocity of the nucleus $j$\ is
negligible in the center of mass system in comparison to the
electron velocity ($|v_{j}-v_{e}|\approx|v_{e}|$), and there is no need to integrate quantities
over $d^{3}n_{j}$. The electron
capture cross section has to be integrated either over a Boltzmann or a Fermi distribution of electrons,
depending on the astrophysical scenario. The electron capture
rates are a function of $T$\ and the electron number density, $n_{e}=Y_{e}\rho N_{A}$ 
\cite{FFN85}. In a completely
ionized plasma, $Y_{e}=\sum_{i}Z_{i}Y_{i}$, i.e.,  the electron abundance is equal to the total
proton abundance in nuclei. Here $Y_{i}%
$\ denotes the abundance of nucleus $i$\ defined by $Y_{i}=n_{i}/(\rho N_{A}%
)$, where $n_{i}$\ is the number density of nuclei per unit volume and $N_{A}%
$\ is Avogadro's number.
Therefore, 
\begin{equation}
{\small r}_{j}{\small =\lambda}_{j,e}{\small (T,\rho Y}_{e}{\small )n}%
_{j}{\small .}
\label{astrophys9}
\end{equation}

This treatment can be generalized for the capture of positrons, which are in a
thermal equilibrium with photons, electrons, and nuclei. At high densities
($\rho>10^{12}$ g/cm$^{3}$) the neutrino scattering cross sections
on nuclei and electrons are large enough  to
thermalize the neutrino distribution. Inverse electron
(neutrino) capture can also occur and the neutrino capture rate can be
expressed similarly to Eqs. \eqref{astrophys7} or \eqref{astrophys9},
integrating over the neutrino distribution. 

\subsubsection{Beta-decay}
For normal decays, like $\beta$
or $\alpha$ decays with half-life $\tau_{1/2}$, we obtain an equation similar to
Eqs. \eqref{astrophys7} or \eqref{astrophys9} with a decay constant
$\lambda_{j}=\ln2/\tau_{1/2}$\ and
\begin{equation}
{r}_{j}=\lambda_{j}{n}_{j}{.}\label{astrophys10}%
\end{equation}

\subsubsection{Charged particles}

Nuclear cross sections for charged particles are suppressed at low
energies due to the Coulomb barrier. The product of the penetration factor and
the Maxwell-Boltzmann (MB) distribution at a given temperature yields an energy 
window in which most of the reactions occur, known as the
Gamow window.

Experimentally, it is more convenient to work with the astrophysical
$S$\ factor
\begin{equation}
S(E)=E\sigma(E)\exp(2\pi\eta),\label{astrophys13a}%
\end{equation}
with $\eta$\ being the Sommerfeld parameter, describing the s-wave
barrier penetration, $\eta=Z_{j}Z_{k}e^{2}/\hbar v$, and energy $E$ and $v$ the relative velocity of the ions. In this case,
the steep increase of the cross section is transformed into a rather
flat, energy dependent function. One can
easily see the two contributions of the velocity distribution and
the penetrability in the integral
\begin{equation}
\left< \sigma\mathrm{v} \right>_{jk}=\left(  \frac{8}{\pi m_{jk}}\right)  ^{1/2}\frac{1}{\left(
kT\right)  ^{3/2}}\int_{0}^{\infty}S(E)\exp\left[  -\frac{E}{kT}-\frac
{b}{E^{1/2}}\right]  .\label{astrophys13}%
\end{equation}
where $b=2\pi\eta E^{1/2}=(2m_{jk})^{1/2}\pi
e^{2}Z_{j}Z_{k}/\hbar $ and $m_{jk}$ is the reduced mass in units of $m_u$ (unit atomic mass).
Experimentally it is very difficult to perform direct measurements of
fusion reactions involving charged particles at very small
energies. The experimental data at higher energies can be guided by a
theoretical 
model for the cross section, which can then be extrapolated down to the
Gamow energy. However, the extrapolation can be inadequate due to the presence
of resonances and subthreshold resonances, for example.

\begin{figure}[tb]
\begin{center}
\includegraphics[
width=5.5in]
{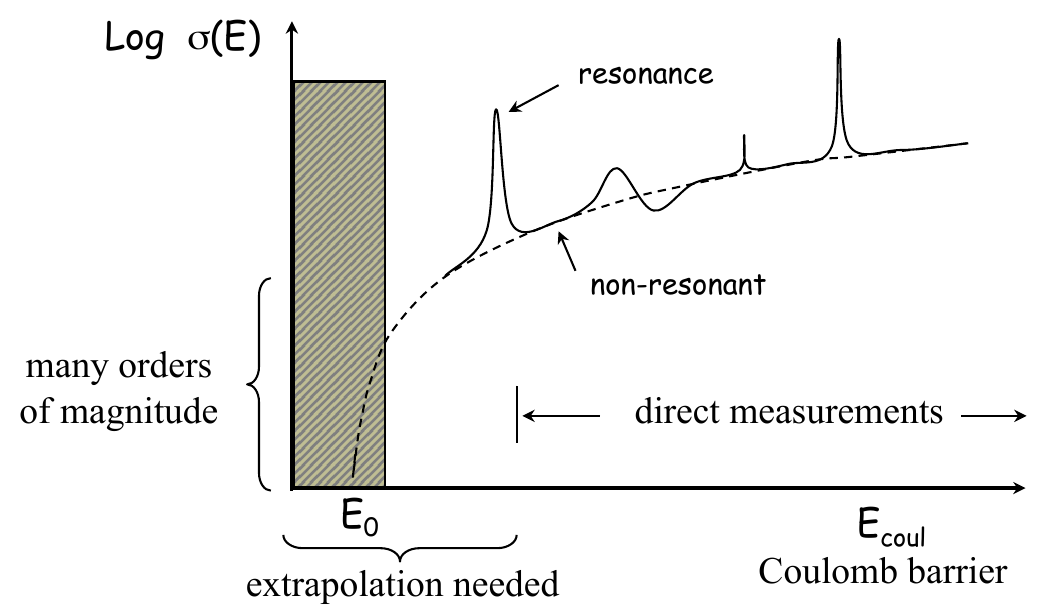}
\caption{Schematic representation of the energy dependence of a fusion
reaction involving charged particles (Courtesy of C. Spitaleri).}  
\label{cross_sec}
\end{center}
\end{figure}

A simple result can be obtained by assuming a constant $S$-factor, i.e., $S(E)=SE_0$.
In this case, the first derivative of the integrand in Eq.
\eqref{astrophys13} yields the location $E_{0}$\ of the Gamow
peak, and the effective width $\Delta$\ of the energy window, i.e.
\begin{align}
E_{0}  & =\left(  \frac{bkT}{2}\right)  ^{2/3}=1.22(Z_{j}^{2}Z_{k}^{2}%
m_{jk}T_{6}^{2})^{1/3}\,\mathrm{keV},\nonumber\\
\Delta & =\left( \frac{16E_{0}kT}{3}\right)^{1/2}=0.749(Z_{j}^{2}Z_{k}^{2}m_{jk}T_{6}^{5}%
)^{1/6}\,\mathrm{keV},\label{astrophys14}%
\end{align}
carrying the dependence on the charges $Z_{j}$,
$Z_{k}$, the reduced mass $m_{jk}$\ of the involved nuclei in units of $m_{u}$, and
the temperature $T_{6}$\ given in 10$^{6}$\ K.

Figure \ref{cross_sec} outlines one of the main challenges in astrophysical reactions
with charged particles. The experimental data can be guided by a theoretical model for the cross
section, which can then be extrapolated to the Gamow energy.  The solid
curve is a theoretical prediction, which supposedly describes the data at
high energies. Its extrapolation to lower energies yields the desired value
of the $S$-factor, or cross section, at the  Gamow energy $E_0$. The
extrapolation can be complicated by the presence of unknown resonances.

\subsubsection{Neutron-induced reactions}

For neutron-induced reactions, the effective energy window for s-wave neutrons 
($l=0$) is given by the location and width of the peak of the MB
distribution function. For $l>0$,  the penetrability of the
centrifugal barrier shifts the effective energy $E_{0}$\ to higher values.
For neutrons with energies less than the height of the centrifugal barrier
one gets \cite{wagoner}
\begin{equation}
E_{0}\approx0.172T_{9}\left(  l+{\frac{1}{2}}\right)  \quad\mathrm{MeV,}%
\ \ \ \ \ \ \ \ \ \ \Delta\approx0.194T_{9}\left(  l+{\frac{1}{2}}\right)
^{1/2}\,\mathrm{MeV},
\end{equation}
Usually, $E_{0}$\ is not much different (in magnitude) from the neutron
separation energy.

\subsection{Reaction networks}

The time evolution of the number densities, $n_i$, of each of the species $i$ in an
astrophysical plasma (at constant density) is obtained by solving equations of the type
\begin{equation}
\left(\frac{{\partial n_{i}}}{{\partial t}}\right)_{\rho=const}%
=\sum_{j} N_{j}^{i} r_{j} +\sum
_{j,k} N_{j,k}^{i} r_{j,k} +\sum_{j,k,l}%
 N_{j,k,l}^{i} r_{j,k,l},\label{astrophys11}%
\end{equation}
where the  
$N^{i}_x$\ can be positive or negative numbers that specify how many
particles of species $i$\ are created or destroyed in a reaction $x$. 
The reactions $x$ fall in three categories:
\begin{enumerate}
\item  decays,
photodisintegrations, electron and positron captures and neutrino-induced
reactions, $r_{j}=\lambda_{j}n_{j}$,  
\item two-particle
reactions, $r_{j,k}= \left<\sigma\mathrm{v}\right>_{j,k}n_{j}n_{k}$, and 
\item three-particle
reactions, $r_{j,k,l}= \left<\sigma\mathrm{v}\right>_{j,l,k}n_{j}n_{k}n_{l}$, like the
triple-$\alpha$ process $\left( \alpha+\alpha+\alpha\right.$
$\longrightarrow$ $\left. ^{12}{\rm C}+\gamma\right)  $. 
\end{enumerate}

The $N^{i}$'s
are given by: 
$$N_{j}^{i}=N_{i}, \ \ \  N_{j,k}^{i}={N_{i}\over \prod_{m=1}^{n_{m}
}|N_{j_{m}}|!}, \ \ \ {\rm and} \ \  N_{j,k,l}^{i}={N_{i}\over \prod_{m=1}^{n_{m}}|N_{j_{m}}|!},$$ 
where
the products in the denominators run over the $n_{m}$\ different species
destroyed in the reaction and avoid double counting 
when identical particles react with each other. 

\begin{figure}[tb]
\begin{center}
\includegraphics[width=5.5in]{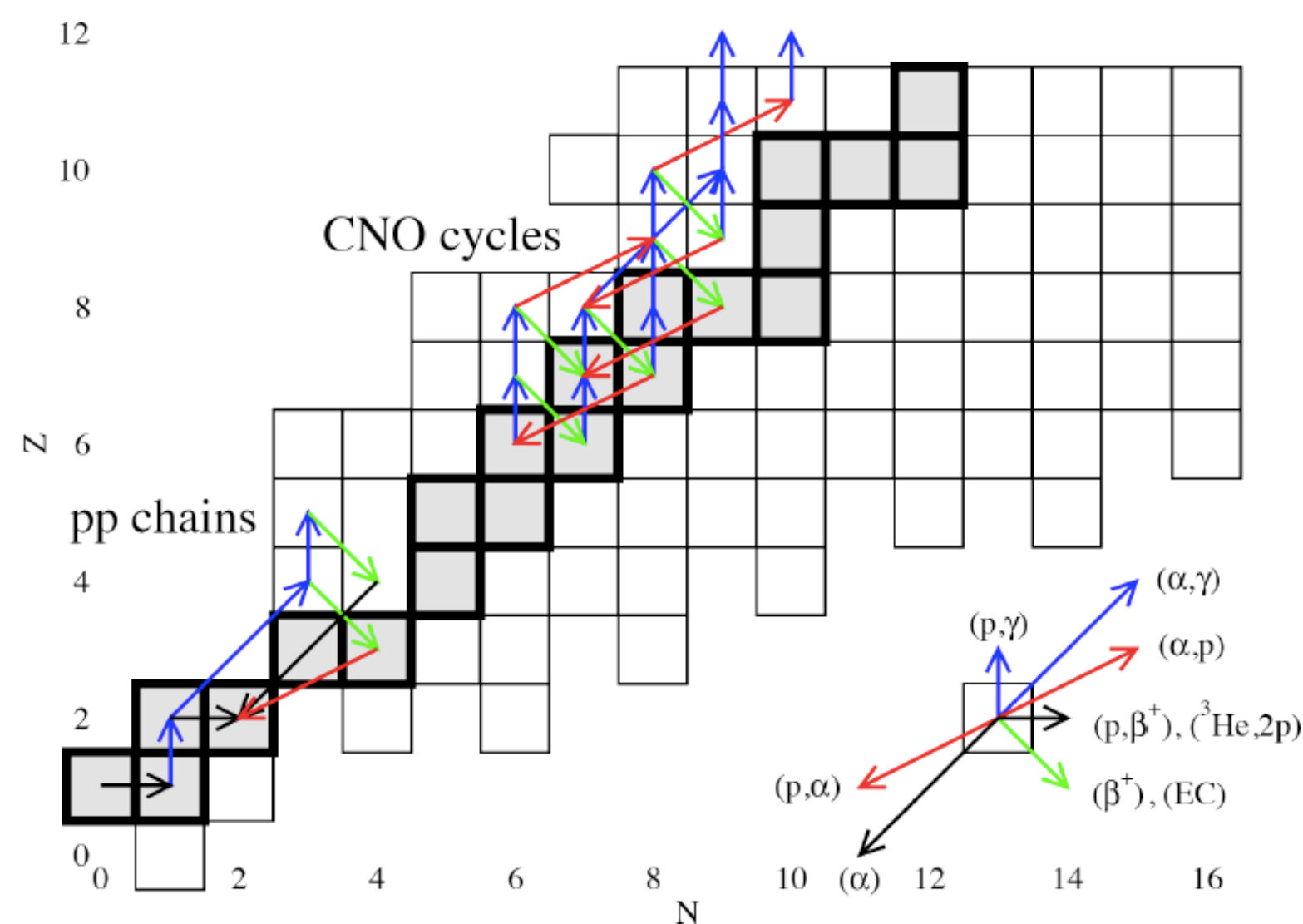}
\caption{Example of reaction networks (pp-chains and CNO-cycles).  A particular nucleus on 
the Segr\`e chart can take different paths along the reaction network, as shown
in the inset at the lower right side. (Courtesy of S.\ Typel) }  
\label{network}
\end{center}
\end{figure}

In terms of the nuclear abundances, $Y_{i}=n_{i}/(\rho
N_{A})$ such that for a nucleus with atomic weight $A_{i}$,
$A_{i}Y_{i}$\ represents the mass fraction of this nucleus,  $\sum
A_{i}Y_{i}=1$ and the reaction network equations can be rewritten as
\begin{equation}
{d {Y}_{i}\over dt}=\sum_{j}N_{j}^{i}\lambda_{j}Y_{j}+\sum_{j,k}N_{j,k}^{i}\rho
N_{A}\left<\sigma\mathrm{v}\right>_{j,k}Y_{j}Y_{k}
+\sum_{j,k,l}N_{j,k,l}^{i}\rho^{2}N_{A}^{2}\left<\sigma\mathrm{v}\right>_{j,l,k}Y_{j}%
Y_{k}Y_{l}.\label{astrophys12}%
\end{equation}

The energy generation per unit volume in a time interval $\Delta t$ is expressed in terms of the 
mass excess $\Delta M_{i}c^{2}$\ of
the participating nuclei
\begin{equation}
\Delta\epsilon=-\sum_{i} \Delta Y_{i} N_{A}%
\Delta M_{i} c^{2} ,\;\;\;\;\;\;\;\;\;{d{\epsilon}\over dt}%
=-\sum_{i} {d{Y}_{i}\over dt}N_{A} \Delta M_{i} c%
^{2} .\label{astrophys15}%
\end{equation}

The solution of the above group of equations allows to deduce the
path for the r process until the heavier elements are reached. 
The relative abundances of elements are also
obtained theoretically by means of these equations by using stellar
models for the initial conditions, as the neutron density and the
temperature. 
Nuclear physics has to contribute with  $\beta$-decay
half-lives, electron and positron capture rates, photo-nuclear and neutrino cross sections. 

Simple examples of
reaction networks are shown in Figure \ref{network} for typical pp-chains and CNO-cycles. 
On the right we show a particular nucleus on the Segr\`e chart
from where  different paths can start along the reaction network. 

\section{Nuclear reaction models}

Explosive nuclear burning in astrophysical environments produces short-lived,
exotic nuclei, which again can be targets for subsequent reactions. In addition,
it 
involves a very large number of stable nuclei, which are still not fully
explored by 
experiments. Thus, it is necessary to be able to predict reaction cross
sections and thermonuclear rates with the aid of theoretical models.
Especially during the hydrostatic burning stages of stars, charged-particle
induced reactions proceed at such low energies that a direct cross-section
measurement is often not possible with existing experimental techniques. Hence
extrapolations down to the stellar energies of the cross sections measured at
the lowest possible energies in the laboratory are usually applied. To be
trustworthy, such extrapolations should have as strong of a 
theoretical foundation as possible. Theory is even more mandatory when excited
nuclei are involved in the entrance channel, or when unstable, very
neutron-rich or neutron-deficient nuclides (many of them being even impossible
to produce with present-day experimental techniques) have to be considered.
Such situations are often encountered in the modelling of explosive
astrophysical scenarios.

\subsection{Potential and DWBA models} 

Potential models
assume that the physically important degrees
of freedom are the relative motion between structureless nuclei in the
entrance and exit channels. The only microscopic information is introduced in terms of
spectroscopic factors
and parameters of the optical potential. The weakness of the models is 
that the nucleus-nucleus potentials adopted for calculating the initial and 
final wavefunctions from the Schr\"{o}dinger equation cannot be unambiguously
defined.
Single-particle wavefunctions  are calculated using nuclear potentials of the form%
\begin{equation}
V(\mathbf{r})=V_{0}(r)+V_{S}(r)\ (\mathbf{l.s})+V_{C}(r) \label{WStot}%
\end{equation}
where $V_{0}(r)$ and $V_{S}(r)$ are the central and spin-orbit interactions,
respectively, and $V_{C}(r)$ is the Coulomb potential of a uniform
distribution of charges. The potentials
$V_{0}(r)$ and $V_{S}(r)$, are usually given in terms of a Woods-Saxon (WS) parameterization.
The parameters of the potentials -- their depth, range and diffuseness,  are chosen to reproduce 
the ground state energy $E_B$ (or the energy of an excited state).  For knockout reactions, they are 
also adjusted to reproduce the orbital radius of the nucleon. Most often, the same parameters
do not reproduce the proper continuum wavefunctions, do not yield location and widths of resonances, etc.
These can be obtained by readjusting the strengths of the potentials, effectively increasing the number of
parameters at hand.  

The WS parameterization is well
suited to describe any reaction of interest, except perhaps for
those cases in which one of the partners is a neutron-rich halo
nucleus. Then, the extended radial dependence leads to
unusual forms for the potentials. Also, for capture reactions in
which the light partner is either a deuteron, triton, $\alpha
$-particle or a heavier nucleus, folding models are more
appropriate. 
The central part of the potential is obtained by a folding of an effective interaction 
with the ground state densities, $\rho_A$
and $\rho_B$, of the nuclei $A$ and $B$:
\begin{equation}
V(\mathbf{r})=\lambda_0\int d^3r_1 d^3 r_2 \rho_A({\bf r}_1) \rho_B({\bf r}_2) v_{eff}({\bf s}) \label{Veff},
\end{equation}
with ${\bf s}= |{\bf r} + {\bf r}_2 - {\bf r}_1|$. $\lambda_0$ is a normalization factor which is close to unity. 

Folding models are based on an effective
nucleon-nucleon interaction, $v_{eff}$, and nuclear densities, $\rho_i$, which are either
obtained experimentally (not really, because only charge densities
can be accurately determined from electron scattering), or
calculated from some microscopic model (typically Hartree-Fock or
relativistic mean field models). The effective interactions as well
as the nuclear densities are subject of intensive theoretical
studies.

Potential models have been applied to all kinds of calculations for nuclear astrophysics. For simplicity, let us 
consider radiative capture reactions  involving a target nucleus and a nucleon. The wavefunctions for the nucleon 
($n$) + nucleus ($x$) system are calculated by
solving the radial Schr\"{o}dinger equation%
\begin{equation}
-\frac{\hbar^{2}}{2m_{nx}}\left[  \frac{d^{2}}{dr^{2}}-\frac{l\left(
l+1\right)  }{r^{2}}\right]  u_{\alpha}\left(  r\right)  + V (r) u_{\alpha
}\left(  r\right)  =E_{\alpha}u_{\alpha}\left(  r\right)  \ . \label{bss}%
\end{equation}
The nucleon $n$, the nucleus $x$, and the $n+x=a$--system have
intrinsic spins labeled by $s=1/2$, $I_{x}$ and $J$, respectively.
The orbital angular momentum for the relative motion of $n+x$ is
described by $l$. Angular momenta are usually coupled as
$\mathbf{l+s}\mathbf{=j}$ and $\mathbf{j+I}_{x}\mathbf{=J}$, where
$\mathbf{J}$ is called the channel spin. In      Eq. \eqref{WStot}, for $V$
one uses $\mathbf{s.l} =\left[ j(j+1)-l(l+1)-3/4\right]  /2$ and
$\alpha$ in      Eq. \eqref{bss} denotes the set of quantum numbers,
$\alpha_{b}=\{E_{b},l_{b},j_{b},J_{b}\}$ for the bound state, and
$\alpha_{c}=\{E_{c},l_{c},j_{c},J_{c}\}$ for the continuum states.

The bound-state wavefunctions are normalized to unity, $\int dr \left\vert
u_{\alpha_{b}}\left(  r\right)  \right\vert ^{2}=1$, whereas the continuum
wavefunctions have boundary conditions at large distances given by
\begin{equation}
u_{\alpha_{c}}(r\rightarrow\infty)=i\sqrt{\frac{m_{nx}}{2\pi k\hbar^{2}}%
}\left[  H_{l}^{(-)}\left(  r\right)  -S_{\alpha_{c}}H_{l}^{(+)}\left(
r\right)  \right]  \ e^{i\sigma_{l}\left(  E\right)  } \label{uE}%
\end{equation}
where $S_{\alpha_{c}}=\exp\left[  2i\delta_{\alpha_{c}}\left(
E\right) \right]  $, with $\delta_{\alpha_{c}}\left(  E\right)  $
and $\sigma _{l}\left(  E\right)  $ being the nuclear and the
Coulomb phase shifts, respectively. In      Eq. \eqref{uE},
$H_{l}^{(\pm)}\left(  r\right)  =G_{l}(r)\pm iF_{l}\left(  r\right)
$, where $F_{l}$ and $G_{l}$ are the regular and irregular Coulomb
wavefunctions. For neutrons, the Coulomb functions reduce to the
usual spherical Bessel functions, $j_{l}\left(  r\right)  $ and
$n_{l}\left(  r\right)  $. With these definitions, the continuum
wavefunctions are normalized as $\left\langle
u_{E_{c}^{\prime}}|u_{E_{c}} \right\rangle =\delta\left(
E_{c}^{\prime}-E_{c}\right)  \delta_{\alpha\alpha^{\prime}}. $

\begin{figure}[tb]
\begin{center}
\includegraphics[
width=4.3in]
{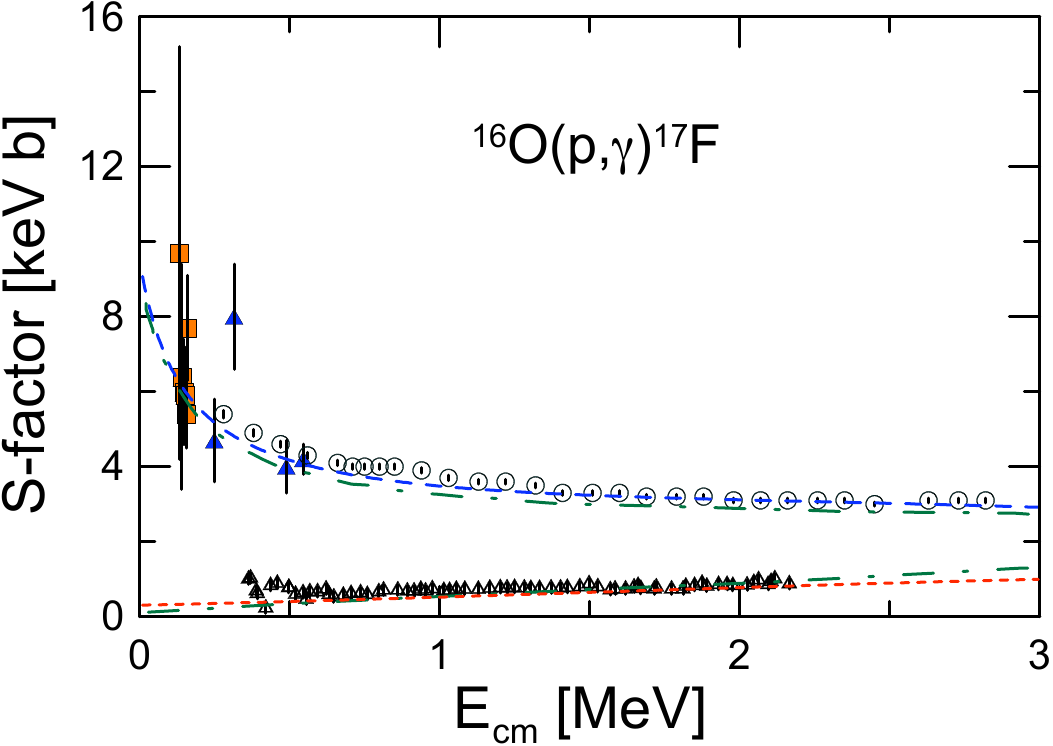}
\caption{Potential model calculation \cite{Jun08} for
the reaction $^{16}$O(p,$\gamma )^{17}$F. The dotted line and the dashed
line are for the capture to the ground state and to the first
excited state respectively. The experimental data are from
Refs. \cite{Rolfs1973,Hes58,Tan59}. The dotted-dashed lines are the result of shell
model calculations published in Ref. \cite{Mor97}.}  
\label{17Op}
\end{center}
\end{figure}

For $n+x\rightarrow a+\gamma$
and $\pi L$ ($\pi=E,(M)=~$electric (magnetic) $L$-pole) transitions,
the cross sections are obtained from
\begin{align}
\sigma_{\pi L,J_{b}}^{\text{rad.cap.}}  & =\frac{(2\pi)^{3}}{k^{2}}\left(
\frac{E_{nx}+E_{b}}{\hbar c}\right)  ^{2L+1}\frac{2(2I_{a}+1)}{(2I_{n}%
+1)(2I_{x}+1)}
\frac{L+1}{L[(2L+1)!!]^{2}}\sum_{J_{c}j_{c}l_{c}}(2J_{c}+1)\nonumber\\
&  \times\left\{
\begin{array}
[c]{ccc}%
j_{c} & J_{c} & I_{x}\\
J_{b} & j_{b} & L
\end{array}
\right\}  ^{2}\ \left\vert \left\langle l_cj_c\left\Vert
\mathcal{O}_{\pi L}\right\Vert l_{b}j_{b} \right\rangle
\right\vert^{2},
\label{respf}%
\end{align}
where $E_{b}$ is the binding energy and $\left\langle
l_cj_c\left\Vert \mathcal{O}_{\pi L}\right\Vert l_{b}j_{b}
\right\rangle$ is the multipole matrix element. For electric multipole transitions,
\begin{align}
\left\langle l_cj_c\left\Vert
\mathcal{O}_{EL}\right\Vert l_{b} j_{b}\right\rangle
&=(-1)^{l_b+l_c-j_c+L-1/2}\frac{e_{L}}{\sqrt{4\pi}}
 \sqrt{(2L+1)(2j_b+1)}
\left(
\begin{array}
[c]{ccc}%
j_{b} & L & j_{c}\\
1/2 & 0 & -1/2
\end{array}
\right)
 \nonumber \\
&\times \int_{0}^{\infty}dr \
r^{L}u_{b}(r)u_{c}(r)
,\label{lol0}%
\end{align}
where  $e_{L}$ is the effective charge, which takes into account the
displacement of the center-of-mass,
$
e_{L}=Z_{n}e\left(  -{m_{n}}/{m_{a}}\right)  ^{L}+Z_{x}e\left(
{m_{x}}/{m_{a}}\right)  ^{L}$.
In comparison with electric dipole transitions, the cross
sections for magnetic dipole transitions are reduced by a factor of
$v^{2}/c^{2}$, where $v$ is the relative velocity of the $n+x$
system. At very low energies, $v\ll c$, $M1$ transitions will be
much smaller than the electric transitions. Only in the case of
sharp resonances, the $M1$ transitions play a significant role.

The total radiative capture cross section is obtained by adding all
multipolarities and final spins of the bound state ($E\equiv E_{nx}$),
\begin{equation}
\sigma^{\text{rad.cap.}} (E)=\sum_{L,J_{b}} (SF)_{J_{b}}\ \sigma^{\text{d.c.}%
}_{L,J_{b}}(E) \ , \label{SFS}%
\end{equation}
where $(SF)_{J_{b}}$ are spectroscopic factors.

As an example, Figure \ref{17Op} shows a potential model calculation \cite{Jun08} for the 
$S$-factor of the $^{16}\text{O}(\text{p},\gamma)^{17}\text{F}$ reaction. The
rate of this reaction influences sensitively the $^{17}$O/$^{16}$O isotopic ratio predicted by
models of massive ($\ge 4M_\odot$) AGB stars, where proton captures
occur at the base of the convective envelope (hot bottom burning).
A fine-tuning of the $^{16}\text{O}(\text{p},\gamma)^{17}\text{F}$  reaction rate may account
for the measured anomalous $^{17}$O/$^{16}$O abundance ratio in small grains
which are formed by the condensation of the material ejected from the surface of
AGB stars via strong stellar winds \cite{Hab04}.
The agreement of the potential model calculation with the experimental data seen in
Figure \ref{17Op} is  very good and comparable to more elaborated calculations
\cite{Mor97}. 

\subsection{Microscopic models} 
In microscopic models, 
nucleons are grouped into clusters and the
completely antisymmetrized relative wavefunctions between the
various clusters are determined by solving the Schr\"{o}dinger
equation for a many-body Hamiltonian with an effective
nucleon-nucleon interaction. 
Typical cluster models are based on 
the Resonating Group Method (RGM) or the Generator Coordinate Method
(GCM). They are based on a set of coupled integro-differential equations of
the form
\begin{equation}
\sum_{\alpha'} \int d^3 r'
\left[
{\mathcal H}^{AB}_{\alpha\alpha'}({\bf r,r'})-E{\mathcal N}^{AB}_{\alpha\alpha'}({\bf r,r'})
\right]
g_{\alpha'}({\bf r'})=0,\label{RGM}
\end{equation}
where 
\[{\mathcal H}^{AB}_{\alpha\alpha'}({\bf r,r'})=\langle \Psi_A(\alpha,{\bf
r})|H| \Psi_B(\alpha',{\bf r'}) \rangle \ \ \ {\rm and} \ \ \ \ 
{\mathcal N}^{AB}_{\alpha\alpha'}({\bf r,r'}) =\langle \Psi_A(\alpha,{\bf r})|
\Psi_B(\alpha',{\bf r'}) \rangle .\] 
In these equations $H$ is the
Hamiltonian for the system of two nuclei (A and B) with the energy
$E$, $\Psi_{A,B}$ is the wavefunction of nucleus A (and B), and
$g_{\alpha}({\bf r})$ is a function to be found by numerical
solution of Eq. \eqref{RGM}, which describes the relative motion of A
and B in channel $\alpha$. Full antisymmetrization between nucleons
of A and B are implicit. 

\begin{figure}[tb]
\begin{center}
\includegraphics[
width=4.3in]
{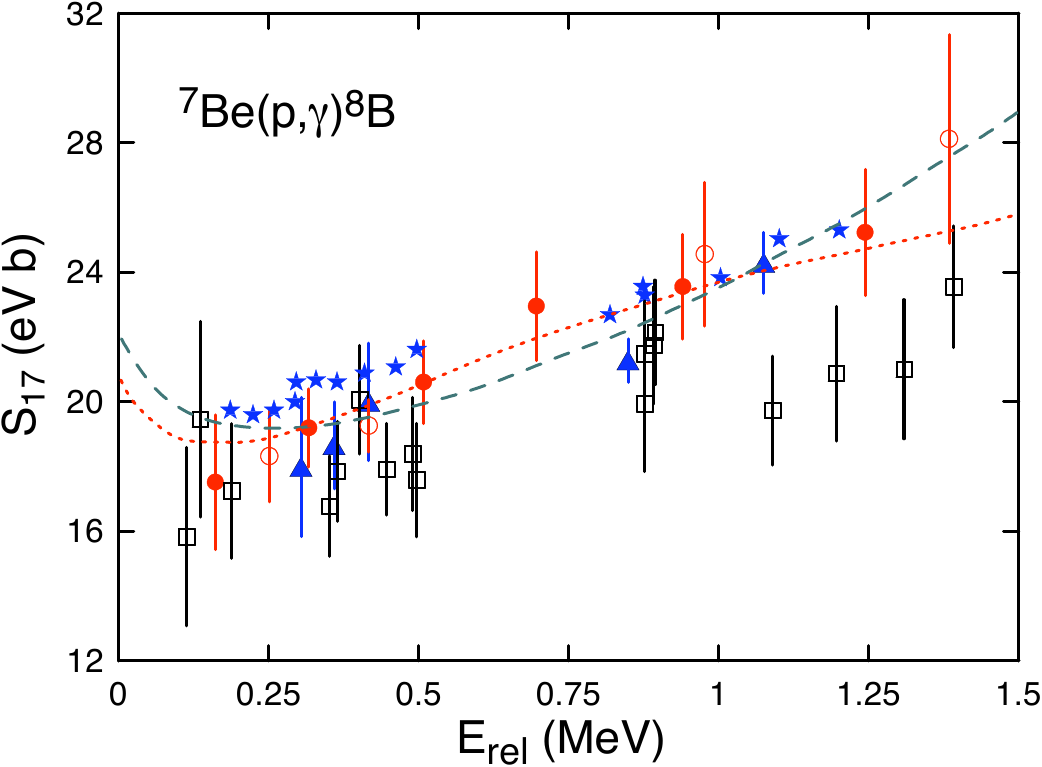}
\caption{Microscopic calculations for the reaction $^{7}$\textrm{Be}%
$(p,\gamma)^{8}\text{B}$. The dashed line is the no-core shell-model calculation of Ref. \cite{Nav06}
and the dotted line is from the resonant group method calculation of Ref. \cite{DB94}. Experimental data are from
Refs. \cite{VCK70,FED83,Baby03,Junghans2003,Iwasa1999,Davids2001a,Schumann2003,Kav69}.}  
\label{7Bep}
\end{center}
\end{figure}

Modern nuclear shell-model calculations,
such as the Monte-Carlo shell model, or the no-core shell model, 
are able to provide the wavefunctions $\Psi_{A,B}$ for light nuclei. 
But so far they cannot describe scattering wavefunctions with a full account of
anti-symmetrization. Moreover, the road to an effective $NN$ interaction
which can simultaneously describe bound and continuum states has not been an
easy one.  Thus, methods based on Eq. \eqref{RGM} seem to be the best way to
obtain scattering wavefunctions needed for astrophysical purposes. Old interactions, 
such as Volkov interactions, are still used for practical purposes.
It is also worth mentioning that this
approach has provided  the best
description of bound, resonant, and
scattering states of nuclear systems \cite{Descouvement08}.

As an example of applications of this method, we again give a radiative capture reaction.
The creation and destruction of $^7$Be in astrophysical environments is
essential for the description of several stellar 
and cosmological processes and is not well understood. $^8$B also plays an
essential role in understanding our Sun. 
High energy $\nu_e$ neutrinos produced by $^8$B decay in
the Sun oscillate into other active species on their way to earth
\cite{Ahm01}. Precise predictions of the production 
rate of $^8$B solar neutrinos are important for testing solar
models, and for limiting the allowed neutrino mixing parameters.
The  most uncertain reaction leading to $^8$B formation in the Sun is the
$^{7}\text{Be}(\text{p},\gamma)^{8}\text{B}$ 
radiative capture reaction \cite{Junghans2003}. Additionally, the Coulomb
dissociation method, discussed later in this review, 
has given some new insights about the electromagnetic matrix elements for this
reaction. Figure  \ref{7Bep} shows 
a comparison of microscopic calculations for the reaction $^{7}$\textrm{Be}%
$(p,\gamma)^{8}\text{B}$ with experimental data. The dashed-dotted line is the
no-core shell-model calculation of Ref. \cite{Nav06} 
and the dotted line is for the resonant group method calculation of
Ref. \cite{DB94}. Experimental data are from 
Refs. \cite{VCK70,FED83,Baby03,Junghans2003,Iwasa1999,Davids2001a,Schumann2003,Kav69}.
It is evident that both, theory and experiment, need improvements for this
important reaction. 

\subsubsection{ Asymptotic normalization coefficients}
Although the potential model works well for many nuclear reactions of interest
in astrophysics, it is often necessary to pursue a more microscopic approach
to reproduce experimental data. Instead of the single-particle wavefunctions one often makes use of overlap
integrals, $I_{b}(\mathbf{r})$, and a many-body wavefunction for the relative
motion, $\Psi_{c}(\mathbf{r})$. Both $I_{b}(\mathbf{r})$ and $\Psi
_{c}(\mathbf{r})$ might be very complicated to calculate, depending on how
elaborate the microscopic model is. The variable $\mathbf{r}$ is the relative
coordinate between the nucleon and the nucleus $x$, with all the intrinsic
coordinates of the nucleons in $x$ being integrated out. The radiative capture
cross sections are obtained from the calculation of $\sigma_{L,J_{b}%
}^{\text{rad.cap.}} \propto|\left<  I_{b}(r)||r^{L}Y_{L}|| \Psi_{c}(r)\right>
|^{2}$. 

\begin{figure}[tb]
\begin{center}
\includegraphics[
width=4.5in]
{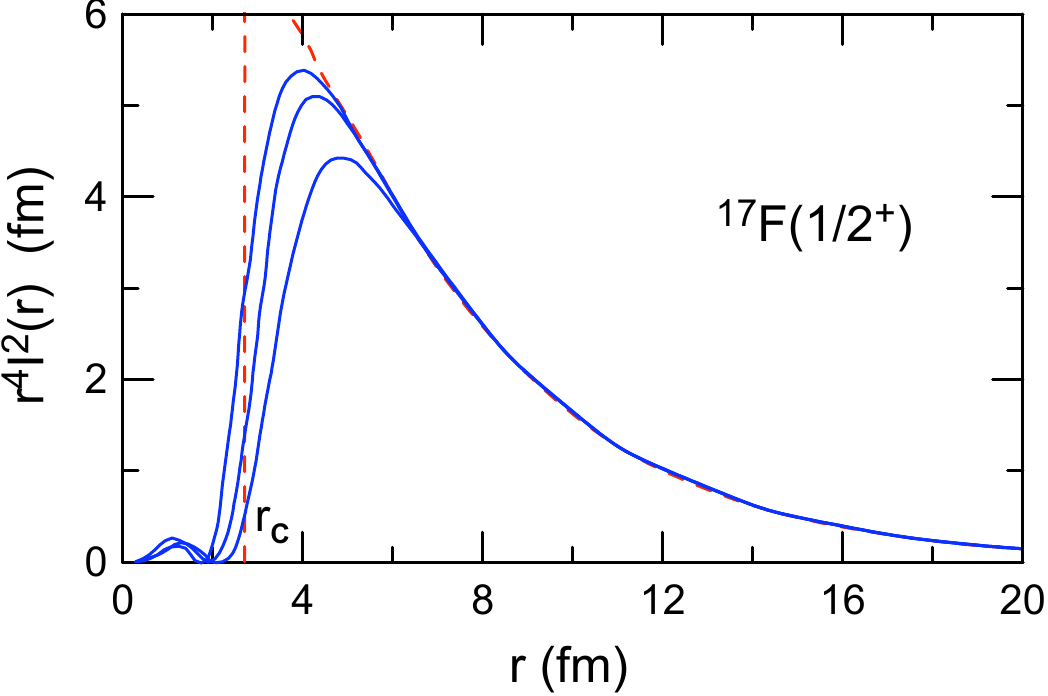}
\caption{Comparison of various radial overlap integrals
$r^4 I^2(r)$ for $^{17}F^*(1/2^+)$  with the normalized
Whittaker function (dashed curve). Most of the contribution
to the rms radius comes from the region outside the core, with radius $r_c$.}  
\label{17Op2}
\end{center}
\end{figure}

The imprints of many-body effects will eventually disappear at large distances
between the nucleon and the nucleus. One thus expects that the overlap
function asymptotically matches the solution of the Schr\"odinger equation
\eqref{bss}, with $V=V_{C}$ for protons and $V=0$ for neutrons. That is, when
$r\rightarrow\infty$,
\begin{align}
I_{b}(r)  &  =C_{1} \frac{W_{-\eta,l_{b}+1/2}(2\kappa r)}{r},
\ \ \ \text{for \ protons}\nonumber\\
&  =C_{2} \sqrt{\frac{2\kappa}{r}}K_{l_{b}+1/2}(\kappa r), \ \ \ \text{for
\ neutrons} \label{whitt}%
\end{align}
where the binding energy of the $n+x$ system is related to $\kappa$ by means
of $E_{b}=\hbar^{2}\kappa^{2}/2m_{nx}$, $W_{p,q}$ is the Whittaker function
and $K_{\mu}$ is the modified Bessel function. In      Eq. \eqref{whitt}, $C_{i}$ is
the asymptotic normalization coefficient (ANC). In Figure \ref{17Op2} we show the comparison 
of the ANC for $^{17}$F ($1/2^+$) as a function of the
distance $r$, with the Whittaker function, Eq. \eqref{whitt}. As can be seen, most of the contribution
to the rms radius comes from the region outside the core.

In the calculation of $\sigma_{L,J_{b}}^{\text{rad.cap.}}$ above, one often
meets the situation in which only the asymptotic part of $I_{b}(r)$ and
$\Psi_{c}(r)$ contributes significantly to the integral over $r$. In these
situations, $\Psi_{c}(r)$ is also well described by a simple two-body
scattering wave (e.g. Coulomb waves). Therefore, the radial integration in
$\sigma_{L,J_{b}}^{\text{rad.cap.}}$ can be done accurately and the only
remaining information from the many-body physics at short distances is
contained in the asymptotic normalization coefficient $C_{i}$, i.e.
$\sigma_{L,J_{b}}^{\text{rad.cap.}}\propto C_{i}^{2}$. We thus run into an
effective theory for radiative capture cross sections, in which the constants
$C_{i}$ carry all information about the short-distance physics, where the
many-body aspects are relevant. It is worthwhile to mention that these
arguments are reasonable for proton capture at very low energies because of
the Coulomb barrier.

The asymptotic normalization coefficients, $C_{\alpha}$, can also be
obtained from the analysis of peripheral transfer and breakup reactions. As the
overlap integral,      Eq. \eqref{whitt}, asymptotically 
becomes a Whittaker function, so does the single-particle
bound-state wavefunction $u_{\alpha}$, calculated with      Eq.
\eqref{bss}. If we label the single-particle ANC by $b_{i}$, then the
relation between the ANC obtained from experiment, or a microscopic
model, with the single-particle ANC given by $(SF)_{i}b_{i}^{2}%
=C_{i}^{2}$ (this becomes clear from     Eq. \eqref{SFS}). The values of
$(SF)_{i}$ and $b_{i}$ obtained with the simple potential model are
useful telltales of the complex short-range many-body physics of
radiative capture reactions. One can also invert this argumentation
and obtain spectroscopic factors if the $C_{i}$ are deduced from a
many-body model, or from experiment, and the $b_{i}$ are calculated
from a single-particle potential model \cite{Xu1994}.

Microscopic calculations of ANCs rely on obtaining the projection, or overlap, of the many-body 
wave functions of nuclei $A$ and $A-1$. 
The overlap integral $\left< (A-1)|A\right>\equiv I_b(r)$ must have
correct asymptotic behavior with 
respect to the variable $r$ which is the distance between the nucleon $N$ and the
c.m. of the nucleus  $A-1$.  
The most common methods are: (a) the resonating group method (RGM), as described
above,  
(b)  the Fadeev method for three-body systems, (c)  a combination of microscopic
cluster method  
and $\mathcal{R}$-matrix approaches, to be discussed later, (d) Green's function
Monte-Carlo method,  
(e) no-core shell model, or (f) hyperspherical functions method. As an example,
early applications  
of the ANC method have obtained $S_{17}(0) = 15.5$ eV b  for the $^7$Be(p,$\gamma$)$^8$B reaction using ANCs calculated with 
$0\hbar \omega$ oscillator wave functions and M3Y(E) effective $NN$ potential as
a model for  
$|A\rangle$ and $\langle (A-1)\bigotimes N|$ \cite{MT90}. The M3Y $NN$ interaction is an
effective interaction constructed as in Eq. 
\eqref{Veff}, with $v_{eff}$ given in terms of sums of (3) Yukawa functions.

\subsubsection{Threshold behavior and the r-process}
The threshold behavior of radiative capture cross sections is fundamental in
nuclear astrophysics because of the small projectile energies in the
thermonuclear region. For example, for neutron capture near the threshold, the
cross section can be written as \cite{BD04}
\[
\sigma_{if}= {\pi\over k^{2}} \left( -4kR {\mathrm{Im}\mathcal{L}_{0}\over
\left\vert \mathcal{L}_{0}\right\vert ^{2}} \right),
\]
where $\mathcal{L}_{0}$ is the logarithmic derivative for the $s$ wave at a
channel radius. Since 
$\mathcal{L}_{0}$ is only weakly dependent on the projectile energy, one
obtains for low energies the well-known $1/v$-behavior.

With increasing neutron energy, higher partial waves with $l>0$ contribute more
significantly to the radiative capture cross section. Thus the product $\sigma
v$ becomes a slowly varying function of the neutron velocity and one can
expand this quantity in terms of $v$ or $\sqrt{E}$ around zero energy,
$$\sigma v=S^{(n)}(0)+\dot{S}^{(n)}(0)\sqrt{E}+\ddot{S}
^{(n)}(0){E\over 2}+\ldots .$$
The quantity $S^{(n)}(E)=\sigma v$ is the astrophysical $S$-factor for
neutron--induced reactions and the dotted quantities represent derivatives
with respect to $E^{1/2}$, i.e., $$\dot{S}^{(n)}=2\sqrt{E}{dS^{(n)}\over
dE} \ \ \ {\rm and} \ \ \  \ddot{S}^{(n)}=4E{d^{2}S^{(n)}\over dE^2}+2
{dS^{(n)}\over dE}.$$ 
The astrophysical
$S$-factor for neutron-induced reactions is different from that
for charged-particle induced reactions. In the astrophysical
$S$-factor for charged--particle induced reactions also the
penetration factor through the Coulomb barrier has to be
considered (Eq. \eqref{astrophys13a}).
Inserting this into Eq.~\eqref{astrophys5}, one obtains for the
reaction rate of
neutron-induced reactions%
\begin{equation}
\left\langle \sigma v\right\rangle =S(0)+\left(  \frac{4}{\pi}\right)
^{\frac{1}{2}}\dot{S}(0)(k_{\mathrm{B}}T)^{\frac{1}{2}}+\frac{3}{4}\ddot
{S}(0)k_{\mathrm{B}}T+\ldots.\label{RRN}%
\end{equation}

In most astrophysical neutron-induced reactions, neutron s-waves will
dominate, resulting in a cross section showing a 1/$v$--behavior (i.e.,
$\sigma(E)\propto1/\sqrt{E}$). In this case, the reaction rate will become
independent of temperature, $R=\mathrm{const}$. Therefore it will suffice to
measure the cross section at one temperature in order to calculate the rates
for a wider range of temperatures. The rate can then be computed very easily
by using%
\begin{equation}
R=\left\langle \sigma v\right\rangle =\left\langle \sigma\right\rangle
_{T}v_{T}=\mathrm{const.}\ ,
\end{equation}
with
$v_{T}=\left(  2kT/m\right)  ^{1/2}$.

The mean lifetime $\tau_{\mathrm{n}}$ of a nucleus against neutron capture,
i.e., the mean time between subsequent neutron captures, is inversely
proportional to the available number of neutrons $n_{\mathrm{n}}$ and the
reaction rate $R_{\mathrm{n\gamma}}$, 
$\tau_{\mathrm{n}}=(n_{\mathrm{n}}R_{\mathrm{n\gamma}})^{-1}$.
If this time is shorter than the $\beta$-decay half-life of the nucleus, it
will be likely to capture a neutron before decaying (r process). In this
manner, more and more neutrons can be captured to build up nuclei along an
isotopic chain until the $\beta$-decay half-life of an isotope finally becomes
shorter than $\tau_{\mathrm{n}}$. With the very high neutron densities
encountered in several astrophysical scenarios, isotopes very far off
stability can be synthesized.

\subsubsection{Halo nuclei}
For low values of the binding energy $|E_{B}|$, e.g. for \textit{halo-nuclei}, the
simple $1/v$-law does not apply anymore. A significant deviation
can be observed if the neutron energy is of the order of the
$|E_{B}|$-value.  For radiative capture to weakly-bound final states, the
bound-state wave function $u_{lj}(r)$ in  
Eq. \eqref{lol0} 
decreases very slowly in the nuclear exterior, so that the
contributions come predominantly from far outside the nuclear
region, i.e., from the nuclear halo. For this
asymptotic region, the scattering and bound wave functions in
Eq.~\eqref{lol0} can be approximated by their asymptotic
expressions
neglecting the nuclear potential,
$u_{l}(kr)\propto j_{l}(kr)$, and  $u_{l_{0}}(r)\propto h_{l_{0}
}^{(+)}(i\xi r)$,
where $j_{l}$ and $h_{l_{0}}^{(+)}$ are the spherical Bessel, and the Hankel
function of the first kind, respectively. The separation energy $|E_{B}|$ in
the exit channel is related to the parameter $\xi$ by $|E_{B}|=\hbar^{2}%
\xi^{2}/(2m_{nx})$.

Performing the calculations of the radial integrals in Eq.
\eqref{lol0}, one readily obtains the energy dependence of the
radiative capture cross section for halo nuclei \cite{BS92}. For
example, for a transition $s \rightarrow p$
it becomes%
\begin{equation}
\sigma_{(E1)}^{(\mathrm{rc})}(\mathrm{s}\rightarrow\mathrm{p})\propto\frac
{1}{\sqrt{E}}\frac{(E+3|E_{B}|)^{2}}{E+|E_{B}|}\ ,\label{qqq1}%
\end{equation}
while a transition $p \rightarrow s$ has the energy dependence
\begin{equation}
\sigma_{(E1)}^{(\mathrm{rc})}(\mathrm{p}\rightarrow\mathrm{s})\propto
\frac{\sqrt{E}}{E+|E_{B}|}\ .\label{qqq2}%
\end{equation}
If $E\ll|E_{B}|$, the conventional energy dependence is recovered. From the
above equations one obtains that the reaction rate is not constant (for s-wave
capture) or proportional to $T$ (for p-wave capture) in the case of small
$|E_{B}|$-values.

In the case of charged particles, $S(E)$ is expected to be a slowly varying
function in energy for non-resonant nuclear reactions. In this case, $S\left(
E\right)  $ can be expanded in a McLaurin series, as was done to obtain Eq. \eqref{RRN}.
Using the expansion in Eq. \eqref{astrophys13} and approximating
the product of the exponentials $\exp\left(
-E/k_{\mathrm{B}}T\right)  $ and $\exp\left[ 2\pi\eta\left(
E\right)  \right]  $ by a Gaussian centered at the energy $E_{0}$,
Eq. \eqref{astrophys13} \ can be evaluated as \cite{rolfs-11}
\begin{equation}
\left\langle \sigma v\right\rangle =\left(  \frac{2}{m_{ab}}\right)
^{1/2}\text{ }\frac{\Delta}{\left(  kT\right)  ^{3/2}}\ S_{\mathrm{eff}%
}\left(  E_{0}\right)  \exp\left(  -\frac{3E_{0}}{kT}\right) \label{svap}%
\end{equation}
with%
\begin{equation}
S_{\mathrm{eff}}\left(  E_{0}\right)  =S\left(  0\right)  \left[  1+\frac
{5}{12\tau}+\frac{\dot{S}\left(  0\right)  }{S\left(  0\right)  }\left(
E_{0}+\frac{35E_{0}}{12\tau}\right)  +\frac{\ddot{S}\left(  0\right)
}{2S\left(  0\right)  }\left(  E_{0}^{2}+\frac{89E_{0}^{2}}{12\tau}\right)
\right]  \ .\label{svap2}%
\end{equation}

The quantity $E_{0}$\ defines the effective mean energy for
thermonuclear fusion and is given by Eq. \eqref{astrophys14}. The
quantity $\tau$ is given
by $
\tau={3E_{0}}/{kT}$,
and $\Delta$ is given by Eq. \eqref{astrophys14}.

\subsubsection{Resonances}
For the case of resonances, where $E_{r}$ is the resonance energy,
we can approximate $\sigma\left(  E\right)  $ by a Breit-Wigner
resonance formula, %
\begin{equation}
\sigma_{\mathrm{r}}(E)=\frac{\pi\hbar^{2}}{2\mu E}\frac{\left(  2J_{R}%
+1\right)  }{(2J_{a}+1)(2J_{b}+1)}\frac{\Gamma_{\mathrm{p}}\Gamma
_{\mathrm{\gamma}}}{\left(  E_{\mathrm{r}}-E\right)  ^{2}+\left(
\Gamma_{\mathrm{tot}}/2\right)  ^{2}}\ ,\label{BW}%
\end{equation}
where $J_{R}$, $J_{a}$, and $J_{b}$ are the spins of the resonance and the
nuclei $a$ and $b$, respectively, and the total width $\Gamma_{\mathrm{tot}} $
is the sum of the particle decay partial width $\Gamma_{\mathrm{p}}$ and the
$\gamma$-ray partial width $\Gamma_{\gamma}$. The particle partial width, or
entrance channel width, $\Gamma_{\mathrm{p}}$, can be expressed in terms of the
single-particle spectroscopic factor $SF_{i}$ and the single-particle width
$\Gamma_{\mathrm{s.p.}}$ of the resonance state, $
\Gamma_{\mathrm{p}}=SF_{i}\times\Gamma_{\mathrm{s.p.}}
$.
The single-particle width $\Gamma_{\mathrm{s.p.}}$ can be
calculated from the scattering phase shifts of a scattering
potential with the potential parameters being determined by
matching the resonance energy.
The $\gamma$ partial widths $\Gamma_{\mathrm{\gamma}}$ are calculated from the
reduced electromagnetic transition probabilities $B(J_{i}\rightarrow J_{f};L)$
which carry the nuclear structure information of the resonance states and the
final bound states. The reduced transition rates are usually computed within
the framework of the nuclear shell model.

Most of the typical transitions are $M1$ or $E2$ transitions. For these, the
relations are
\begin{equation}
\Gamma_{\mathrm{E2}}[\mathrm{eV}]=8.13\times10^{-7}\ E_{\gamma}^{5}%
\mathrm{\ [{MeV}]\ \ }B(E2\mathrm{)\ [{e^{2}fm^{4}}]}\label{gl1}%
\end{equation}
and
\begin{equation}
\Gamma_{\mathrm{M1}}[\mathrm{eV}]=1.16\times10^{-2}\ E_{\gamma}^{3}%
\mathrm{\ [{MeV}]\ }B(M1)\mathrm{\ }[\mu_{N}^{2}]\mathrm{\ .}\label{gl2_7}%
\end{equation}

For the case of narrow resonances, with width $\Gamma\ll E_{r}$, the
Maxwellian exponent $\exp\left(  -E/kT\right)  $ can be taken out
of the integral, and one finds%
\begin{equation}
\left\langle \sigma v\right\rangle =\left(  \frac{2\pi}{m_{ab}kT}\right)
^{3/2}\hbar^{2}\left(  \omega\gamma\right)  _{R}\exp\left(  -\frac{E_{r}}%
{kT}\right)  \ ,\label{svres}%
\end{equation}
where the resonance strength is defined by%
\begin{equation}
\left(  \omega\gamma\right)  _{R}=\frac{2J_{R}+1}{(2J_{a}+1)(2J_{b}%
+1)}\ \left(  1+\delta_{ab}\right)  \ \frac{\Gamma_{\mathrm{p}}\ \Gamma
_{\gamma}}{\Gamma_{\mathrm{tot}}}.\label{svres2}%
\end{equation}

For broad resonances, Eq. \eqref{astrophys13} is usually calculated
numerically. An interference term has to be added. The total
capture cross section is then given by~\cite{rol74}
\begin{equation}
\sigma(E)=\sigma_{\mathrm{nr}}(E)+\sigma_{\mathrm{r}}(E)+2\left[
\sigma_{\mathrm{nr}}(E)\sigma_{\mathrm{r}}(E)\right]  ^{1/2}\mathrm{cos}%
[\delta_{\mathrm{R}}(E)]\ .\label{eq-int}%
\end{equation}
In this equation $\delta_{\mathrm{R}}(E)$ is the resonance phase
shift. Only the contributions
with the same angular momentum of the incoming wave interfere in
Eq.~\eqref{eq-int}.

\subsection{$\mathcal{R}$-matrix theory}\label{sec:Rmatrix}
Reaction rates dominated
by the contributions from a few resonant or bound states are often
extrapolated to energies of astrophysical interest in terms of $\mathcal{R}$\textit{- or }$K$\textit{-matrix}
fits. The appeal of
these methods rests on the fact that analytical expressions can be
derived from underlying formal reaction theories
that allow for a rather simple parametrization of the data. However, the
relation between the parameters of the $\mathcal{R}$-matrix model and the
experimental data (resonance energies and widths) is only quite
indirect. The $K$-matrix formalism solves this problem, but
suffers from other drawbacks \cite{Barker94}.

\subsection{Elastic and inelastic scattering reactions}
In the $\mathcal{R}$-matrix formalism,  the
eigenstates of the nuclear Hamiltonian in the interior region of a nucleus are denoted by
$X_{\lambda}$, with energy $E_{\lambda}$, and are required to satisfy
the boundary condition
$$r{dX_{\lambda}\over dr}+bX_{\lambda}=0\
$$
at the channel radius $r=R$, where the constant $b$ is a real number. The true
nuclear wavefunction $\Psi$\ for the compound system is not stationary, but
since the $X_{\lambda}$ form a complete set, it is possible to expand $\Psi$ in
terms of $X_{\lambda}$, i.e. $$
\Psi=\sum_{\lambda}A_{\lambda}X_{\lambda}
, \ \ \ {\rm where} \ \ \  A_{\lambda}=\int_{0}^{R}X_{\lambda}\ \Psi\ dr.$$
The differential equations for $\Psi$\ and $X_{\lambda}$\ are (for s-wave
neutrons)%
\begin{align}
-\frac{\hbar^{2}}{2m}\frac{d^{2}\Psi}{dr^{2}}+V\Psi & =E\Psi\label{5.28}\\
-\frac{\hbar^{2}}{2m}\frac{d^{2}X_{\lambda}}{dr^{2}}+VX_{\lambda}  &
=E_{\lambda}X_{\lambda}\ ,\ \ \ \ \ r\leq R\ .\label{5.29}%
\end{align}

\begin{figure}[tb]
\begin{center}
\includegraphics[
width=4.3in]
{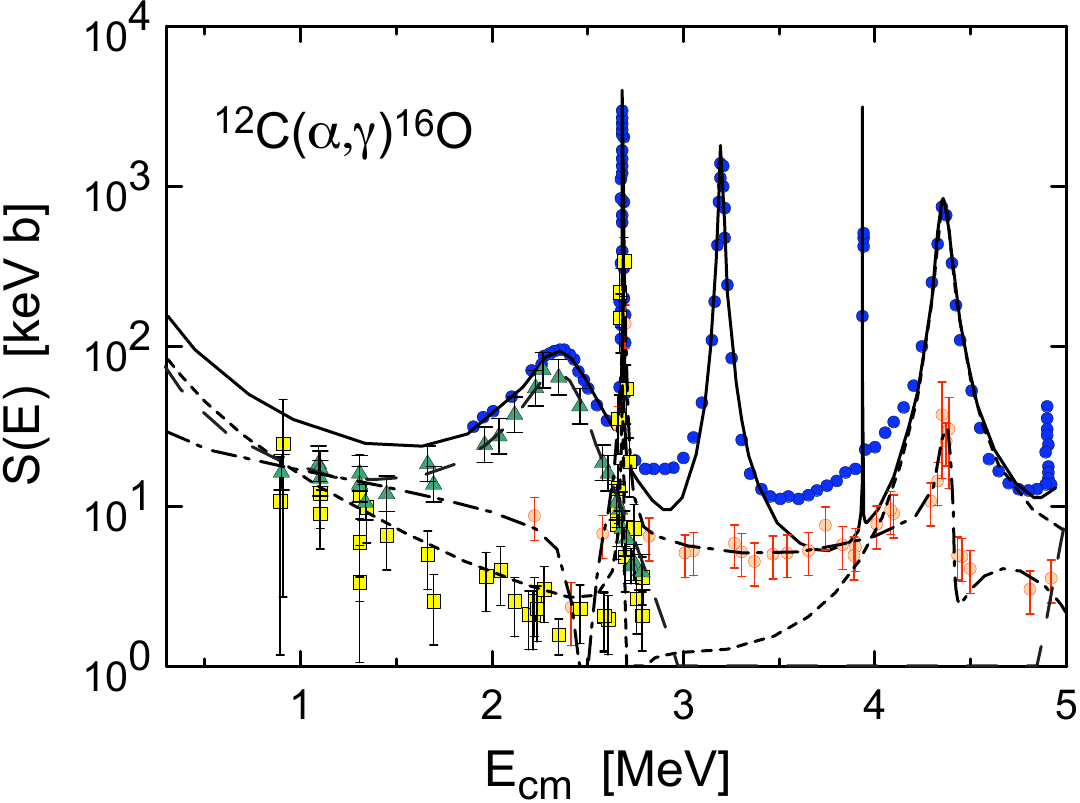}
\caption{Total $S$ factor data (filled-in circles) \cite{Schu05} for $^{12}$C($\alpha,\gamma$)$^{16}$O 
compared with E1 (open triangles) and E2 (open squares) contributions \cite{Assu06,Mat06}. The solid 
line represents the sum of the single amplitudes of
an $\mathcal{R}$-matrix fit \cite{Hes58} (the dotted and dashed lines are the $E1$ and $E2$
amplitudes, respectively). In 
addition, the $\mathcal{R}$ matrix fit of \cite{Kun02} to their  data
(dotted-dashed line) is shown. (Adapted from Ref. \cite{Str08}).}  
\label{R_matrix}
\end{center}
\end{figure}

Multiplying Eq. (\ref{5.28}) by $\Psi$\ and Eq. (\ref{5.29}) by $X_{\lambda}$,
subtracting and integrating,  we have
\[
A_{\lambda}={1\over  E-E_{\lambda}}\frac{\hbar^{2}}{2mR}
X_{\lambda}\left(  R\right)  \left[  R\Psi^{\prime}(R)+b\Psi(R)\right]  \ ,
\]
where the prime indicates the differentiation with respect to $r$.
This result, together with the definition of $\Psi$, gives
\begin{equation}
\Psi(R)=\mathcal{R}\left[  R\Psi^{\prime}(R)+b\Psi(R)\right] \label{5.30}
\end{equation}
where the function $\mathcal{R}$ relates the value of $\Psi(R)$\ at the
surface to its derivative at the surface:
\begin{equation}
\mathcal{R=}\frac{\hbar^{2}}{2mR}\sum_{\lambda}\frac{X_{\lambda}\left(
R\right)  X_{\lambda}\left(  R\right)  }{E_{\lambda}-E}\ .\label{rmat}
\end{equation}

Rearranging Eq. (\ref{5.30}) we have
$R{\Psi^{\prime}(R)}/{\Psi(R)}={(1-b\mathcal{R})}/{\mathcal{R}%
},\label{rcons}%
$
which is just the logarithmic derivative $\mathcal{L}^{I}$ which can be used to
determine  
the $S$-matrix element $S_{0}
$\ in terms of the $\mathcal{R}$\ function. This gives%
\[
S_{0}=\left[  1+\frac{2ikR\mathcal{R}}{1-(b+ikR)\mathcal{R}}\right]
e^{-2ikR}\ .
\]

Finally, we assume that $E$ is near to a particular $E_{\lambda},$\ say
$E_{\alpha}$, neglect all terms $\lambda\neq\alpha$\ in Eq. (\ref{rmat}), and
define
$$\Gamma_{\alpha}={\hbar^{2}k\over m}X_{\alpha}^{2}\left(  R\right)
, \ \ \ {\rm and}  \ \ \  \Delta_{\alpha}=-{b\over 2kR}\Gamma_{\alpha},
$$
so that the $S$-matrix element becomes%
\begin{equation}
S_{0}=\left[  1+\frac{i\Gamma_{\alpha}}{\left(  E_{\alpha}+\Delta_{\alpha
}-E\right)  -i\Gamma_{\alpha}/2}\right]  e^{-2ikR}\label{s0}%
\end{equation}
and the scattering cross section is%
\begin{equation}
\sigma_{\mathrm{sc}}=\frac{\pi}{k^{2}}\left\vert e^{2ikR}-1+\frac
{i\Gamma_{\alpha}}{\left(  E_{\alpha}+\Delta_{\alpha}-E\right)  -i\Gamma
_{\alpha}/2}\right\vert ^{2}\ .\label{scat}%
\end{equation}

We see that the procedure of imposing the boundary
conditions at the channel radius leads to isolated s-wave resonances of
Breit-Wigner form. If the constant $b$ is non-zero, the
position of the maximum in the cross section is shifted. The level shift does
not appear in the simple form of the Breit-Wigner formula because $E_{\alpha
}+\Delta_{\alpha}$\ is defined as the resonance energy. In general, a nucleus
can decay through many channels and when the formalism is extended to take
this into account, the\ $\mathcal{R}$-function becomes a matrix. In this
$\mathcal{R}$-matrix theory the constant $b$ is real and $X_{\lambda}\left(
R\right)  $ and $E_{\lambda}$\ can be chosen to be real so that the eigenvalue
problem is Hermitian \cite{WE47}.

The $\mathcal{R}$-matrix theory  can be easily generalized to
account for higher partial waves and spin-channels. If we define the reduced
width by
$\gamma_{\lambda}^{2}={\hbar^{2}X_{\lambda}^{2}\left(  R\right)  }/{2mR}
$, which is a property of a particular state and not dependent of the scattering
energy $E$ of the scattering system, we can write
\[
\mathcal{R}_{\alpha\alpha^{\prime}}=\sum_{\lambda}\frac{\gamma_{\lambda
\alpha^{\prime}}\gamma_{\lambda\alpha}}{E_{\lambda}-E}\ ,
\]
where $\alpha$ is the channel label. $\gamma_{\lambda\alpha},$ $E_{\lambda}$,
and $b$ are treated as parameters in fitting the experimental data. If we
write the wavefunction for any channel as $\Psi\sim I+S_{\alpha}O$, where $I$ and $O$ 
are incoming and outgoing waves, Eq. (\ref{5.30}) means%
\[
R\frac{I^{\prime}\left(  R\right)  +S_{\alpha}O^{\prime}\left(  R\right)
}{I\left(  R\right)  +S_{\alpha}O\left(  R\right)  }=\frac{1-b\mathcal{R}%
}{\mathcal{R}}.
\]
Thus, as in Eq. (\ref{s0}), the $S$-matrix is related to the $\mathcal{R}$-matrix and
from the above relation we obtain that,
\begin{equation}
S_{\alpha}=\frac{I\left(  R\right)  }{O\left(  R\right)  }\left[
\frac{1-\left(  \mathcal{L}^{I}\right)  ^{\ast}\mathcal{R}}{1-\mathcal{L}%
^{I}\mathcal{R}}\right]  .
\end{equation}

The total cross sections for states with angular momenta and spins given by
$l$, $s$ and $J$ is%
\begin{equation}
\sigma_{\alpha\alpha^{\prime}}=\frac{\pi}{k_{\alpha}^{2}}\sum_{JJ^{\prime
}ll^{\prime}ss^{\prime}}g_{J}\left\vert S_{\alpha Jls,\alpha^{\prime}%
J^{\prime}l^{\prime}s^{\prime}}\right\vert ^{2}\ ,\ \ \ \ \ \alpha\neq
\alpha^{\prime},
\end{equation}
where $g_{J}$ are spin geometric factors.

In the statistical model, it can be argued that because the $S$-matrix
elements vary rapidly with energy, the statistical assumption implies that
there is a random phase relation between the different components of the
$S$-matrix. The process of energy averaging then eliminates the cross terms
and gives%
\begin{equation}
\sigma_{\mathrm{abs}}=\sum_{\alpha^{\prime}\neq\alpha,J^{\prime}l^{\prime
}s^{\prime}}\sigma_{\alpha\alpha^{\prime}}=\frac{\pi}{k_{\alpha}^{2}}%
\sum_{Jls}g_{J}\left[  1-\left\vert S_{\alpha Jls}\right\vert ^{2}\right]
=\frac{\pi}{k_{\alpha}^{2}}\sum_{Jls}g_{J}T_{ls}^{J}\left(  \alpha\right)  \ ,
\end{equation}
where the symmetry properties of the $S$-matrix in the form
$\sum_{\alpha^{\prime}\neq\alpha,{\mathcal J}^\prime}S_{\alpha
{\mathcal J},\alpha^{\prime}{\mathcal J}^\prime}\ S_{\alpha {\mathcal J},\alpha
^{\prime}{\mathcal J}^\prime}^{\ast}=1$ 
with $ {\mathcal J}= Jls$, ${\mathcal J}^\prime=J^{\prime}l^{\prime}s^{\prime}$ 
have been used, 
and we have introduced the general definition of the transmission coefficient%
\begin{equation}
T_{ls}^{J}\left(  \alpha\right)  =1-\left\vert S_{\alpha Jls}\right\vert ^{2}.\label{tramiss}
\end{equation}

\subsection{Radiative capture reactions}
Consider an $\mathcal{R}$-matrix calculation of the radiative capture reaction $n + x \rightarrow a +
\gamma$ to a state of
nucleus $a$ with a given spin $J_f$. The cross section  can be written as \cite{Bar91} 
$ \sigma_{J_f}=\sum_{J_i} \sigma_{J_iJ_f}$, with
\begin{equation}
\label{eq7}
\sigma_{J_iJ_f}=\frac{\pi}{k^2}\frac{2J_i+1}{(2J_n+1)(2J_x+1)}\sum_{Il_i}|T_{Il_iJ_fJ_i}|^2.
\end{equation}
Here, $J_i$ is the total angular momentum of the colliding nuclei $n$
and $x$ in the initial state, $J_n$ and $J_x$ are their spins, and $I$, $k$, and $l_i$ are 
their channel spin, wave number
and orbital angular momentum in the initial state. $T_{Il_iJ_fJ_i}$
is the transition amplitude from the initial continuum state
($J_i,I,l_i$) to the final bound state ($J_f,I$). In the one-level,
one-channel approximation, the resonant amplitude for the capture
into the resonance with energy $E_{R_n}$ and spin $J_i$, and
subsequent decay into the bound state with the spin $J_f$ can be
expressed as
\begin{equation}
\label{eq8}
T^R_{Il_iJ_fJ_i}=-ie^{i(\sigma_{l_i}-\phi_{l_i})}\frac{[\Gamma^{J_i}_{bIl_i}(E)\Gamma^{J_i}_{\gamma
J_f}(E)]^{1/2}}{E - E_{R_n} + i \frac{\Gamma_{J_i}}{2}}.
\end{equation}
Here we assume that the boundary parameter is equal to the shift
function at resonance energy and $\phi_{l_i}$ is the hard-sphere
phase shift in the $l_i$th partial wave,
\begin{equation}
\phi_{l_i}=\arctan\Big{[}\frac{F_{l_i}(k,r_c)}{G_{l_i}(k,r_c)}\Big{]},
\end{equation}
where $F_{l_i}$ and $G_{l_i}$ are the regular and irregular
 Coulomb functions, $r_c$ is the channel radius, and $\sigma_{l_i}$ is the Coulomb 
 phase factor, $\sigma_{l_i}=\sum_{k=1}^{l_i}\arctan(\eta_i/k)$,
where $\eta_i$ is the Sommerfeld parameter.
$\Gamma^{J_i}_{nIl_i}(E)$ is the observable partial width of the
resonance in the channel $n$ + $x$, $\Gamma^{J_i}_{\gamma J_f}(E)$
is the observable radiative width for the decay of the given
resonance into the bound state with the spin $J_f$, and
$\Gamma_{J_i} \approx \sum\limits_I \Gamma^{J_i}_{nIl_i}$ is the
observable total width of the resonance level. The energy dependence
of the partial widths is determined by
\begin{equation}
\label{eq10}
\Gamma^{J_i}_{nIl_i}(E)=\frac{P_{l_i}(E)}{P_{l_i}(E_{R_n})}\Gamma^{J_i}_{nIl_i}(E_{R_n})
\end{equation}
and
\begin{equation}
\label{eq11} \Gamma^{J_i}_{\gamma
J_f}(E)=\left(\frac{E+\varepsilon_f}{E_{R_n}+\varepsilon_f}\right)^{2L+1}\Gamma^{J_i}_{\gamma
J_f}(E_{R_n}),
\end{equation}
where $\Gamma^{J_i}_{nIl_i}(E_{R_n})$ and $\Gamma^{J_i}_{\gamma
J_f}(E_{R_n})$ are the experimental partial and radiative widths,
$\varepsilon_f$ is the binding energy of the bound state in
nucleus $a$, and $L$ is the multipolarity of the $\gamma$-ray transition. The
penetrability $P_{l_i}(E)$ is expressed as
\begin{equation}
\label{eq12}
P_{l_i}(E)=\frac{kr_c}{F^2_{l_i}(k,r_c)+G^2_{l_i}(k,r_c)}.
\end{equation}

 The non-resonant amplitude can be calculated by
 \begin{eqnarray}
\label{eq13}
T^{NR}_{Il_iJ_fJ_i}&=&-(2)^{3/2}i^{l_i+L-l_f+1}e^{i(\sigma_{l_i}-\phi_{l_i})}\frac{(\mu_{nx}k_\gamma
r_c)^{L+1/2}}{\hbar
k} e_L
\sqrt{\frac{(L+1)(2L+1)}{L[(2L+1)!!]^2}}
C_{J_fIl_f}F_{l_i}(k,r_c)\nonumber\\
&&\times G_{l_i}(k,r_c)W_{l_f}(2\kappa
r_c)\sqrt{P_{l_i}}(l_i0L0|l_f0) U(Ll_fJ_iI;l_iJ_f)J^\prime_L(l_il_f),
\end{eqnarray}
where, $e_L$ is the effective charge, $U$ is a geometric coefficient, and
\begin{equation}
\label{eq14} J^\prime_L(l_il_f)=
\frac{1}{r_c^{L+1}}\int^\infty_{r_c} dr \ r\frac{W_{l_f}(2\kappa
r)}{W_{l_f}(2\kappa
r_c)}\left[\frac{F_{l_i}(k,r)}{F_{l_i}(k,r_c)}-\frac{G_{l_i}(k,r)}{G_{l_i}(k,r_c)}\right].
\end{equation}
$W_l(2\kappa r)$ is the Whittaker hypergeometric function,
$\kappa$ = $\sqrt{2\mu_{nx}\varepsilon_f}$ and $l_f$ are the wave
number and relative orbital angular momentum of the bound state, and
$k_\gamma$ = $(E+\varepsilon_f)$/$\hbar c$ is the wave number of the
emitted photon.

The non-resonant amplitude contains the radial integral ranging only
from the channel radius $r_c$ to infinity since the internal
contribution is contained within the resonant part. Furthermore, the
$\mathcal{R}$-matrix boundary condition at the channel radius $r_c$ implies that
the scattering of particles in the initial state is given by the
hard sphere phase. Hence, the problems related to the interior
contribution and the choice of incident channel optical parameters
do not occur. Therefore, the direct capture cross section only
depends on the ANC and the channel radius $r_c$.

The $\mathcal{R}$-matrix method described above can be extended to the analysis
of other types of reactions, e.g. transfer reactions \cite{Des04}.
The goal of the $\mathcal{R}$-matrix method is to parameterize some experimentally known quantities,
such as cross sections or phase shifts, with a small number of parameters, which
are then used  to extrapolate the cross section down to astrophysical
energies. One example is given in  Figure \ref{R_matrix} which shows the experimental data 
and $\mathcal{R}$-matrix fits for the cross section of the reaction
$^{12}$C($\alpha,\gamma$)$^{16}$O cross section, of relevance to helium
burning \cite{Str08}.

\subsection{Statistical models} A large fraction of the
reactions of interest proceed through compound systems that exhibit high
enough level densities for statistical methods to provide a reliable
description of the reaction mechanism. The theoretical  treatment of nuclear
reactions leading to formation and decay of compound nuclei was developed by
Ewing and Weisskopf \cite{WW40}, based on two ideas: (a) the compound
nucleus formation independence hypothesis as proposed by Niels Bohr \cite{Bo36},
and (b) the reciprocity theorem, or time-reversal properties of the underlying
Hamiltonian. This allows one to relate capture and decay cross sections, usually
expressed in terms of transmission probabilities, defined in
Eq. \eqref{tramiss}.    

Later, the Ewing-Weisskopf theory was extended to include angular momentum dependence by 
Hauser and Feshbach \cite{HH52}. The
Hauser-Feshbach (HF) model has been widely used with considerable
success in nuclear astrophysics. Explosive burning in supernovae involves in general intermediate mass
and heavy nuclei. Due to a large nucleon number, they have intrinsically a high
density of excited states. A high level density in the compound nucleus at the
appropriate excitation energy allows for the use of the statistical-model
approach for compound nuclear reactions \cite{HH52} which averages over
resonances. 

A high level density in the compound nucleus also allows for the use of
averaged transmission coefficients $T$, which do not reflect 
resonance behavior, but rather describe absorption via an
imaginary part of the (optical) nucleon-nucleus potential as
described in Ref. \cite{MW79}. This leads to the expression
derived 
\begin{align}
\sigma_{i}^{\mu\nu}(j,o;E_{ij})  & ={\frac{{\pi\hbar^{2}/(2\mu_{ij}E_{ij})}%
}{{(2J_{i}^{\mu}+1)(2J_{j}+1)}}}\label{nastro7}\\
\times & \sum_{J,\pi}(2J+1){\frac{{T_{j}^{\mu}(E,J,\pi,E_{i}^{\mu},J_{i}^{\mu
},\pi_{i}^{\mu})T_{o}^{\nu}(E,J,\pi,E_{m}^{\nu},J_{m}^{\nu},\pi_{m}^{\nu})}%
}{{T_{tot}(E,J,\pi)}}}\nonumber
\end{align}
for the reaction $i^{\mu}(j,o)m^{\nu}$\ from the target state $i^{\mu}$\ to
the excited state $m^{\nu}$\ of the final nucleus, with a center of mass
energy $E_{ij}$\ and reduced mass $\mu_{ij}$. $J$\ denotes the spin, $E$\ the
corresponding excitation energy in the compound nucleus, and $\pi$\ the parity
of excited states. When these properties are used without subscripts they
describe the compound nucleus, subscripts refer to states of the participating
nuclei in the reaction $i^{\mu}(j,o)m^{\nu}$\ and superscripts indicate the
specific excited states. Experiments measure $\sum_{\nu}\sigma_{i}^{0\nu
}(j,o;E_{ij})$, summed over all excited states of the final nucleus, with the
target in the ground state. Target states $\mu$\ in an astrophysical plasma
are thermally populated and the astrophysical cross section $\sigma_{i}^{\ast
}(j,o)$\ is given by
\begin{equation}
\sigma_{i}^{\ast}(j,o;E_{ij})={\frac{\sum_{\mu}(2J_{i}^{\mu}+1)\exp
(-E_{i}^{\mu}/kT)\sum_{\nu}\sigma_{i}^{\mu\nu}(j,o;E_{ij})}{\sum_{\mu}%
(2J_{i}^{\mu}+1)\exp(-E_{i}^{\mu}/kT)}}.\label{csstar}%
\end{equation}
The summation over $\nu$\ replaces $T_{o}^{\nu}(E,J,\pi)$\ in Eq.
\eqref{nastro7} by the total transmission coefficient
\begin{align}
T_{o}(E,J,\pi)  & =\sum_{\nu=0}^{\nu_{m}}T_{o}^{\nu}(E,J,\pi,E_{m}^{\nu}%
,J_{m}^{\nu},\pi_{m}^{\nu})\nonumber\\
+  & \int\limits_{E_{m}^{\nu_{m}}}^{E-S_{m,o}}\sum_{J_{m},\pi_{m}}%
T_{o}(E,J,\pi,E_{m},J_{m},\pi_{m})\rho(E_{m},J_{m},\pi_{m})dE_{m}%
.\label{Tlong}%
\end{align}
Here $S_{m,o}$\ is the channel separation energy, and the
summation over excited states above the highest experimentally
known state $\nu_{m}$\ is changed to an integration over the level
density $\rho$. The summation over target states $\mu$\ in Eq.
\eqref{csstar} has to be generalized accordingly.

\begin{figure}[tb]
\begin{center}
\includegraphics[
width=4.3in]
{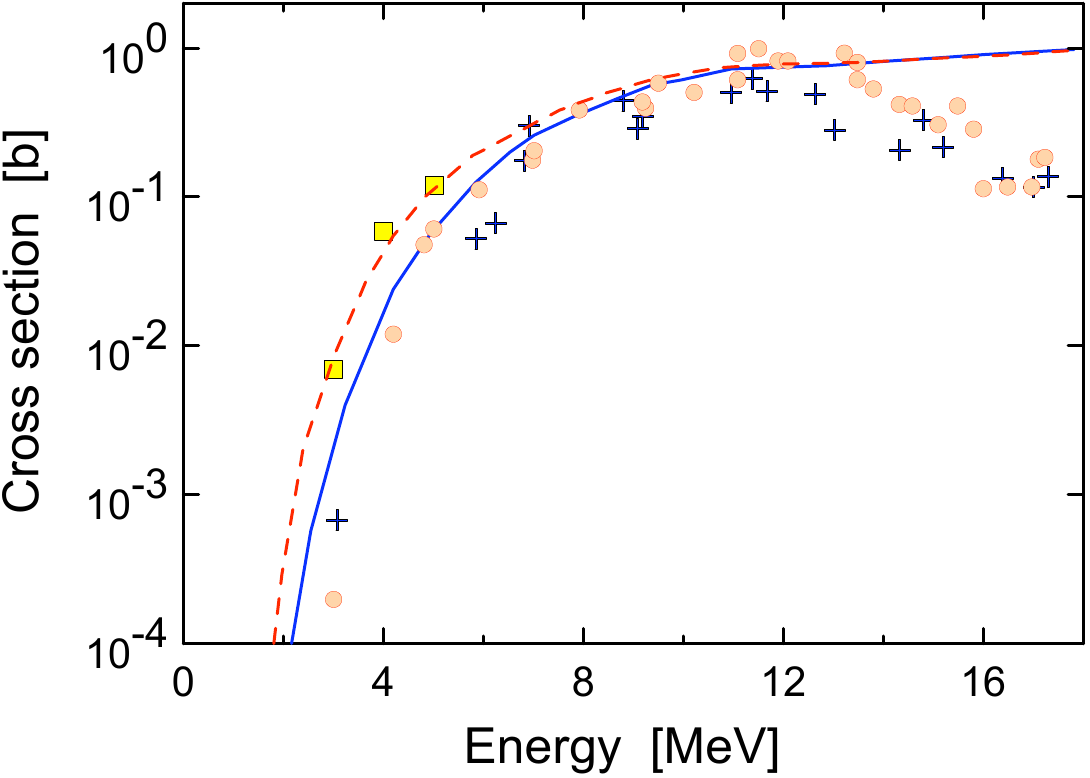}
\caption{Cross-section data for two sets of $^{75}As(p,n)$ measurements
(squares: \cite{Kai79}; circles: \cite{Mus88}) in comparison with Hauser-Feshbach
predictions (solid line: \cite{RT01}). Also shown is the experimental
cross section of $^{85}Rb(p,n)$ (crosses with error bars: \cite{Kas02}) 
in comparison with HF predictions (dashed line: \cite{RT01}).(Adapted from Ref. \cite{Rap06}).}  
\label{statistical}
\end{center}
\end{figure}

The important ingredients of statistical-model calculations, as indicated in
the above equations, are the particle and $\gamma$-transmission coefficients
$T$\ and the level density of excited states $\rho$. Therefore, the
reliability of such calculations is determined by the accuracy with which
these components can be evaluated (often for unstable nuclei).

Figure \ref{statistical} -- adapted from Ref. \cite{Rap06} -- shows the
cross-section data for  
two sets of $^{75}$As$(p,n)$ measurements
(squares: \cite{Kai79}; circles: \cite{Mus88}) in comparison with Hauser-Feshbach
predictions (solid line: \cite{RT01}). Also shown is the experimental
cross section of $^{85}$Rb$(p,n)$ (crosses with error bars: \cite{Kas02}) 
in comparison with HF predictions (dashed line: \cite{RT01}). The experimental
results are on average lower than the HF predictions.

\subsection{Spin-isospin response}

Beta-decay, electron capture and neutrino scattering involve similar operators and nuclear 
matrix elelments. We thus consider only the case of neutrino scattering. In the following, $p_\ell\equiv\{{\bf p}_\ell,
E_\ell\}$ and $q_\nu\equiv\{{\bf q},E_{\nu}\}$ are
the lepton and the neutrino momenta,
$ k = P_i-P_f\equiv \{{\bf k},k_\emptyset \},$
 is the momentum transfer,  $P_i$ and $P_f$ are
momenta of the initial and final nucleus, ${\rm M}$ is the nucleon
 mass, ${\rm m}_\ell$ is the mass of the charged
lepton, and $g_{\scriptscriptstyle V}$, $g_{\scriptscriptstyle A}$, $g_{\scriptscriptstyle M}$ and $g_{\scriptscriptstyle
P}$ are, respectively, the vector, axial-vector, weak-magnetism
and pseudoscalar effective dimensionless coupling constants. Their
numerical values are typically given by
$ g_{\scriptscriptstyle V}=1$, $g_{\scriptscriptstyle A}=1.26$,
$g_{\scriptscriptstyle M}=\kappa_p-\kappa_n=3.70$, and $g_{\scriptscriptstyle P}= g_{\scriptscriptstyle
A}(2\mathrm{M} \mathrm{m}_\ell )/(k^{2}+\mathrm{m}_\pi^2)$.

For the neutrino-nucleus reaction the momentum transfer is
$k=p_\ell-q_\nu$, and the corresponding cross section reads 
\[\sigma
(E_\ell,J_f) = \frac{|{\bf p}_\ell| E_\ell}{2\pi} F(Z\pm1,E_\ell)
\int_{-1}^1
d(\cos\theta){\mathcal T}_{\sigma}({\rm q},J_f),
\label{23}\]
where $$F(Z\pm1,E_\ell)=  {2\pi \eta \over  \exp(2 \pi \eta) -1}, \ \ \ {\rm with} \ \ \ \eta = {Z_\pm Z_e \alpha \over v_\ell},$$ is  the Fermi function
($Z_\pm=Z+1$, for neutrino, and $Z-1$, for antineutrino), $\theta\equiv
\hat{\bf q}\cdot\hat{\bf p}$ is the angle between the incident
neutrino and ejected lepton, and the transition amplitude for initial (final) angular momentum $J_i$ ($J_f$) is
\[
{\mathcal T}_{\sigma}(\kappa,J_f)= \frac{1}{2J_i+1} \sum_{ s_\ell,s_\nu }\sum_{M_i, M_f }
\left|\left<{J_fM_f}\left|H_{{ {W}}}\right|{J_iM_i}\right>\right|^{2}.
\label{24}\]
One can cast the transition amplitude in
the compact form~\cite{Krm05}\footnote{The indices $\emptyset$ and $z$ denote the 
time-component and the third-component
of four-vectors, respectively.}
\begin{eqnarray}
{\mathcal T}_{\sigma}(\kappa,J_f)&=&\frac{4\pi
G^2}{2J_i+1}\sum_{{\sf J}}\Bigg[ |\left<{J_f}||{\sf O}_{\emptyset{\sf
J}}||{J_i}\right>|^2{\mathcal L}_{\emptyset} 
+\sum_{\sf {M}=0\pm 1}|\left<{J_f}||{\sf O}_{{\sf M J}}||{J_i}\right>|^2{\mathcal L}_{\sf M}
 \label{25}\nonumber \\
&-&2\Re\left(|\left<{J_f}||{\sf O}_{\emptyset{\sf J}}||{J_i}\right>
\left<{J_f}||{\sf O}_{0{\sf J}}||{J_i}\right>\right){\mathcal L}_{\emptyset z}\Bigg],
\end{eqnarray}
where $G=(3.04545\pm 0.00006){\times} 10^{-12}$ is
the Fermi coupling constant (in natural units), and
\begin{eqnarray}
{\mathcal L}_{\emptyset}&=&1+\frac{|{\bf p}|\cos\theta}{E_\ell},
\ \ \ 
{\mathcal L}_{\emptyset
z}=\left(\frac{q_z}{E_\nu}+\frac{p_z}{E_\ell}\right),
\nonumber \\
{\mathcal L}_{0}\equiv {\mathcal L}_z&=&1+\frac{2q_zp_z}{E_\ell
E_\nu}-\frac{|{\bf p}|\cos\theta}{E_\ell},
\ \ \ \
{\mathcal L}_{\pm1}=1-\frac{q_zp_z}{E_\ell E_\nu}\pm
\left(\frac{q_z}{E_\nu}-\frac{p_z}{E_\ell}\right) S_1,
\label{26}
\end{eqnarray}
 with 
\begin{equation} q_z={\hat k}\cdot
{\bf q}=\frac{E_\nu(|{\bf p}|\cos\theta-E_\nu)}{\kappa},
\ \ \ \ 
p_z={\hat k}\cdot {\bf p}=\frac{|{\bf p}|(|{\bf p}|-E_\nu\cos\theta)}{\kappa},
\label{27} 
\end{equation}
being the $z$-components of the neutrino and lepton
momenta, and $S_1=\pm 1$ for neutrino scattering and antineutrino 
scattering, respectively.

Explicitly, the operators in \eqref{25} are
\begin{eqnarray}
{\sf O}_{\emptyset{\sf J}}&=&g_{\scriptscriptstyle{V}}{\mathcal M}_{\sf J}^{\scriptscriptstyle V}
+2i\overline{g}_{\mbox{\tiny A}}{\mathcal M}^{\scriptscriptstyle A}_{\sf J}
+i(\overline{g}_{\mbox{\tiny A}}+\overline{g}_{\mbox{\tiny P1}}){\mathcal M}^{\scriptscriptstyle A}_{z{\sf J}},
\nonumber \\
{\sf O}_{{\sf M}{\sf J}} &=&i(\delta_{{\sf M}z}\overline{g}_{\mbox{\tiny P2}}-g_{\mbox{\tiny A}} +
{\sf M} \overline{g}_{\mbox{\tiny W}}){\mathcal M}^{\scriptscriptstyle A}_{{\sf M}{\sf J}}
+2\overline{g}_{\mbox{\tiny V}}{\mathcal M}^{\scriptscriptstyle V}_{{\sf M}{\sf J}}-\delta_{{\sf M} z}\overline{g}_{\mbox{\tiny V}}{\mathcal M}_{\sf J}^{\scriptscriptstyle V},
\label{15}
\end{eqnarray}
where $\hat{\bf k}={\bf k}/\kappa$, $\kappa \equiv|{\bf k}|$, and the following short notation
\begin{equation}
\overline{g}_{\mbox{\tiny V}}={g}_{\mbox{\tiny V}}\frac{\kappa}{2{\rm M}};~
\overline{g}_{\mbox{\tiny A}}={g}_{\mbox{\tiny A}}\frac{\kappa}{2{\rm M}};~ \overline{g}_{\mbox{\tiny W}}=({g}_{\mbox{\tiny V}}
+{g}_{\mbox{\tiny M}})\frac{\kappa}{2{\rm M}};
\overline{g}_{\mbox{\tiny P1}}={g}_{\mbox{\tiny P}}\frac{\kappa}{2{\rm M}}\frac{q_\emptyset}{{\rm m}_\ell};~
\overline{g}_{\mbox{\tiny P2}}={g}_{\mbox{\tiny P}}\frac{\kappa}{2{\rm M}}\frac{\kappa}{{\rm m}_\ell},
\label{8}
\end{equation}
has also been introduced. The elementary operators are given by
\begin{eqnarray}
{\mathcal M}^{\scriptscriptstyle V}_{\sf J}&=&j_{\sf J}(\rho) Y_{{\sf J}}(\hat{\bf r});
\ \ \
{\mathcal M}^{\scriptscriptstyle A}_{\sf J}=
{\kappa}^{-1}j_{\sf J}(\rho)Y_{\sf J}(\hat{\bf r})(\mbox{\boldmath$\sigma$}\cdot\mbox{\boldmath$\nabla$});
\nonumber\\
{\mathcal M}^{\scriptscriptstyle A}_{{\sf MJ}}&=&\sum_{{\sf L}}i^{\sf  J-L-1}\ F_{{\sf MLJ}}j_{\sf L}(\rho)
\left[Y_{{\sf L}}(\hat{\bf r})\otimes{\mbox{\boldmath$\sigma$}}\right]_{{\sf J}};
\label{16}\ 
{\mathcal M}^{\scriptscriptstyle V}_{{\sf MJ}}={\kappa}^{-1}\sum_{{\sf L}}i^{\sf  J-L-1}
F_{{\sf MLJ}}j_{\sf L}(\rho) [Y_{\sf L}(\hat{\bf r})\otimes\mbox{\boldmath$\nabla$}]_{{\sf J}},\nonumber \\
\end{eqnarray}
where $\rho=\kappa r$.
Notice that the initial and final states of the matrix elements in Eq. \eqref{25} involve an isospin unit change, 
and implicitly contain isospin operators $\tau_\pm$.

\subsubsection{Fermi and Gamow-Teller matrix elements}

Most reactions in typical stellar scenarios involve small momentum transfer such that 
$\rho \ll 1$. In this case, the angular dependence of the above operators
becomes irrelevant. Using
$j_m(\rho)\sim \delta_{m0}$ for the spherical Bessel functions in
Eqs. (\ref{16}) and after some algebra one can
show that  (for charged-current)
\begin{equation}
{\mathcal T}_{\sigma}(\kappa,J_f)\sim {\mathcal C} \left[ |\left<{J_f}||\Sigma_{k=1}^A 
\tau_\pm(k)||{J_i}\right>|^2
+g_{\tiny A}^2|\left<{J_f}||\Sigma_{k=1}^A \sigma(k)
\tau_\pm(k)||{J_i}\right>|^2\right]
 \label{GT},
\end{equation}
where ${\mathcal C}$ is a function depending on the lepton and neutrino energies.
The $\tau_+$ operator corresponds to $\beta^-$ decay and the
$\tau_-$ to $\beta^+$ decay, so that 
$\tau_+$ $\mid$ n$\rangle$ = $\mid$ p$\rangle$ and 
$\tau_-$$\mid$ p$\rangle$ = $\mid$ n$\rangle$, changing a neutron
into a proton and vice-versa.

The spin-independent and spin-dependent operators appearing
on the right-hand-side of the above equation are known as the Fermi and Gamow-Teller operators.
The Fermi operator is the isospin raising/lowering operator:
in the limit of good isospin, which typically is good to 5\% or
better in the description of low-lying nuclear states,
it can only connect states within the same isospin multiplet.
That is, it is capable of exciting only one state, the state
identical to the initial state in terms of space and spin,
but with $(T,M_T) = (T_i, M_{Ti} \pm 1)$ for $\beta^-$ and
$\beta^+$ decay, respectively. 

\begin{figure}[t]
\begin{center}
{\includegraphics[width=8cm]{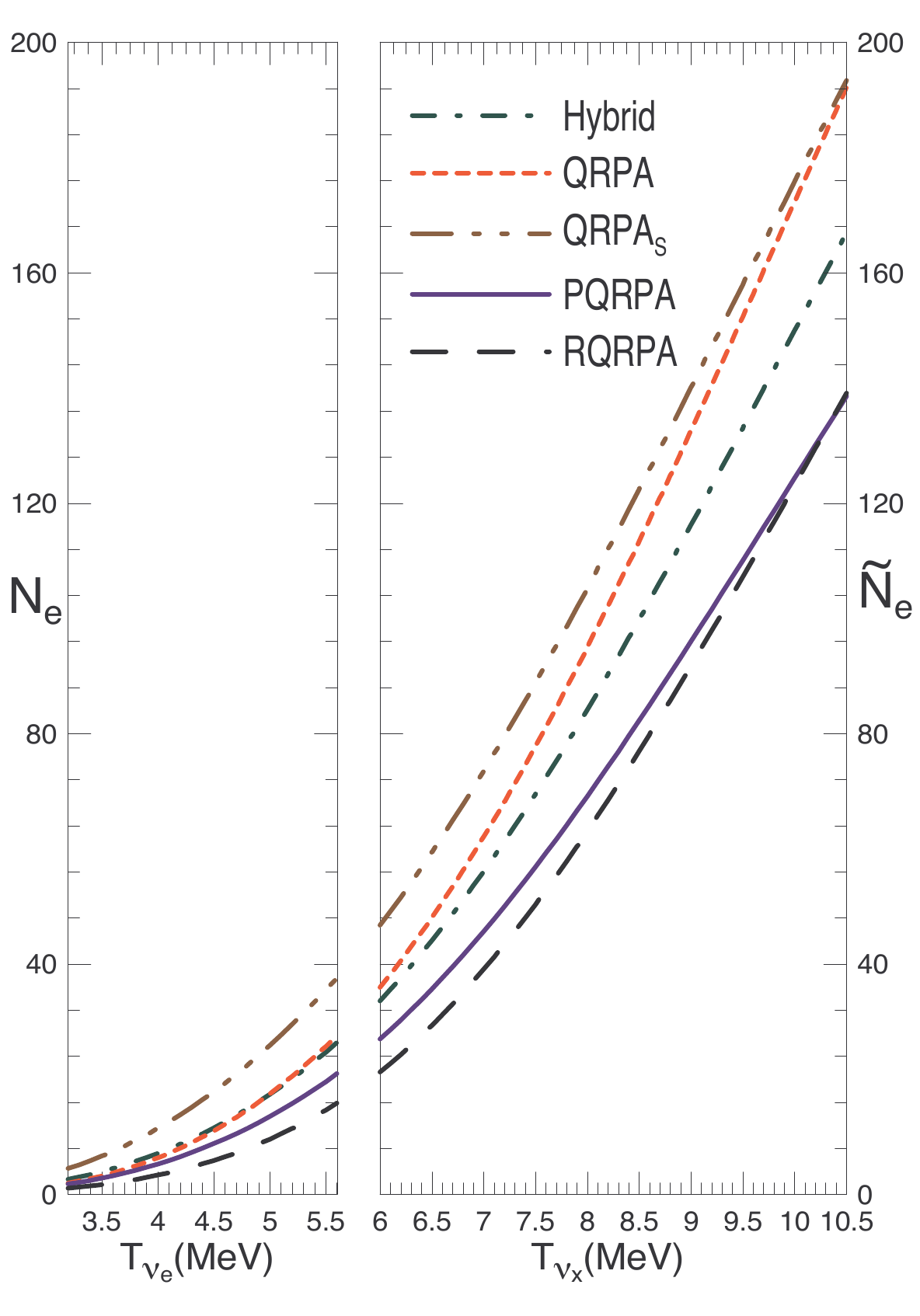}}
\end{center}
\vspace{-0.5cm}
\caption{
Number of events obtained from the convolution of the neutrino fluxes with
the cross section obtained with different nuclear structure
models: Hybrid (dashed-dot line)~\cite{Kol01},
quantum phase approximation (QRPA) (dashed line), QRPA$_S$ (dashed-dot dot
line)~\cite{Laz07}, projected QRPA (PQRPA) (solid line) \cite{SB08}, and
renormalized QRPA (RQRPA) (dashed line)~\cite{Paa07}. (Adapted from Ref. \cite{SB08}). 
}\label{neutfig}
\end{figure}

Eq. \eqref{GT} is only appropriate at the lowest-order expansion in $\rho$, when the nucleus
responds like an elementary particle.  Then we can
characterize its response by its macroscopic quantum numbers,
the spin and charge.  The next-to-leading-order term of the expansion in powers of $\rho$ 
probes the nucleus  at shorter
length scales. The
operators in Eq. \eqref{16} are obtained by an expansion of the plane-wave
lepton wavefunction, which, 
for not too large $\bf{k}$, becomes $\exp(i {\bf k} \cdot {\bf r}) \sim 1 + i {\bf k} \cdot {\bf r} $. 
Thus, the next term in the expansion includes a ``first forbidden'' term
$ \sum_{i=1}^A {\bf r}_i \tau_3(i) $
and similarly for the spin operator
$ \sum_{i=1}^A [{\bf r}_i \otimes \mbox{\boldmath$\sigma$}(i)]_{J=0,1,2} \tau_3(i) $.
These operators generate collective radial excitations,
leading to the so-called ``giant resonance'' excitations, with a typical excitation energy of
10-25 MeV.  They tend to exhaust  the Thomas-Reiche-Kuhn sum rule, 
\begin{equation} 
\sum_f | \langle f | \sum_{i=1}^A r(i) \tau_3(i) | i \rangle |^2
\sim {N Z \over A} \sim {A \over 4} \label{TRK}
\end{equation}
where the sum extends over a complete set of final nuclear states.
The first-forbidden operators tend to dominate the cross sections
for scattering the high energy supernova neutrinos ($\nu_{\mu}$s
and $\nu_\tau$s), with $E_\nu \sim$ 25 MeV, off light nuclei.
It also follows from Eq. \eqref{TRK} that the cross sections per
target nucleon are roughly constant. This conclusion changes when high energy neutrinos,
with high energy transfers, are considered.

The number of events detected for supernova explosions can be calculated as,
\[
N_{\alpha}=N_t \int_0^\infty  {\mathcal F}_\alpha(E_\nu) \cdot \sigma(E_\nu)
\cdot \epsilon(E_\nu) dE_\nu,
\label{4}\]
where the index $\alpha=\nu_e,{\bar \nu}_e,\nu_x$ and
$(\nu_x=\nu_\tau,\nu_\mu,{\bar \nu}_\mu,{\bar \nu}_\tau)$
indicates the neutrino or antineutrino type,
$N_t$ is the number of target nuclei, ${\mathcal F}_\alpha(E_\nu)$ is the
neutrino flux, $\sigma(E_\nu)$ is the neutrino-nucleus cross section,
$\epsilon(E_\nu)$ is the detection efficiency, and $E_\nu$ is the neutrino energy.

In Figure \ref{neutfig} (from reference \cite{SB08}), we show calculations  for the number of events 
of detected supernova neutrinos due to $(\nu_e+{\bar \nu}_e)$ interactions
on $^{56}$Fe, such as those performed in the KARMEN collaboration \cite{KAR98}.
The $N_{e}$ and ${\tilde N}_{e}$ are calculated as a function of
the neutrino temperatures $T_{\nu_e}$ and $T_{\nu_x}$, folding $\sigma_e(E_\nu)$
from different nuclear structure models with the neutrino fluxes
${\mathcal F}_{\nu_e}(E_\nu,T_{\nu_e})$ and ${\mathcal F}_{\nu_x}(E_\nu,T_{\nu_x})$,
respectively \cite{Aga07}. The fluxes depend on the distance to the supernova,
the neutrino energy, and the neutrino effective temperature. The number of events are
obtained from the convolution of the neutrino fluxes with the cross section
obtained with different nuclear-structure models: Hybrid (dashed-dot line)~\cite{Kol01},
quantum phase approximation (QRPA) (dashed line), QRPA$_S$ (dashed-dot dot
line)~\cite{Laz07}, projected QRPA (PQRPA) (solid line) \cite{SB08}, and
renormalized QRPA (RQRPA) (dashed line)~\cite{Paa07}. One clearly sees that the
differences between the calculated cross sections with different nuclear models
increase as a function of the neutrino temperatures. This is an example of the
limitations of nuclear models in describing weak-interaction processes in stars.

\subsection{Field theories}
Field theories adopt a completely independent approach for nuclear physics
calculations in which the concept of nuclear potentials is not used. The basic
method of field theories is to start with a Lagrangian for the fields. From this
Lagrangian one can ``read'' the Feynman diagrams and make practical
calculations, not without bypassing well-known complications such as
regularization and renormalization. Quantum chromodynamics (QCD) is the proper
quantum field theory for nuclear physics. But it is a very hard task to bridge
the physics from QCD to the one in low-energy nuclear processes.  Effective
field theory (EFT) tries to help in this construction by making use of the
concept of the separation of scales. One can form small expansion 
parameters from the ratios of short and long distance scales, defined by
\begin{equation}
\epsilon=\frac{\rm short \ distance \ scales}{\rm long \ distance \ scales}
\end{equation}
and try to explain physical observables in terms of powers of $\epsilon$.

In low-energy nuclear processes, the characteristic momenta are much smaller than
the mass of the pion, which is the lightest hadron that mediates the strong interaction.
In this regime, one often uses the pionless effective field theory, in which
pions are treated as heavy particles and  are integrated out of the theory
\cite{KSW96}. In this theory, the dynamical degrees of freedom are nucleons and
the pion and the delta resonance degrees of freedom 
are hidden in the contact interactions between nucleons. The scales of the problem are
the nucleon-nucleon scattering length, $a$, the
binding energy, $B$, and the typical nucleon momentum $k$ in the center-of-mass frame.
Then, the nucleon-nucleon interactions are calculated perturbatively with the small expansion parameter 
\begin{equation}
p={(1/a,B,k)\over \Lambda}
\end{equation}
which is the ratio of the light to the heavy scale. The heavy scale $\Lambda$ is set by the
pion mass ($m_\pi \sim 140$ MeV).

The pionless
effective Lagrangian will only involve the nucleon field
$\Psi^T=(p,n)$ and its derivatives. It must obey the symmetries
observed in strong interactions at low energies, such as parity,
time-reversal, and Galilean invariance. The Lagrangian can then be
written as a series of local operators with increasing dimensions.
In the limit where the energy goes to zero, the interactions of
lowest dimension dominate. To leading order ($LO$), the relevant
Lagrangian ($\hbar=c=1$) is given by
\begin{equation}
{\mathcal L}=\Psi^\dagger \left( i\partial_t+{\nabla^2\over
2m}\right)\Psi - C_0(\Psi^T {\mathcal P}\Psi)(\Psi^T{\mathcal
P}\Psi)^\dagger,
\end{equation}
where $m$ is the nucleon mass \cite{CRS99}. The projection operators
${\mathcal P}$ enforce the correct spin and isospin quantum numbers in
the channels under investigation. For spin-singlet interactions
${\mathcal P}_i=\sigma_2\tau_2\tau_i/\sqrt{8}$, while for spin-triplet
interactions ${\mathcal P}_i=\sigma_2\sigma_i\tau_2/\sqrt{8}$. 

\begin{figure}[t]
\begin{center}
{\includegraphics[width=12cm]{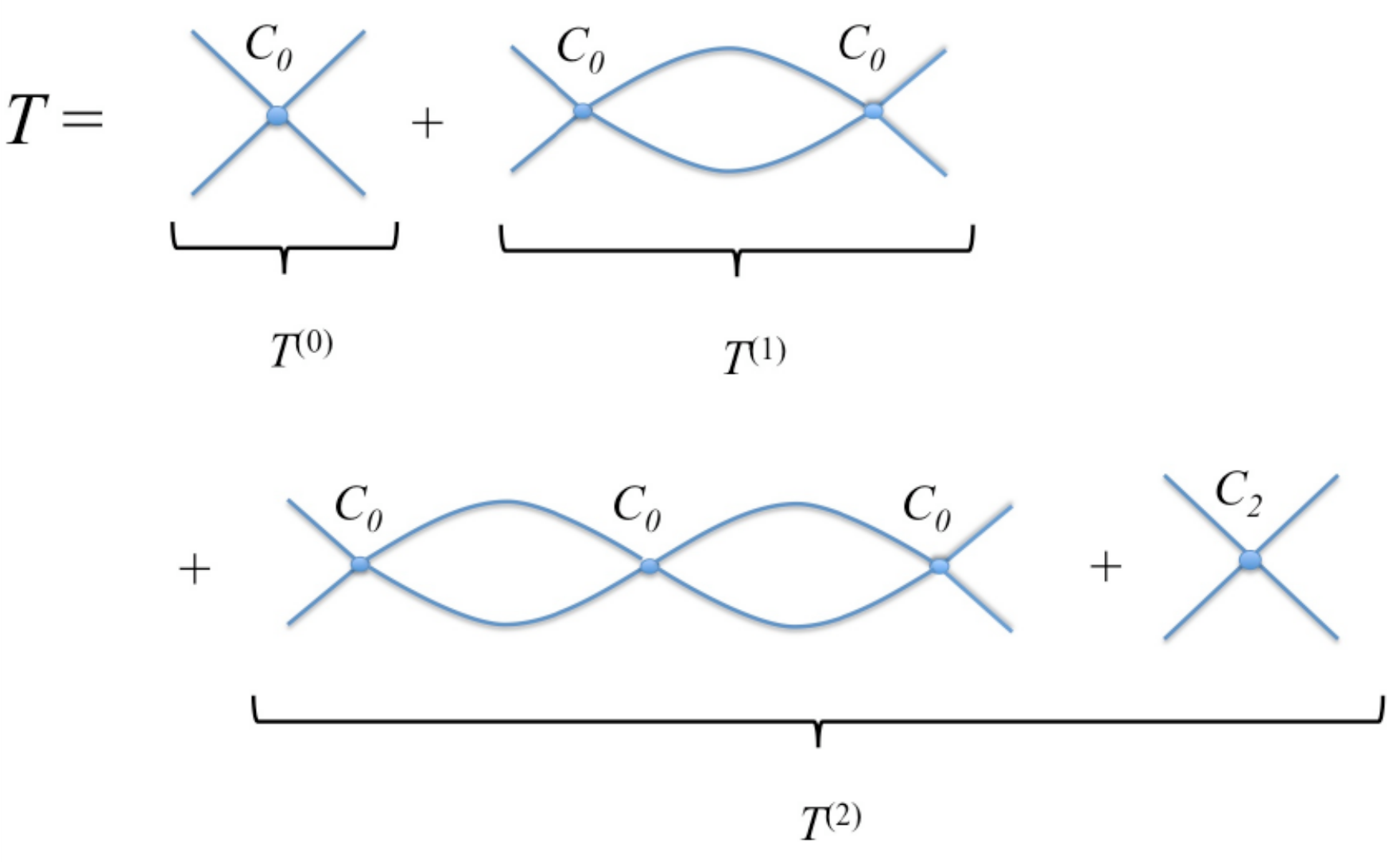}}
\end{center}
\vspace{-0.5cm}
\caption{\label{Teft}
Feynman diagram series for $NN$-scattering in pionless effective field
theory.}
\end{figure}

The Feynman-diagram rules can be directly ``read'' from the
Lagrangian at hand. In the case that the scattering length $a$ is
large, i.e., $a\gg 1/\Lambda$, as it is in the nucleon-nucleon
system, the full scattering amplitude $T$ is obtained from an
infinite sum of such Feynman diagrams (see Figure \ref{Teft}), leading to a
geometric series 
that can be written analytically as
\begin{equation}
T(p)={C_0 \over {1-C_0J(p)}}, \ \ \ \ \  \ \  J(p)=\int {d^3q\over
{(2\pi)^3}} {1\over {E-{\bf q}^2/m+i\epsilon}},
\end{equation}
where $E=p^2/m$ is the total center-of-mass energy (see, e.g., Ref.
\cite{BK02}). The integral is linearly divergent but is finite using {\it
dimensional regularization}. One gets $$J=-(\mu+ip){m\over 4\pi} ,$$ where
$\mu$ is the regularization parameter. The scattering amplitude
$T(p)$ has then the same structure as the $s$-wave partial-wave
amplitude, $$T=-{4\pi\over p\cot\delta-ip},$$ and one obtains the
effective range expansion for the phase shift $\delta$,
$$p\cot\delta=-{1\over a}+r_0{p^2\over 2}+\cdots$$ in the zero-momentum limit when
the coupling constant takes the renormalized value
\begin{equation}
C_0(\mu)={4\pi\over m}{1\over {(1/a-\mu)}}.
\end{equation}

In leading order, we see that the effective range vanishes, or
$r_0=0$. The small inverse $1/a$ scattering length is given by the
difference between two large quantities. For example, in
proton-neutron scattering we have $a_{pn}=-23.7$ fm in the {\it pn}
spin-singlet channel. Choosing the value $\mu=m_\pi$ for the
regularization parameter, one obtains $C_0 = 3.54$ fm$^2$. Physical
results should be independent of the exact value of the
renormalization mass $\mu$ as long as $1/a < \mu\ll m_\pi$.

The theory is now ready for practical applications. For example,
this procedure has been applied to obtain the electromagnetic form
factor of the deuteron, the electromagnetic polarizability and the
Compton scattering cross section for the deuteron, the radiative
neutron capture on protons, and the continuum structure of halo
nuclei. Based on the same effective field theory, the three-nucleon
system and neutron-deuteron scattering have been investigated
\cite{BK02}. Better agreement with data can be obtained in higher orders
(next-to-leading order [$NLO$], next-to-next-to leading order
[$N^2LO$], etc.). For nuclear processes involving momenta $p$ comparable to
$m_\pi$, the starting effective, pionfull, Lagrangian is more complicated. But
the basic field theoretic method remains the same. The EFT unifies
single-particle approaches in a model-independent framework, with
the added power counting that allows for an a priori estimate of
errors. Concepts of quantum field theory, such as
regularization and renormalization, are key ingredients of the
theory. 

\begin{figure}[t]
\begin{center}
{\includegraphics[width=12cm]{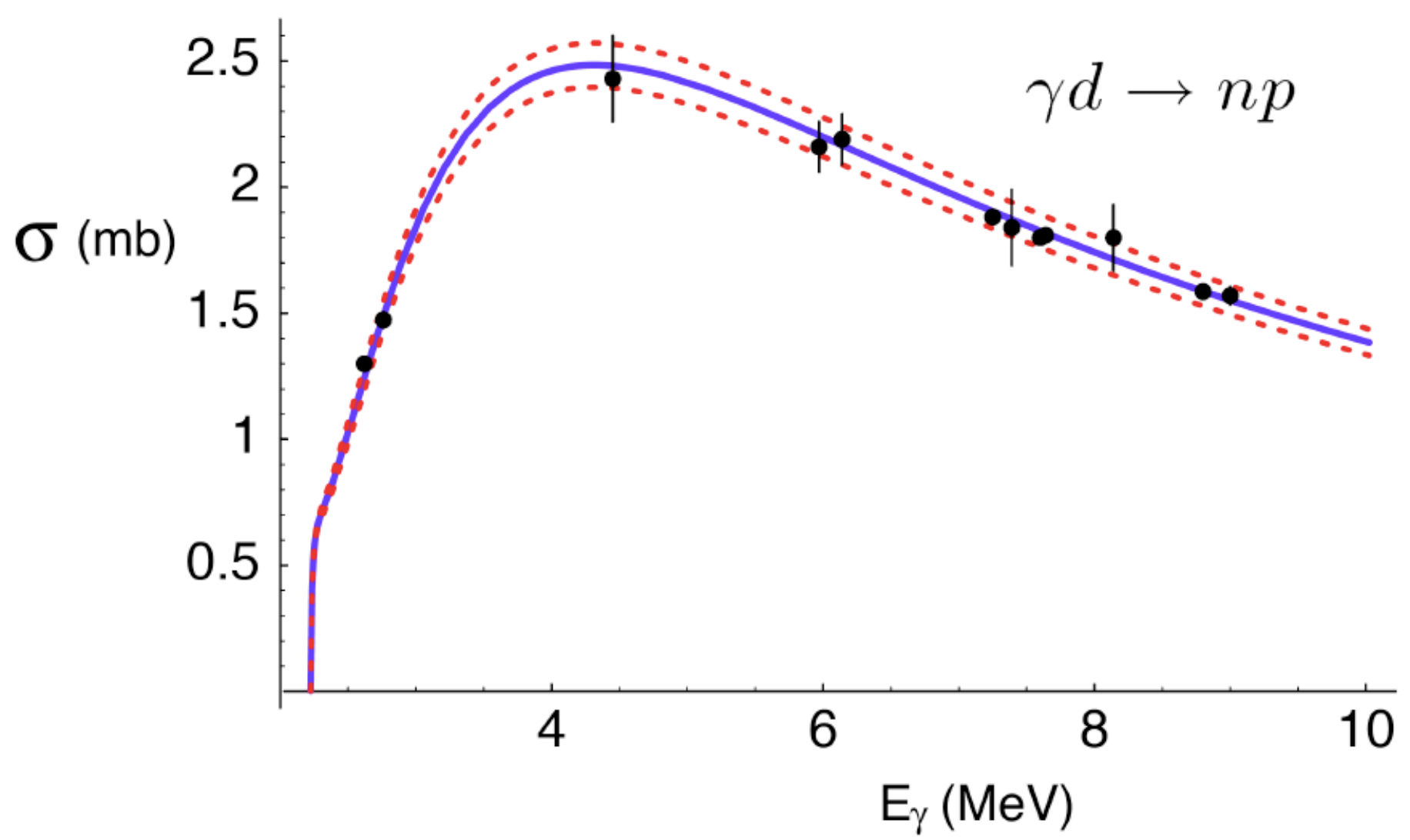}}
\end{center}
\vspace{-0.5cm}
\caption{
The cross section for 
$d \rightarrow np$. The curves correspond
to EFT cacluations for cold $np \rightarrow d$
and the dashed lines denote
the a 3\% theoretical uncertainty. Ref. \cite{Rup00} has further reduced this
uncertainty to below 
1\%. (Adapted from Ref. \cite{Rup00}).}\label{np}
\end{figure}

In nuclear astrophysics, this theory has been applied to $ np\rightarrow d\gamma$
for big-bang nucleosynthesis \cite{Chen99,Rup00}; $\nu d$ reactions for
supernovae physics \cite{KR99} and the solar $pp$ fusion process
\cite{But01}. EFT has also been used to deduce observables in reactions with
halo nuclei  and loosely bound states, with promising applications to
astrophysics \cite{BHK02,BHK03,HWK08}. 
So far, perhaps the most enlightening application of EFT for nuclear physics is
the  $np \rightarrow d \gamma$ cross section, specially because  there is no
data at the energies of relevance 
for the big bang nucleosynthesis (BBN). EFT has provided a calculation with 1\%
error \cite{Rup00} in the energy range relevant to BBN. The EFT predictions also 
agree with a very recent measurement of the inverse process in the same energy
region (see Figure \ref{np}). 

\section{Effects of environment electrons}

The form of the astrophysical $S$ factor given in Eq. \eqref{astrophys13a}
assumes that the electric charges 
of nuclei are ``bare'' charges. However, neither at very low laboratory
energies, nor in stellar environments this is the case. In stars, the
bare Coulomb interaction between the nuclei is screened by the
electrons in the plasma surrounding them. If one measures reaction
rates in the laboratory, using atomic targets (always), then atomic
electrons screen as well. But the screening is different from the
screening in the stellar plasma. Therefore we discuss these two problems
separately in the following subsections.

\subsection{Stellar electron screening problem}

In
astrophysical plasmas with high densities and/or low temperatures, effects of
electron screening are very important, as will be discussed later. 
This means that the reacting
nuclei, due to the background of electrons and nuclei, feel a different
Coulomb repulsion than in the case of bare nuclei. Under most conditions (with
non-vanishing temperatures), the generalized reaction-rate integral can be
separated into the traditional expression without screening \eqref{astrophys4}
and a screening factor
\begin{equation}
 \left<\sigma\mathrm{v}\right>_{j,k}^{\ast}=f_{scr}(Z_{j},Z_{k},\rho,T,Y_{i})
 \left<\sigma\mathrm{v}\right>_{j,k}, 
\end{equation}
in terms of the nuclear abundances, defined in section 2.2.

This screening factor is dependent on the charge of the involved particles,
the density, temperature, and the composition of the plasma.  At high densities
and low temperatures, screening factors
can enhance reactions by many orders of magnitude and lead to pycnonuclear ignition.

Consider a concentration of
negative and positive charges with neutral total charge, that is,
$ \sum_i Z_iec_{i0} =0$,
where $c_{i0}$ is the spatially uniform concentration of positive ($i = +$) or
negative ($i = -$) charges. Because of the interaction between the charges, these 
concentrations are no more spatially uniform, with smaller charges tending to concentrate around larger charges.

The concentrations around the charges are populated
according to the statistical distribution of the individual charge
energies in the presence of a Coulomb field $V(r)$, yet to be found.
Assuming Boltzmann statistics, this argumentation implies that
\begin{equation}
c_+(r)=c_{+0}\exp\left[-{Z_+eV(r)\over kT}\right] \ \ {\rm and} \ \
\ c_-(r)=c_{-0}\exp\left[-{Z_-eV(r)\over kT}\right] .
\end{equation}
If the ion close to which we are considering the screening is
positive, then $V(r)>0$ and $ c_+(r)<c_{+0}$, or $c_-(r)>c_{-0}$,
and the reverse is true if $V(r)<0$.

The charge density at position $r$ is given by
\begin{equation}
\rho(r)= \sum_i Z_iec_{i}=\sum_i Z_iec_{i0}\exp\left[-{Z_ieV(r)\over
kT}\right].
\end{equation}
If $ Z_ieV(r)/ kT \ll 1 \ (${\it weak \ screening}), then
$ \rho(r)=-(e^2V(r)/kT) \sum_iZ_i^2 c_{i0}$.

To obtain the potential $V(r)$ one has to  solve the
Poisson equation for the potential $V(r)$ which, for the above charge
distribution, becomes
\[
-{1\over r^2} {d\over dr} \left[ r^2{ dV\over dr}\right] = 4\pi
\rho(r)=\left({1\over R_D}\right)^2V,
\]
where the {\it Debye radius} $R_D$ is defined by \begin{equation}
    R_D^2={kT\over 4\pi e^2\sum_i Z_i^2c_{i0}} .
\end{equation}
Since $V(r) \rightarrow 0$ as $r \rightarrow•Ž\infty$, the solution
of this equation is $V(r)=(A/r) \exp(-r/R_D)$. The normalization
constant is fixed by the condition  $V(r) \rightarrow Z_ie/r$ as $r
\rightarrow•Ž0$. Thus, \begin{equation} V(r)={Z_ie\over r}
\exp\left(-{r\over R_D}\right)
\end{equation}

Screening modifies the Coulomb potential between the nuclear radius
$R$ and the classical turning point $R_0$, and consequently modifies
the barrier penetration. For weak screening $R_D \gg R, R_0$. In
other words, we can expand $V(r)$ around $r=0$. To first order, the
barrier energy for an incoming projectile with charge $Z_2e$ is
$
V(r)=Z_1Z_2e^2/ r +U(r)
$,
where the Debye-Hueckel {\it screening potential},
$U(r)=U(0)=const.$, is given by
$
U_0=-Z_1Z_2e^2/ R_D
$.

The impact of the screening potential on the barrier penetrability
and therefore on the astrophysical reaction rates can be
approximated through a screening factor $f=
\exp\left({U_0/ kT}\right)$,
which, in the weak screening limit, becomes $f \simeq 1-U_0/kT$.

In summary, for the weak screening limit, the reaction rates are
modified according to \begin{equation} \langle \sigma v
\rangle_{screened} = f\langle \sigma v \rangle_{bare}
\end{equation}
where
\begin{equation}
f=1+0.188{Z_1Z_2\rho^{1/2}\xi^{1/2}\over T_6^{3/2}},  \ \ \ {\rm
where} \ \ \xi =\sum_i (Z_i^2 +Z_i)^2 Y_i .
\end{equation}

\begin{figure}[tb]
\begin{center}
\includegraphics[
width=4in]
{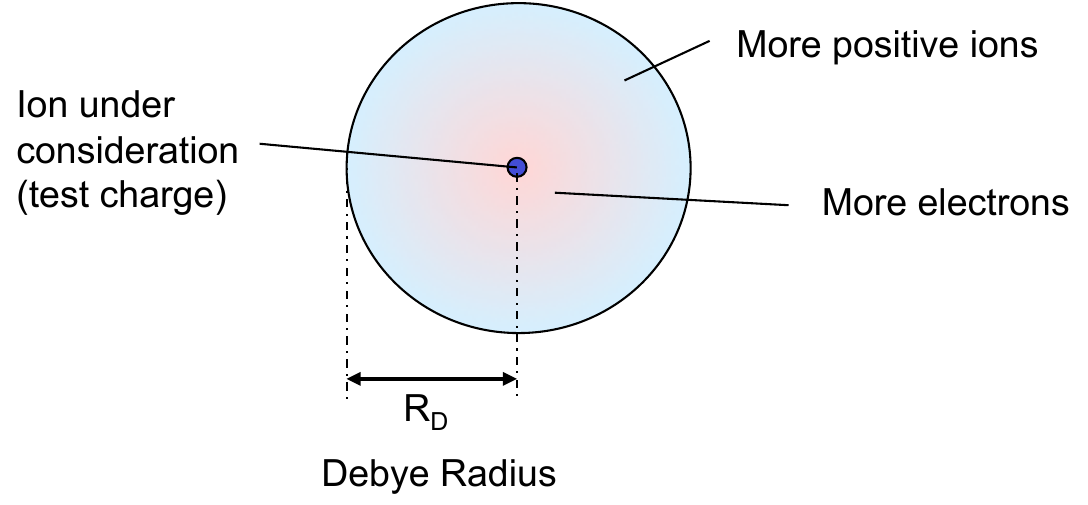}
\caption{Scematic view of the Debye-H\"uckel sphere. An ion at the center of the sphere 
is surrounded by a cloud of ions, with the ions of opposite charge (electrons) agglomerating closer to it.}  
\label{debye}
\end{center}
\end{figure}

The Debye-H\"uckel model (Figure \ref{debye}) is an 
ideal plasma since the average interaction energy between particles 
is smaller than the average kinetic energy of a particle  \cite{Kob95,Bai95}. In the case 
of {\it strong screening}, in low density plasmas, the potential energy cannot
not be described by the Debye-H\"uckel  
model since the probability of finding other charged particles in a Debye
sphere almost vanishes.  
For a strongly coupled plasma, the ion-sphere model \cite{Sal54,Fuj04}
is more suitable.  
The ion-sphere model is equivalent to the {\it Wigner-Seitz sphere} used in
condensed-matter theory.  
It assumes an ion having $Z_b$ bound electrons, positioned at $r=0$, and $Z_f$
free electrons ($Z_b+Z_f=Z$) occupying the rest of the ion sphere volume. The
plasma  effects are taken into account by confining 
the ion and the $Z_f$ electrons inside the ion sphere. To obtain the potential
$V(r)$ one adds to the bare ion potential, 
$ V=Ze^2/r$, the potential due to bound electrons, $V_b$, and that due to free
electrons, $V_f$. A Slater type, or Kohn-Sham type, exchange-potential is  
also added.  To obtain the bound and free electron densities one solves the
Schr\"odinger equation, or Dirac equation,  
with $V(r)$. From this, one builds the bound and free electron densities which are
then used to calculate the new potentials $V_b$ and $V_f$. This process is done
iteratively until convergence is reached.  

The plasma density enters the ion-sphere model through
the boundary conditions imposed on the potential, that is, through the
neutrality conditions of the ion-sphere.  
Approximate schemes to obtain the ion-sphere potential have been developed. A
widely used approximation for the potential energy of a single free electron
electron inside the ion-spehre is given by 
 \begin{equation} 
 V({\bf r},{\bf r}')=\bigg(-\frac{Ze^2}{r}+ \frac{e^2}{\vert{\bf r-r}'\vert} \bigg)
 \bigg[1-{r\over 2R_i}\bigg(3-\frac{r^2}{R_i^2}\bigg)\bigg]\theta(R_i-r), \label{ionsph}
 \end{equation} 
where ${\bf r}$ and ${\bf r}'$ are the positions of the bound electron and the
projectile ion, respectively,  
$\theta(R_i-r)$ is the step function, the ion-sphere radius $R_i(=[3(Z-1)/4\pi
n_{{e}}]^{1/3})$ is given by the plasma electron  
density $n_e$ since the total charge within the ion-sphere is neutral. This
hydrogenic ion-sphere potential \eqref{ionsph} can be generalized to $Z_f$ free
electrons inside the ion-sphere.   
The ion-sphere model also has its limitations. For instance, charge transfer
processes in collisions  
between positive ions in strongly coupled  plasmas has not been fully explored
and could modify the range of validity of the model  
 \cite{Jun05}.

For screening in plasmas with intermediate densities, i.e., when $n_eR\approx
1$, where $R=R_i$ or $R=R_D$,  more complicated models  are  necessary and are still
under theoretical scrutiny.  This is based on the simple observation that in the
stars along the main sequence, there is only about 1-3 ions within the Debye
sphere. Thus, in principle,  the Debye screening model should not be applicable
to screening in these  environments. Also, static models as the Debye-H\"uckel
and ion-sphere models do not contain dynamical effects due to the fast motion of
free electrons. Dynamical fluctuations, due to the fast motion of the electrons,
and non-spherical effects could modify the screening in non-static models. The
possibility of dynamic effects was first mentioned  in Ref. \cite{Mit77} and
studied in Ref. \cite{CSK88}. But, the existence  
of dynamic effects was criticized since the reacting particles are in
thermodynamic equilibrium and hence such an effect is not 
expected \cite{BS97}. According to Ref. \cite{GB98}, higher order effects, beyond
the Debye-H\"uckel approximation, modify 
the screening enhancement in solar fusion reactions by only a very small amount of about 1\%.
This conclusion is not in accord with results obtained by other authors \cite{SS01,Jun05}.

More recently, there have been additional claims that mean field models cease to
be valid 
under the conditions prevailing in stellar cores in general and in the Sun,
because particle fluctuations  within the Debye-H\"uckel sphere are percent-wise  
large.  These claims have been substantiated with molecular dynamics
calculations  \cite{SS01}. However, it has not been 
pursed further and has not been verified independently. This certainly deserves
further theoretical studies.

\subsection{Laboratory atomic screening problem}

\begin{figure}[tb]
\begin{center}
\includegraphics[
width=4in]
{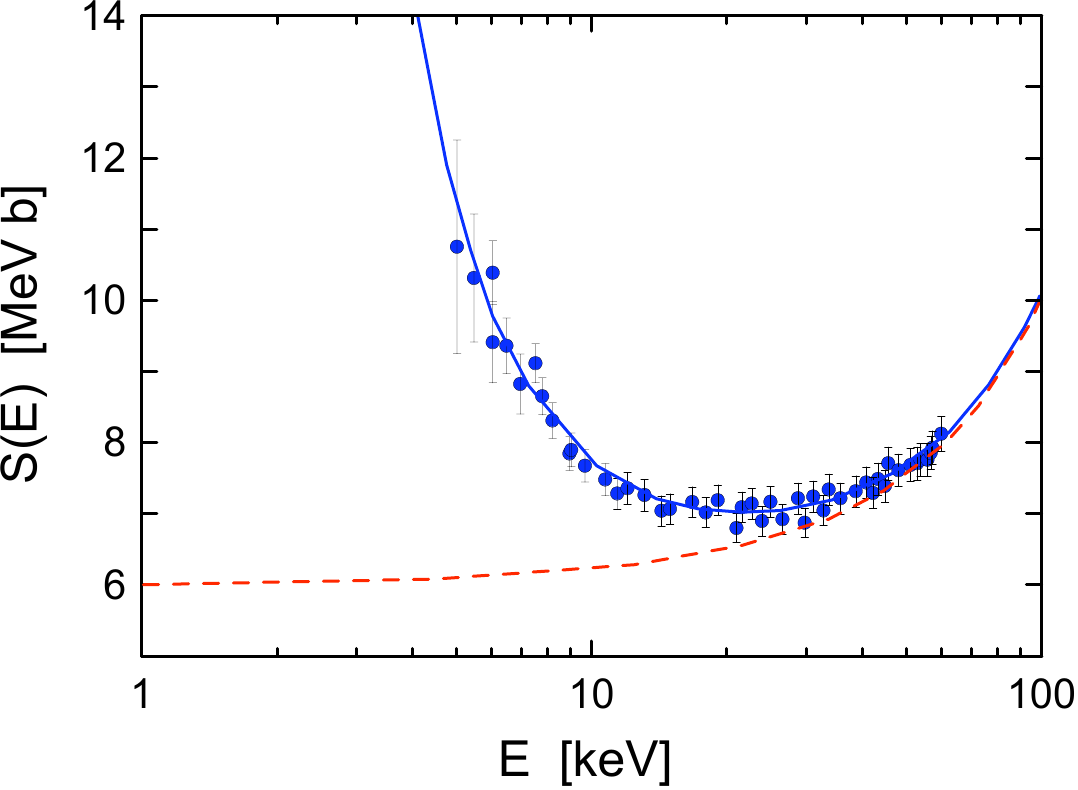}
\caption{$S$-factor data for the $^3$He($d,p$)$^4$He reaction from Ref. \cite{Ali01}.  The dashed
curve represents the $S$-factor for bare nuclei and the solid curve that for screened nuclei with
$U_e = 219$ eV.}  
\label{screening1}
\end{center}
\end{figure}

Laboratory screening has been studied in more detail 
experimentally, as one can control different charge states of the
projectile+target system in the laboratory \cite{Ass87,Rol95,Rol01}.
Experimental techniques improve steadily and one can measure fusion cross sections
at increasingly lower energies where the screened Coulomb potential
can be significantly smaller than the bare Coulomb potential. The deviation from the
bare Coulomb potential  is seen as an increase in the
astrophysical $S$-factor extracted at the lowest energies (see Figure
\ref{screening1}). This 
enhancement has been experimentally observed for a large number of
systems \cite {Eng88,Eng92,Ang93,Pre94,Gre95}. The screening effects
of the atomic electrons can be calculated \cite{Ass87} in the
adiabatic approximation at the lowest energies and in the sudden
approximation at higher energies with a smooth transition in between
\cite{Sho93}.

In the adiabatic approximation one assumes that the velocities of
the electrons in the target are much larger than the relative motion
between the projectile and the target nucleus. In this case, the
electronic cloud adjusts to the ground-state of a ``molecule''
consisting of two nuclei separated by a time-dependent distance
$R(t)$, at each time instant $t.$ Since the closest-approach
distance between the nuclei is much smaller than typical atomic
cloud sizes, the binding energy of the electrons will be given by
the ground-state energy of the $Z_P+Z_T$ atom, $B(Z_P+Z_T)$. Energy
conservation
implies that the relative energy between the nuclei increases by $%
U_e=B(Z_P+Z_T)-B(Z_T)$. This energy increment enhances the fusion
probability because the tunneling probability through the Coulomb barrier
between the nuclei increases accordingly. In other words, the fusion cross
section measured at 
laboratory energy $E$ represents in fact a fusion cross section at
energy $E+U_e$, with $U_e$ being known as the {\it screening
potential. }Using eq. \eqref{astrophys13a}, one gets for non-resonant reactions
\begin{equation}
\sigma \;(E+U_e)= \exp \left[ \pi \eta (E)\frac{U_e}E\right] \
\sigma (E)\;,  \label{sig1}
\end{equation}
where we assumed that the factor $S(E)/E$ varies much slower with
$E$, as compared to the energy dependence of $\exp \left[ -2\pi \eta
(E)\right] $ .

The exponential factor on the right-hand-side of eq. \eqref{sig1} is the
enhancement factor due to screening by the atomic electrons in
the target. For light systems, the velocity of the atomic electrons
is comparable to the relative motion between the nuclei. Thus, a
dynamical calculation is more appropriate to study the effect of atomic
screening \cite{Sho93}. However, the screening potential $U_e$
obtained from a dynamical calculation cannot exceed that obtained in
the adiabatic approximation because the dynamical calculation
includes atomic excitations, which reduce the energy transferred from
the electronic binding to the relative motion. The adiabatic approximation is thus
the {\it upper limit} of the enhancement due to laboratory screening.

The experimental value of $U_e$ needed to reproduce the experimental data
by using Eq. \eqref{sig1} are systematically larger than the  adiabatic model by
a factor of 2  \cite{Ass87,Rol95,Rol01}.  For example, the
cross section of the $^3$He($d,p$)$^4$He reaction was studied over a wide range of
energies \cite{Ali01}: the results led to $U_e =
219\pm 15$ eV, significantly larger than the adiabatic limit from atomic physics,
$U_{ad} = 119$ eV. Many theoretical attempts to
explain this puzzle have been carried out (see, e.g.,
refs. \cite{Sho93,BBH97,FZ99,HB02,Fio03}).  
The fusion cross sections change exponentially with a small variation of the
relative energy between 
the nuclei. Many small effects have been considered theoretically and, as shown
in Ref. \cite{BBH97}, 
they are not able to explain the differences between the experimental and theoretical
values of $U_e$. The calculated fractional change  in the cross sections
involving light nuclei at  
astrophysical energies are: (a)   vacuum polarization ($10^{-2}$), (b)
relativity ($10^{-3}$), (c) Bremsstrahlung ($10^{-3}$),  
(d) atomic polarization ($10^{-5}$) and nuclear polarization ($10^{-10}$) \cite{BBH97}. 
In Ref. \cite{Rol01} effects due
to thermal motion, vibrations inside atomic, molecular or crystal system, and
due to finite beam energy width were considered. All these effects are marginal
at the energies, which are presently measurable (at the level of $10^{-3}$, or below). 

A possible solution of the laboratory screening problem was proposed
in Refs. \cite{LSBR96,BFMH96}.
Experimentalists often use the extrapolation of the {\it
Andersen-Ziegler} tables \cite{AZ77} to obtain the average value of
the projectile energy due to stopping in the target material. The
stopping is due to ionization, electron-exchange, and other atomic
mechanisms. However, the extrapolation is challenged by theoretical
calculations, which predict a lower stopping. Smaller stopping was
indeed verified experimentally \cite{Rol01}. At very low energies,
it is thought that the stopping mechanism is mainly due to electron
exchange between projectile and target. This has been studied in
Ref. \cite{BD00} in the simplest situation; proton+hydrogen
collisions. The calculated stopping power was added to the nuclear
stopping power mechanism, i.e. to the energy loss by the Coulomb
repulsion between the nuclei (Rutherford scattering). The obtained stopping power is
shown to be smaller than the extrapolations from the Andersen-Ziegler, as
verified experimentally by \cite{GS91}. 

The stopping power in atomic He$^{+}+$He
collisions using the two-center molecular orbital basis was calculated in
Ref. \cite{Ber04}, and a good  agreement with the data of Ref.
\cite{GS91} at low energies was obtained. In particular, it was found that a
threshold effect exists, sharply decreasing the stopping power at lower energies
due to the disappearance of resonant tunneling in the electron-exchange
mechanism. The agreement with the data disappears when the nuclear recoil is
included. In fact, 
an unexpected ``quenching'' of the nuclear recoil  was observed 
experimentally in Ref. \cite{Form}, for stopping of deuteron  projectiles on
deuteron targets. But this cannot be explained in 
terms of a threshold cutoff effect and it  seems to 
violate a basic principle of nature, as the nuclear recoil is due to Coulomb
repulsion (Rutherford scattering) between 
projectile and target atoms. Energy loss due to Rutherford straggling is though
to be well described theoretically \cite{AZ77}.  
The fusion reaction $d(d,p)t$ was recently studied in deuterated metals and
insulators, i.e. for 58 samples across the periodic 
table, where a dramatic increase was observed for all the metals
\cite{Rai02}. The experimentally determined values of the screening energy are
about one order of magnitude larger than the value achieved in a gas-target
experiment and significantly larger than the theoretical predictions \cite{Cze01}.   
This result has not been proven independently.

The present status of the laboratory screening of nuclear fusion reactions is
rather confusing, and  more research is obviously needed. However, for the solar
fusion reaction chains, the effects  of laboratory and stellar screening by
electrons seem to be under control.

\section{Solutions with indirect methods}

\subsection{Elastic scattering}\label{sec:elastic_th}

Elastic scattering of nuclei is sensitive to their matter
distribution. This is due to the dependence of the optical potential on the
matter distribution of nuclei. Folding optical potentials, often used in the
analysis of experiments, depend on the nuclear projectile, $P$, and target, $T$,
densities as 
\begin{equation}
U_{opt}=(1+i\alpha)\int \rho_P({\bf r}_1) \rho_T ({\bf r}_2) v_{eff}({\bf
  r}_1,{\bf r}_2)d^3r_1d^3r_2, \end{equation} 
where $v_{eff}$ is the effective nucleon-nucleon interaction and $\alpha$ is a
parameter to fit the imaginary normalization of the optical potential. Nuclear
densities are a basic input in theoretical calculations of astrophysical
reactions at low energies.   

Elastic scattering in high-energy collisions essentially measures the Fourier
transform of the matter distribution. Considering for simplicity the
one-dimensional case, for light nuclei one has 
\begin{equation}
\int e^{iqx} \rho(x) dx \sim \int dx {e^{iqx} \over a^2+x^2} =
{\pi\over a} \ e^{-qa}, \end{equation}
where $q=2k\sin \theta/2$, for a
c.m. momentum $k$, and a scattering angle $\theta$. For heavy nuclei, 
the density $\rho$ is better described by a Fermi function, and
\begin{equation}
\int dx {e^{iqx} \over 1+e^{(x-R)/a}} \sim (4\pi)  \sin (qR) \ e^{-\pi q
a},
\end{equation}
for $R>>a$, and $%
qa>>1$. A similar result emerges from the elastic scattering amplitude, $F(q)$,
for the momentum transfer $q$, calculated in the eikonal approximation (see
below), i.e., 
\[
f_{\rm el}(q) \sim \int db b J_0(qb) [1-e^{i\chi(b)}] \sim \int db b {J_0(qb)
  \over 1+\exp[({b-R\over a})]} 
\sim {R\over q} J_1(qR) \exp(-\pi q a),
\] 
where $J_n$ are Bessel functions of order $n$.
Thus, the distance between minima in elastic scattering
cross sections measures the nuclear size, while its exponential
decay dependence reflects the surface diffuseness. 

\begin{figure}[t]
\begin{center}
{\includegraphics[width=9cm]{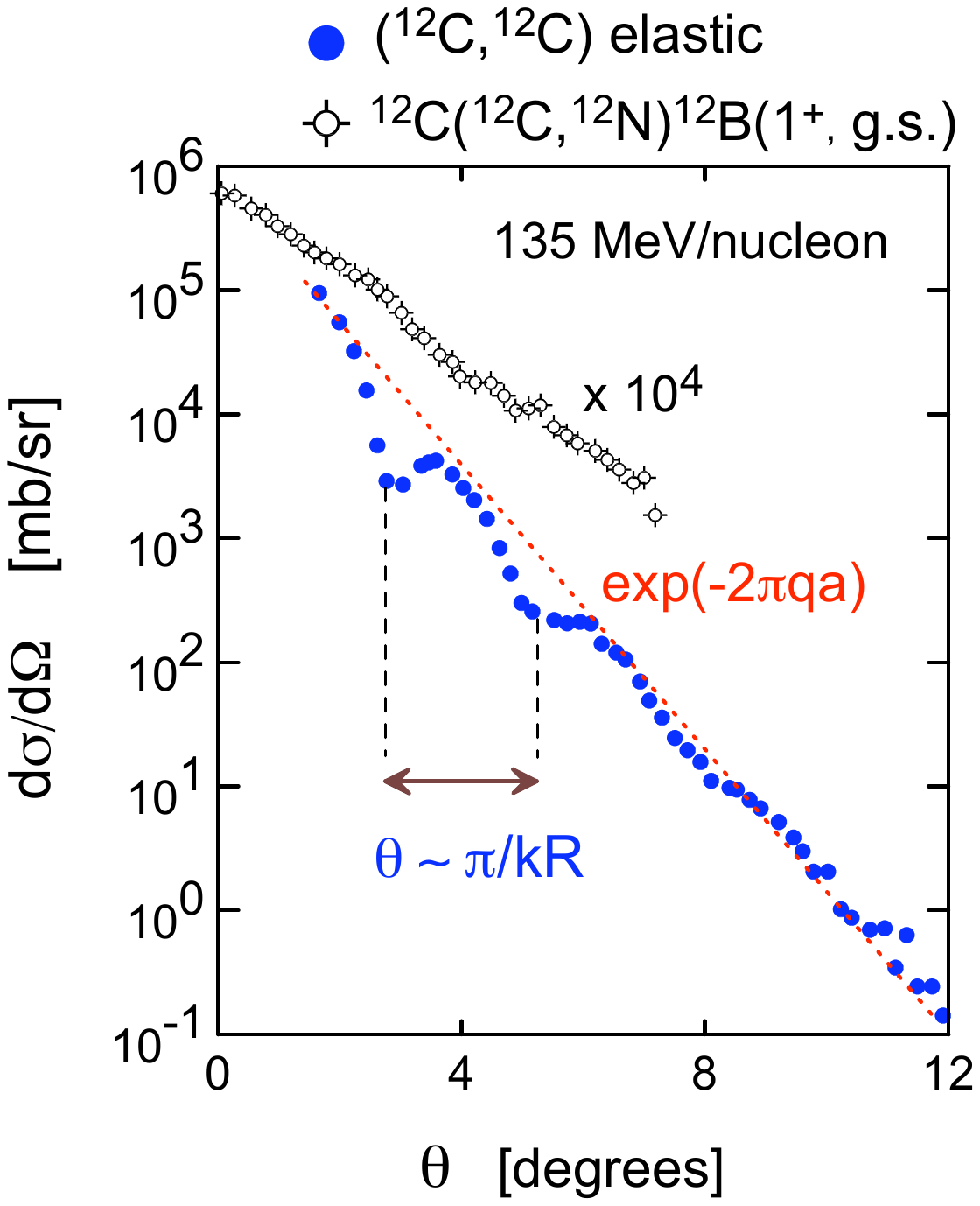}}
\end{center}
\vspace{-0.5cm}
\caption{\label{elastic}
Elastic scattering data for ($^{12}$C,$^{12}$C) is shown \cite{Ichi94,Ichi94b},
together with data for the inelastic scattering for the excitation of
Gamow-Teller states in the $^{12}$C($^{12}$C,$^{12}$N)$^{12}$B reaction at $E =
135$~MeV/nucleon.   
}
\end{figure}

During recent years, elastic proton scattering has been one of the
major sources of information on the matter distribution of unstable
nuclei at radioactive beam facilities. Information on the matter distribution of
many 
nuclei important for the nucleosynthesis in inhomogeneous Big Bang
and in r-processes scenarios could also be obtained from elastic
scattering experiments. Due to the loosely-bound character and small
excitation energies of many of these nuclei, high energy resolution
is often necessary. But the data deviate from this simple behavior as soon as
the energy transfer, or 
$Q$-value, differs from zero, as manifest in the inelastic scattering data.

For low-energy nucleus-nucleus scattering, the distorted wave Born approximation (DWBA) amplitude
is given by
\begin{equation}
f_{\mathrm{DWBA}}\;(\theta )=-\frac{\mu }{2\pi \hbar ^{2}}\int
\chi _{\beta}^{(-)^{\ast }}(\mathbf{k}^{\prime },\mathbf{r}\,)U_{opt}^{\alpha \beta} ({\bf r}) 
\chi _{\alpha}^{(+)}(\mathbf{k},\mathbf{r}\,)d^{3}r, \label{DWBA}
\end{equation}
where $\mu$ is the reduced mass,  $\chi _{\beta}^{(-)}$ ($\chi _{\alpha}^{(+)}$) are incoming (outgoing)
distorted waves in  channel $\alpha$ ($\beta$),  and $U_{opt}^{\alpha \beta}$ is the optical potential. 

Reactions with secondary beams have been studied at relatively high
energies, $\,E_{\mathrm{lab}}\gtrsim 30$~MeV/nucleon. The distorted waves can be
approximated by eikonal waves. This is 
valid for small-angle scattering and the scattering amplitude in the eikonal
approximation is  
\begin{equation}
f_{\text{el}}(\theta )=ik\int bdb\;J_{0}(qb)\left[ 1-e^{i\phi (b)}\right]
\label{fel(thet}
\end{equation}
where $k=\sqrt{2\mu E}/\hbar$ is the relative momentum, and
\begin{equation}
\phi (b)=\phi _{C}(b)+\phi _{N}(b)\;,\;\;\;\;\;\;\phi_{N}(b)=-\frac{1}{%
\hbar \mathrm{v}}\int_{-\infty }^{\infty }\,dz\,U\left[ \sqrt{b^{2}+z^{2}}%
\right] \quad ,
\label{phieik5}
\end{equation}
is the nuclear eikonal phase and $\chi _{C}(b)$ is the Coulomb eikonal phase
$$\phi_C(b)={2Z_1Z_2e^2 \over \hbar v}  \ln\,(kb).$$

Optical  potentials are usually parameterized in the form 
\begin{equation}
U_{N}(r) =-V_{R}\,f_{V}(r)-iW_{R}\,f_{W}(r)+4ia_{I}\,V_{I}\,\frac{d}{dr}%
f_{W}(r)  
+ 2\left( \frac{\hbar }{m_{\pi }c}\right) ^{2}\,\frac{1}{r} \frac{d}{dr}
\left[ V_{S}\,f_{S}(r)\right] \,(\mathbf{l\cdot s})+V_{\mathrm{coul.}}
\label{UN(r)=-}
\end{equation}
where 
\begin{equation}
f_{i}(r)={1\over  1+\exp [(r-R_{i})/a_{i}]}
\end{equation}
for $\,i=V,W\,$ and $\,S\,$; with $\,R_{i}=r_{i}\,A^{1/3}\,$. The first
(second) term is the usual real (imaginary) part of the optical potential.
The third term is peaked at the surface of the nucleus and is used to
simulate a stronger absorption of the incoming nucleon at the surface of the
nucleus. It is a correction due to the Pauli blocking effect. 
The last term in Eq.~\eqref{UN(r)=-} is a spin-orbit correction.  It causes
interference between the scattering from opposite sides of the nucleus.

Sometimes one has to go beyond the optical model description of inelastic
scattering and  
introduce the effects of  polarization potentials.  Under the action of a small
interaction, a wavefunction is modified in lowest order to 
\begin{equation}
|\psi _{n}^{\prime }\rangle =|\psi _{n}\rangle +\sum_{m\neq n}\,\frac{%
\langle \psi _{m}|U|\psi _{n}\rangle }{E_{n}-E_{m}}\,|\psi _{m}\rangle
\label{psinpol}
\end{equation}
If we assume that $\,|\psi _{n}\rangle \,$ is the ground state, this equation
says that during the action of the potential $\,U$, the wavefunction
acquires small components from excited states. At the end of this process,
the wavefunction can return to its initial state again. The modification of
the wavefunction during the action of the potential is called 
``polarization''. 

While elastic scattering data is considered a simpler way to access the sizes,
density profile, and other geometric features of nuclei,   
inelastic scattering requires many more pieces of information about the
intrinsic properties of nuclei. This is shown in Figure \ref{elastic},  
where the elastic scattering data for ($^{12}$C,$^{12}$C) is shown
\cite{Ichi94,Ichi94b} together with data 
for the inelastic scattering for the excitation of spin-dipole states in the
$^{12}$C($^{12}$C,$^{12}$N)$^{12}$B reaction at $E = 135$~MeV/nucleon.  
The beautiful exponential decrease of the cross section with the nuclear
diffuseness is clearly seen for the elastic scattering data. The  
inelastic data is much more sensitive to the models used to describe the nuclear
excitation. Sometimes, further complications, such as  
polarization effects, must be taken into account. The coupling to other
inelastic channels also has to be considered. This is often necessary and  
complicates the nice feature of elastic scattering as a probe of the nuclear
geometry and density profiles.   

\subsection{Coulomb excitation and dissociation}

In low-energy collisions, the theory of Coulomb excitation is very
well understood \cite{AW75} and has been used to analyze experiments on multiple
excitations and reorientation effects~\cite{Cline1988}. At the other end of the
beam energy scale -- Coulomb excitation of intermediate-energy or relativistic
heavy ions -- the kinematics is characterized by straight-line trajectories
\cite{WA79}. In the experiment, the selection of impact parameters $b$ exceeding
the sum of the radii of the two colliding nuclei by several fm (via restrictions
on scattering angles) keeps the colliding systems at ``safe'' distances,
minimizing the nuclear contribution to the excitation also in reactions above
the Coulomb barrier. Most Coulomb-excitation experiments at rare-isotope beam
facilities to date have been performed at intermediate bombarding energies of
50-100 MeV/nucleon.  It has been a very successful tool to extract precious
information on electromagnetic properties of nuclear transitions with relevance
to nuclear structure as well as nuclear astrophysics \cite{Glas98}. 

\subsubsection{Coulomb excitation}

\begin{figure}[t]
\begin{center}
{\includegraphics[width=9cm]{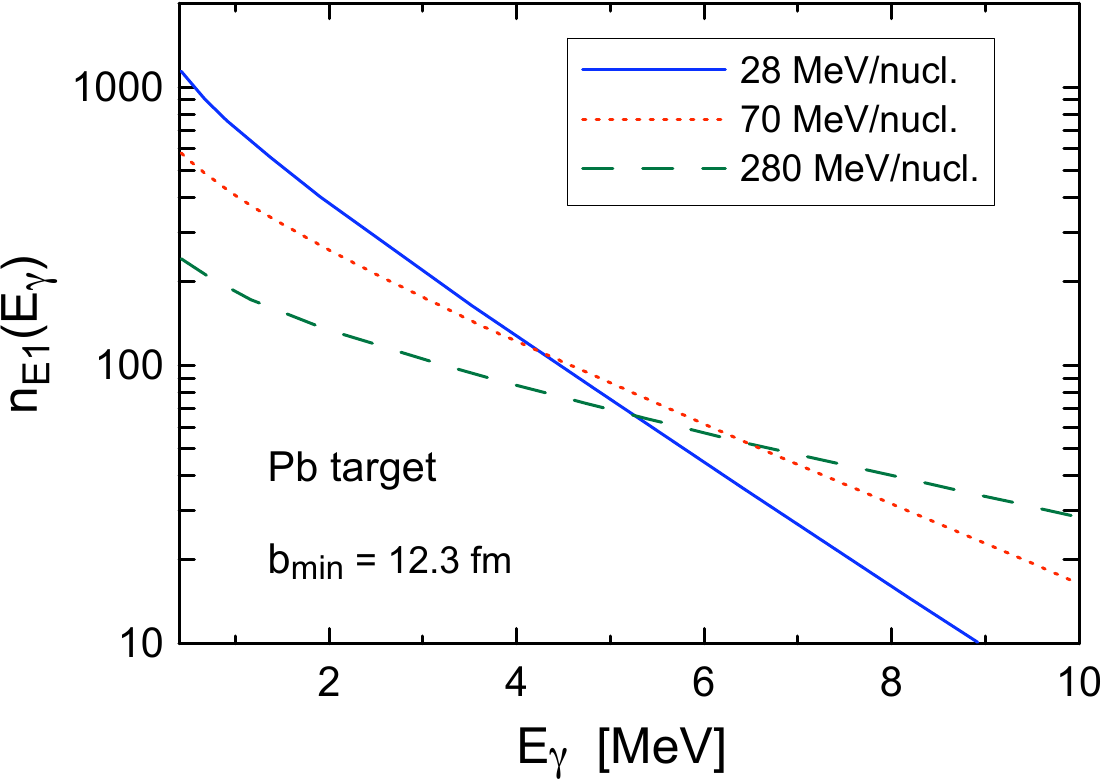}}
\end{center}
\vspace{-0.5cm}
\caption{
Total number of virtual photons for the $E1$ multipolarity, ``as seen'' by a
projectile passing by a lead target  at impact parameters $b_{min}=12.3$ fm and
larger, for three typical bombarding energies. (Adapted from T.\ Glasmacher~ \cite{Glas98}).} 
\label{vpn}
\end{figure}

Following multipole expansion of the electromagnetic field of a nucleus with charge $Z_2$ as it
evolves along a classical Rutherford   
trajectory, and with first order time-dependent perturbation theory, the
Coulomb excitation cross 
section is given by \cite{AW75}
\begin{equation}
{\frac{d\sigma_{i\rightarrow f}}{d\Omega}}=\left(
\frac{d\sigma}{d\Omega }\right)
_{\mathrm{el}}\frac{16\pi^{2}Z_{2}^{2}e^{2}}{\hbar^{2}} \sum
_{\pi\lambda\mu}{\frac{B(\pi\lambda,I_{i}\rightarrow
I_{f})}{(2\lambda +1)^{3}}}\mid
S(\pi\lambda,\mu)\mid^{2},\label{cross_2}
\end{equation}
where
$B(\pi\lambda,I_{i}\rightarrow I_{f})$ is the reduced transition
probability of the projectile nucleus, $\pi\lambda=E1,\ E2,$
$M1,\ldots$ is the multipolarity of the excitation, and
$\mu=-\lambda,-\lambda+1,\ldots,\lambda$.
The orbital integrals $S(\pi\lambda,\mu$) contain the information on
the dynamics of the reaction \cite{Ber88}. Inclusion of absorption
effects in $S(\pi\lambda,\mu$) due to the imaginary part of an
optical nucleus-nucleus potential were worked out in Ref.
\cite{BN93}. These orbital integrals depend on the Lorentz factor
$\gamma=(1-v^{2}/c^{2})^{-1/2}$, with $c$ being the speed of light,
on the multipolarity $\pi\lambda\mu$, and on the adiabacity
parameter $\xi (b)=\omega_{fi}b/\gamma v<1$, where
$\omega_{fi}=\left( E_{f}-E_{i}\right) /\hbar=E_x/\hbar$ is the excitation
energy (in units of $\hbar$) and $b$ is the impact parameter.

\begin{figure}[t]
\begin{center}
{\includegraphics[width=10cm]{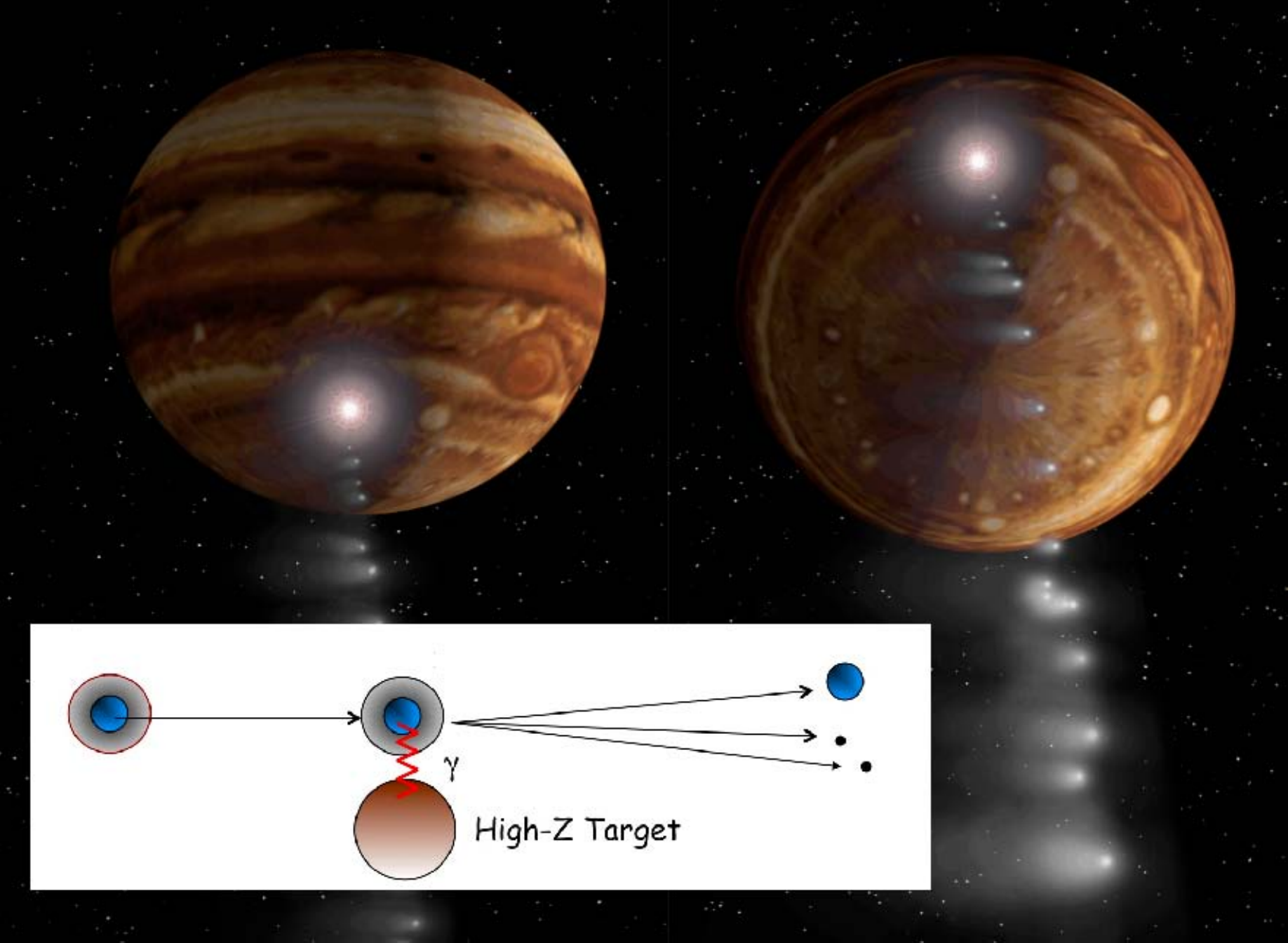}}
\end{center}
\caption{\label{comet}
Jupiter and comet Shoemaker-Levy 9, as imaged by the
Hubble Space Telescope (HST), on May 18, 1994, when the giant planet was at a
distance of 420 million miles (670 million km) from Earth. The gravitational interaction
of Jupiter with the comet has broken it up into many pieces (picture from
NASA). Much stronger tidal forces 
occur when nuclei come close to each other due to their mutual electromagnetic
fields.  The insert shows
the dissociation of a fast nuclear projectile passing by the Coulomb field of a
high-$Z$ projectile.}
\end{figure}

Because, the Coulomb excitation process is an external process, i.e., it occurs
when the nucleons from one nucleus  
are outside the nuclear matter distribution from the other nucleus, the
matrix elements for Coulomb excitation are the  
same as those for excitation by real photons (except for $E0$ Coulomb
excitations, which are extremely small). Therefore,  
Coulomb excitation cross sections are directly related to the
photonuclear cross sections by means of the equation \cite{Ber88} 
\begin{equation}
\frac{d\sigma _{C}\left( E_{x}\right) }{dE_{x}}=\sum_{E\lambda }\frac{%
n_{E\lambda }\left( E_{x}\right) }{E_{x}}\sigma _{E\lambda }^{\gamma }\left(
E_{x}\right) +\sum_{M\lambda }\frac{n_{M\lambda }\left( E_{x}\right) }{E_{x}}%
\sigma _{M\lambda }^{\gamma }\left( E_{x}\right) \;,  \label{sigmac}
\end{equation}
where $\sigma {_{\pi \lambda }^{\gamma }}$ $\left( E_{x}\right) \;$are the
photonuclear cross sections for the multipolarity $\pi \lambda $  and $E_{x}$ is the excitation energy.

The photonuclear cross sections are related to the reduced matrix elements,
for the excitation energy $E_{x}$, through the relation \cite{Ber88} 
\begin{equation}
\sigma _{\gamma }^{\pi
\lambda }(E_{x})=\frac{(2\pi )^{3}(\lambda +1)%
}{\lambda \left[ (2\lambda +1)!!\right] ^{2}}\left( \frac{E_{x}}{\hbar c}%
\right) ^{2\lambda -1}\frac{dB\left( \pi \lambda, E_{x}\right)}{dE_{x}}  \label{(1.2)}
\end{equation}
where $dB/dE_{x}$ are the electromagnetic response
functions, such that $$B(\pi\lambda,I_{i}\rightarrow I_{f})=\int dE_x \ {dB\left(
  \pi \lambda, E_{x}\right)\over dE_x}.$$ 
For differential cross sections one obtains 
\begin{equation}
\frac{d\sigma_C (E_{x})}{d\Omega }=\frac{1}{E_{x}}\sum\limits_{\pi \lambda }%
\frac{dn_{\pi \lambda }}{d\Omega }(E_{x},\theta )\sigma _{\gamma }^{\pi
\lambda }(E_{x}),  \label{(1.6)}
\end{equation}
where $\Omega$ denotes to the solid scattering angle.

Due to the use of high-energy projectiles in radioactive beam facilities, it is
important to account for the strong absorption  
properly, as it occurs at small impact parameters. The eikonal formalism developed in Ref.
\cite{BN93} is appropriate for this purpose.  The virtual photon numbers in Eq. \eqref{sigmac} become
\begin{equation}
n_{\pi \lambda }(E_x )=Z_{1}^{2}\alpha \ {\frac{\lambda \bigl[(2\lambda
+1)!!\bigr]^{2}}{(2\pi )^{3}\ (\lambda +1)}}\ \sum_{m}\ |G_{\pi \lambda
m}|^{2}\ g_{m}(E_x)\;,  \label{n}
\end{equation}
and 
\begin{equation}
g_{m}(E_x )=2\pi \ \biggl({\frac{E_x }{\gamma \hbar v}}\biggr)^{2}\ \int db\
b\ K_{m}^{2}\biggl({\frac{E_x b}{\gamma \hbar v}}\biggr)\ \exp \bigl\{-2\ \phi
_{I}(b)\bigr\}\;,  \label{g}
\end{equation}
where $\phi _{I}(b)$ is the imaginary part of the eikonal phase $\phi (b)$, Eq. \eqref{phieik5}.
The functions $G_{\pi \lambda
m}(c/v)$ are given in Ref. \cite{WA79}.

In Figure \ref{vpn} we show a calculation (with $E_\gamma \equiv E_x$) of the
virtual photons for the $E1$ multipolarity, ``as seen'' by a  
projectile passing by a lead target at impact parameters equal to and exceeding
$b=12.3$ fm, for three  typical bombarding energies. As the projectile energy  
increases, more virtual photons of large energy are available for the
reaction. This increases the number of states accessed in the excitation
process. 

\subsubsection{Coulomb dissociation}\label{sec:CD}

\begin{figure}[t]
\begin{center}
{\includegraphics[width=10cm]{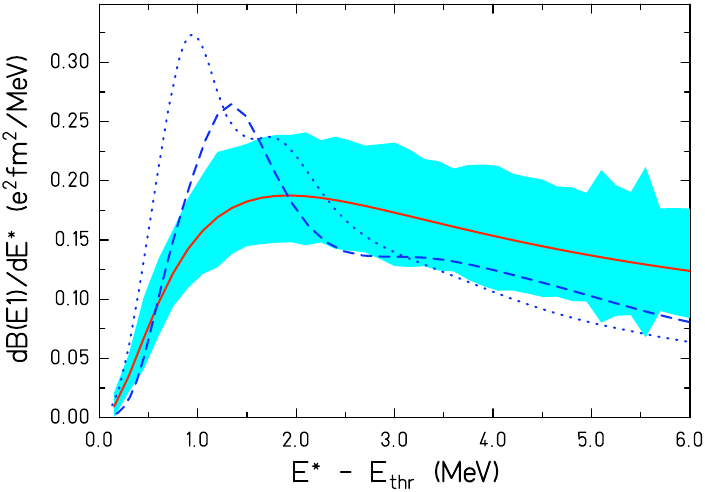}}
{\includegraphics[width=10cm]{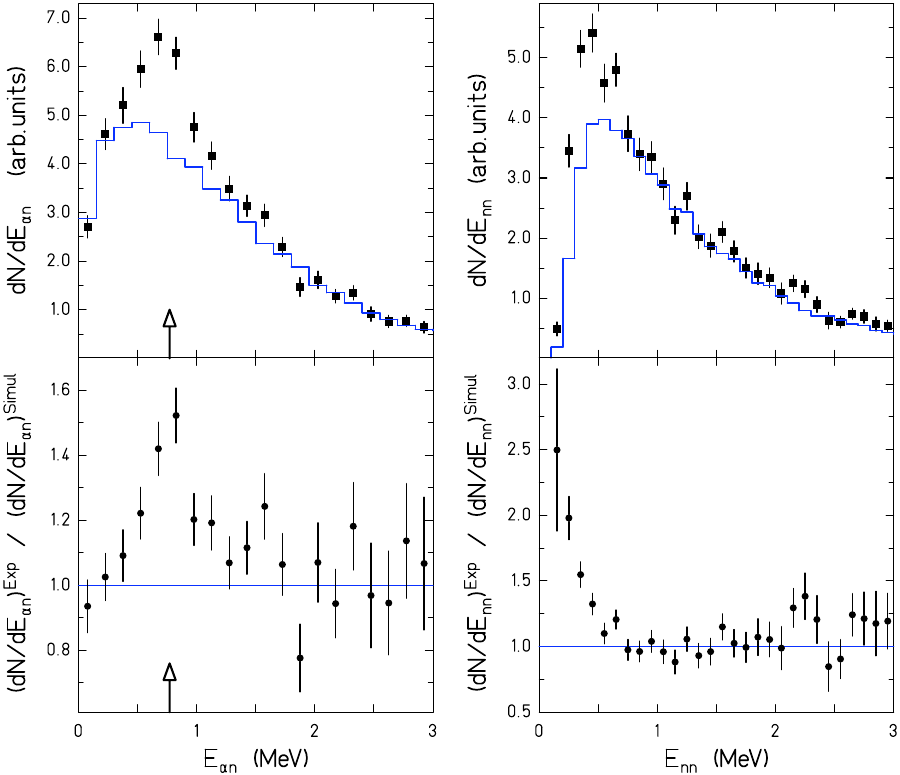}}
\end{center}
\vspace{-0.5cm}
\caption{
Upper figure: Electric dipole response function for $^6$He. The shaded area
represents the experimental results from a Coulomb dissociation experiment
\cite{aumann99}. The dashed and dotted lines correspond to results from  
three-body decay models from Refs. \cite{Danilin98,Cobis97}. Lower figure:
Measurement of two-body correlations in the three-body decay of $^6$He. The
lower panels show the ratio between the measured $\alpha-n$ and $n$-$n$
relative-energy spectra (upper panels) and the spectra simulated (histograms)
according to standard phasespace distributions \cite{aumann99}. (Courtesy of
T. Aumann).}  
\label{aumannfig}
\end{figure}

Coulomb dissociation is a process  analogous to what happened to the comet
Shoemaker-Levy as it disintegrated during its approximation  
to Jupiter in 1994 (see Figure \ref{comet}).  Approximately 1.5 to 2.2 hours
after closest approach, the comet (which was presumably a  
single body at the time) was broken apart by tidal forces into at least 21
pieces. The pieces continued to orbit Jupiter with a period of approximately 2
years. Due to gravitational forces from the Sun, which changed the orbits
slightly on the next approach to Jupiter, the pieces impacted the planet. 
Much stronger tidal forces occur when nuclei come close to each other due to
their mutual electromagnetic field. This leads to their dissociation,  
especially for weakly-bound nuclei (see insert to Figure \ref{comet}).

The idea behind the Coulomb dissociation method is relatively simple. The
(differential, or angle-integrated) Coulomb breakup cross 
section for $a+A\rightarrow b+c+A$ follows from Eq.
\eqref{cross_2}. It can be rewritten as
\begin{equation}
{d\sigma_{C}^{\pi\lambda }(E_\gamma)\over
d\Omega}=F^{\pi\lambda}(E_\gamma;\theta;\phi)\ .\ \sigma_{\gamma+a\
\rightarrow\ b+c}^{\pi\lambda}(E_\gamma),\label{CDmeth}
\end{equation}
where $E_\gamma$ is the energy transferred from the relative motion to
the breakup, and $\sigma_{\gamma+a\ \rightarrow\
b+c}^{\pi\lambda}(E_\gamma)$ is the photo-dissociation cross section for
the multipolarity ${\pi\lambda}$ and photon energy $E_\gamma$.  Time reversal
allows one to deduce the radiative
capture cross section $b+c\rightarrow a+\gamma$ from $\sigma_{\gamma+a\ \rightarrow\ b+c}%
^{\pi\lambda}(E_\gamma)$,
\begin{equation}
\sigma^{\pi\lambda}_{b+c\rightarrow a+\gamma}(E_\gamma)=
{2(2j_a+1)\over (2j_b+1)(2j_c+1)}{k^2\over k_\gamma^2}\sigma_{\gamma+a\ \rightarrow\ b+c}%
^{\pi\lambda}(E_\gamma),
\label{CDmeth5}
\end{equation}
where $k^2=2m_{bc}(E_\gamma-S)$ with $S$ equal to the separation energy, and $k_\gamma=E_\gamma/\hbar c$.

 This method was proposed in Ref. \cite{Baur1986} and has
been tested 
successfully in a number of reactions of interest to astrophysics.
The most celebrated case is the reaction
$^{7}$Be$(p,\gamma)^{8}$B, first studied in Ref. \cite{Motobayashi1994},
followed by 
numerous experiments in the last decade (see Section~\ref{sec:CDexp} for the
details of Coulomb dissociation experiments).   

The two-neutron capture on $^4$He could perhaps play a role in the post-collapse
phase in type-II supernovae. The bottleneck in this nucleosynthesis scenario
is the formation of nuclei with $A \ge 9$ from nucleons and
$\alpha$-particles. In principle, the reaction $^4$He$(2n,\gamma)^6$He could  be
relevant in bridging the instability gap at $A=5$, although it is believed that this
reaction cannot compete with the ($\alpha n,\gamma$) process in a type-II
supernova scenario. Experiments with Coulomb dissociation have been used to
study this question, as shown in the example presented in Figure
\ref{aumannfig}. The upper part of the figure  displays the  electric dipole
response function for $^6$He. The shaded areas represent the experimental
results from a Coulomb dissociation experiment \cite{aumann99}. The dashed and
dotted lines correspond to results from  
three-body decay models from Refs. \cite{Danilin98,Cobis97}. In the lower figure 
we show the measurement of two-body correlations in the three-body decay
of $^6$He. The lower panels display the ratio between the measured $\alpha$-n and
$n$-$n$ relative-energy spectra (upper panels) and the spectra simulated
(histograms) according to standard phasespace distributions
\cite{aumann99}. From the analysis of this experiment it was found that 10\% of
the dissociation cross section proceeds via the formation of $^5$He. A rough
estimate yields 1.6 mb MeV for the photoabsorption cross section for
$^6$He$(\gamma,n)^5$He, which agrees with theoretical calculations
\cite{Efros96}. From this experiment one concludes that the cross sections for
formation of $^5$He and $^6$He via one (two) neutron capture by $^4$He are not
large enough to compete with the  ($\alpha$n, $\gamma$) capture process (for
more details, see ref. \cite{Aumann2006}). Nonetheless, this and the previously
mentioned examples, show the relevance of the Coulomb dissociation method to
assess some of the basic questions of relevance for nuclear astrophysics.  

\subsubsection{Higher-order effects and nuclear dissociation}

Eq. \eqref{CDmeth} is based on first-order perturbation theory. It
further assumes that the nuclear contribution to the breakup is small,
or that it can be separated under certain experimental conditions.
The contribution of the nuclear breakup has been examined by several
authors (see, e.g. \cite{BN93}). For example, $^8$B has a small proton separation
energy ($\approx 140$ keV). For such loosely-bound systems it was shown that
multiple-step, or higher-order effects, are 
important \cite{BB93,Bert05}. These effects occur by means of
continuum-continuum transitions. The role of higher multipolarities
(e.g., $E2$ contributions \cite{EB96} in the 
breakup reactions and the coupling to high-lying states)
also has to be investigated carefully. Investigations related to the effects of
relativity have attracted much theoretical interest as well 
\cite{Bert03,Sch08,Esb08,OB09}. 

\subsection{Transfer reactions}\label{sec:trf_th}

Transfer reactions, $A(a,b)B$, are effective when a momentum matching
exists between the transferred particle and the internal particles
in the nucleus. Therefore, beam energies should be in the range of a few
10 MeV/nucleon \cite{angela}.
One of the many advantages of using transfer reaction techniques
as surrogates for direct measurements in nuclear astrophysics is to avoid the
treatment of the screening problem \cite{Con07}.

\subsubsection{Trojan horse method}

\begin{figure}[t]
\begin{center}
{\includegraphics[width=15cm]{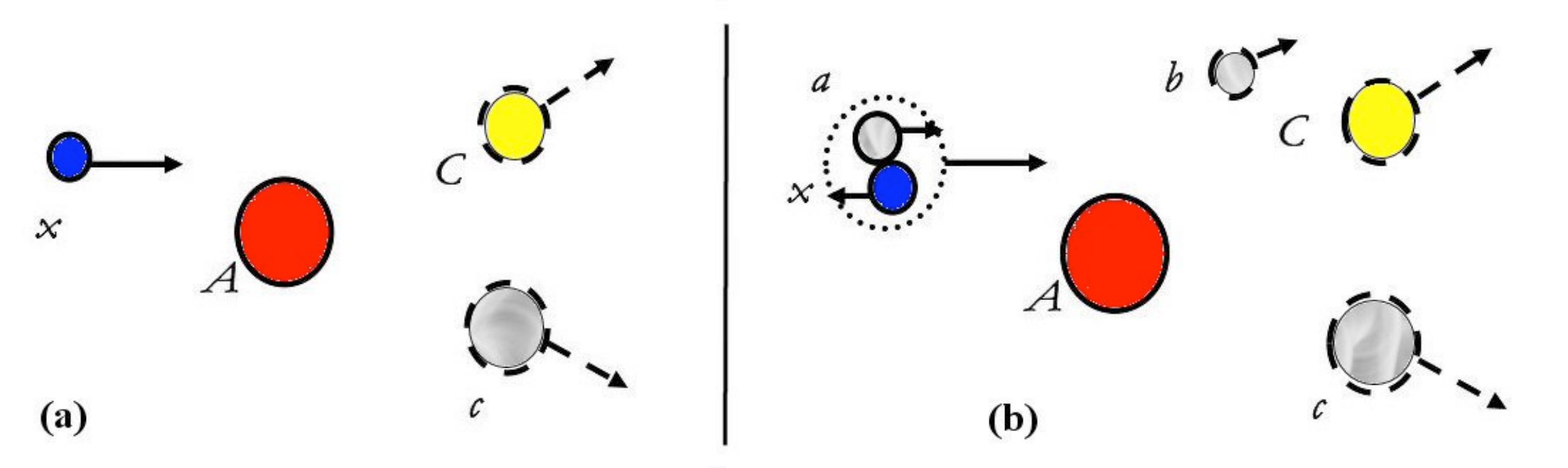}}
\end{center}
\caption{\label{trojan}
(a) Charged-particle
reactions at low energies, $ A+x\rightarrow c+C$, are strongly hindered by
Coulomb repulsion. (b) But, if particle $x$ is 
brought into the reaction zone of nucleus $A$
inside a projectile $a = (b+x)$ with velocity $v_a$, it can induce
a reactions at low energies, corresponding to $v_x
\sim v_a - v_{Fermi} \ll v_a$.}
\end{figure}

Low-energy reaction cross sections of astrophysical interest can be extracted
directly from breakup reactions $A+a \rightarrow b+c+C$ by means
of the {\it Trojan horse} method, as proposed in Ref. 
\cite{Bau86} (see Figure \ref{trojan}). If the Fermi momentum of the particle
$x$ inside $a=(b+x)$ compensates for the initial projectile velocity $v_a$, the
low-energy reaction $A+x=C+c$ is induced at very low (even
vanishing) relative energy between $A$ and $x$ (for more details, see Refs. \cite{Typ00,TyB03}). 

Without loss of generality, we neglect the effects of spin, and
write the cross section for $A+x\rightarrow c+C$ at an energy $E_x$ close to the
threshold as 
\begin{equation}
\sigma_{A+x\rightarrow B+c} ={\pi\over k_x^2}\sum_l (2l+1)|S_{lx}|^2 ,
\end{equation} 
where $k_x$ is the wave number in the incident channel, and  $S_{lx}$ is the
matrix element for the l-th partial wave.

In DWBA, the cross section for the breakup reaction $A+a \rightarrow b+c+C$
is\footnote{In a simple two-body reaction, $T$ is  
related to the scattering amplitude, Eq. \eqref{DWBA}, by $f=-\mu T/(2\pi \hbar
^{2})$.} 
\begin{equation}
{d^3\over d\Omega_b d\Omega_c dE_b} ={m_am_bm_c\over (2\pi)^5\hbar^6}
{k_bk_c\over q_a} \left|\sum_{lm} T_{lm}({\bf k}_a, 
{\bf k}_b, {\bf k}_x) S_{lc} Y_{lm}({\bf k_c})\right|^2, \label{threebody}
\end{equation}
with  
\begin{equation} T_{lm}= \left<\chi_b^{(-)} Y_{lm} f_l
\left|V_{bx}\right|\chi_a^{+}\phi_{bx}\right>,
\end{equation}
where the $\chi$'s denote the scattering waves and the interaction between
$b$ and $x$ is denoted by $V_{bx}$. The internal projectile wavefunction
corresponding to this potential is $\phi_{bx}$. 

The radial motion of particles $x+A$ is governed by the function $f_l$ which
asymptotically is given by
\[
f_l\sim {1\over k_xr} \left[ G_l(\eta,k_xr)+iF_l(\eta,k_xr)\right],
\]
where $F_l$ and $G_l$ are the regular
and irregular Coulomb wavefunctions, respectively.

The threshold behavior of $E_x$ for the reaction  $A+x=C+c$ is well known: since
$|S_{lx}|\sim \exp(-2\pi \eta)$. Thus, 
\begin{equation}
\sigma_{A+x\rightarrow C+c} \sim {1\over k_x^2}\sum_l (2l+1) \exp(-2\pi \eta),
\end{equation} 
independent of the partial wave $l$. Thus, it follows that
$\sigma_{A+x\rightarrow B+c}\sim (1/k_x^2)\ \exp(-2\pi \eta)$. 
In
addition to the threshold behavior of $S_{lx}$, the 
three-body cross section given by Eq. \eqref{threebody} is also governed by the
threshold behavior of $f_l(r)$. This  is 
determined from the behavior of the irregular Coulomb function
$G_l(\eta,k_xr)$ for $k_x \rightarrow 0$ ($\eta \rightarrow \infty$).
The combined threshold behavior of $f_l(r)$,
which for $r\rightarrow \infty$ is given by 
\[ f_{l_x}\sim
(k_xr)^{1/2} \ \exp(\pi \eta ) \ K_{2l+1}(\xi),
\] 
where $K_l$ denotes
the modified Bessel function of the second kind. The
quantity $\xi$ is independent of $k_x$ and is given by
$\xi=(8r/a_B)^{1/2}$, where $a_B=\hbar^2/mZ_AZ_xe^2$ is the Bohr
length. 

Based on the arguments given above, the cross section for the reaction  $A+a
\rightarrow b+c+C$  
close to the threshold of $A+x$, with $x$ ``hidden'' in $a$, is given by
\cite{Bau86} 
\begin{equation} 
{d^3 \sigma \over d\Omega_b
d\Omega_c dE_b} ({E_x \rightarrow 0}) \approx {\rm const.}
\end{equation}
That is, the coincidence cross section tends to be a constant which, in
general, will be different from zero. This is in striking contrast to the
threshold behavior of the two particle reaction $A+x=c+C$. The
strong barrier penetration effect on the charged particle reaction
cross section is canceled completely by the behavior of the factor
$T_{lm}$ for $\eta \rightarrow \infty$. 
Thus, from a measurement of reaction   $A+a \rightarrow b+c+C$  and a
theoretical 
calculation of the factors $T_{lm}$ in Eq. \eqref{threebody}, one extracts
the matrix elements $S_{lc}$ needed for  $A+x=c+C$.
Basically, this technique
extends the method of transfer reactions to continuum states. Very
successful results using this technique have been reported by many 
authors, e.g. Ref. \cite{Spit03,Tri06,Con07}. The problem with the method is
that the $x+A$ scattering is off-the-energy shell.  
The initial- and final-state interactions should be taken into account to get a
correct absolute value of the extracted astrophysical factor. 
 
The main characteristic of the Trojan horse method is the suppression of both
Coulomb barrier and screening effects in the off-shell 
two-body cross section \cite{Typ00,TyB03}. However, the quasifree $A + x$ process can
occur at very low, sub-Coulomb energies, even lower than the simple composition
of projectile velocity + 
Fermi velocity, thanks to the role of the
$(A,x)$-binding energy in compensating for the $A + a$ relative motion. This is
a different approach to the Trojan horse method compared to the original idea of 
Ref. \cite{Bau86}. In particular, in plane-wave impulse approximation (PWIA) the 
cross section of the $A + a \rightarrow c + C + b$  three-body reaction can be
factorized as given by  
\begin{equation}
{d^3\over d\Omega_c d\Omega_C dE_c} =KF \left({d\sigma \over
    d\Omega}\right)_{off} 
\left| \Phi({\bf k}_{xb}) \right|^2, \label{threebody2}
\end{equation}
where $(d\sigma / d\Omega)_{off}$ is the off-energy-shell differential cross
section for the two-body reaction $A + x \rightarrow c + C $ and $KF$ is a
kinematical factor given by  
\begin{equation}
KF={\mu_{Aa}m_c\over (2\pi)^5 \hbar^7} {k_Ck_c^3\over k_{Aa}}
\left[ \left( {{\bf k}_{Bx} \over \mu_{Bx}} - {{\bf k}_{Cc}\over
      m_c}\right)\cdot {{\bf k}_c\over k_c}\right]^{-1} 
, \label{threebody3}
\end{equation}
where $B=A+x$. In Eq. \eqref{threebody2}, $\Phi({\bf k}_{xb})$ is the Fourier
transform of the bound-state wavefunction of $a=b+x$. Of course, the plane-wave
impulse approximation is an idealistic approximation and 
corrections to this method are certainly necessary. For example: (a) the plane
wave might be replaced by a Coulomb, or distorted, wave in the final channel;
(b) the initial and final channels might be treated differently; (c)
higher-order processes might be relevant and so on.  These corrections have been
discussed, e.g. in Ref. \cite{Typ00,TyB03}.  

\begin{figure}[t]
\begin{center}
{\includegraphics[width=12cm]{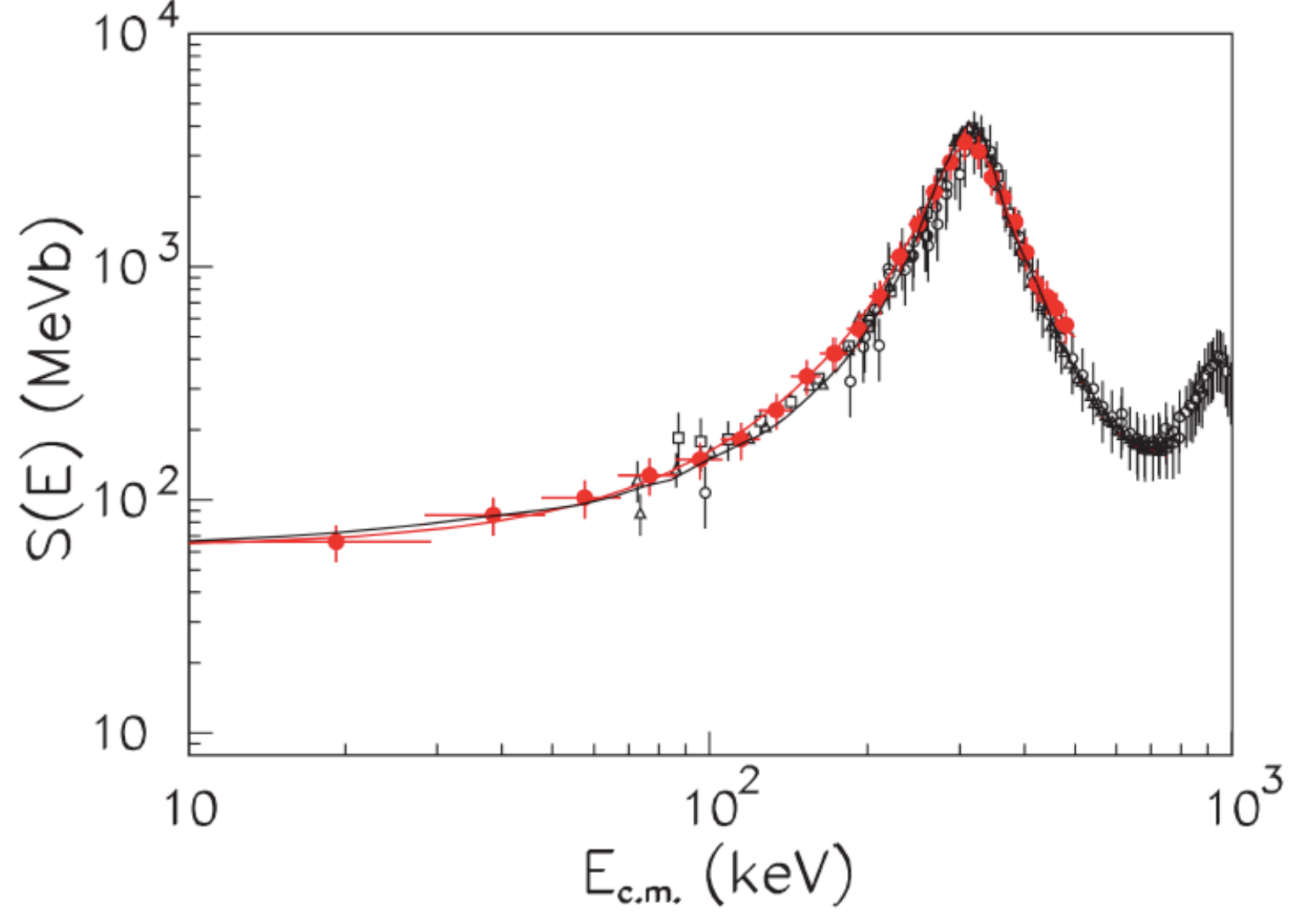}}
\end{center}
\caption{
$S$-factor obtained with the Trojan horse method (full red dots)  for the
reaction   
$^{15}$N$(p,\alpha)^{12}$C (from Ref. \cite{Con07}) using the
$^2$H($^{15}$N,$\alpha^{12}$C$)n$  
reaction at $E_{lab} = 60$ MeV.  The direct data from 
Refs. \cite{Red82,Sch52,Zys79} are also shown as open symbols (circles, squares,
and triangles, respectively). The red line represents a fit to the data. For
comparison, a Breit-Wigner parametrization is also displayed as the black line
\cite{Red82}.}\label{trojan2} 
\end{figure}  

Figure \ref{trojan2} shows the $S$-factor obtained with the Trojan horse method
(full red dots)  for the reaction   
$^{15}$N(p,$\alpha$)$^{12}$C (from Ref. \cite{Con07})  using the
$^2$H($^{15}$N,$\alpha^{12}$C)n  
reaction at $E_{lab} = 60$ MeV.  This reaction is important for nucleosynthesis
in AGB stars.  The direct data from  
Refs. \cite{Red82,Sch52,Zys79} are also shown as open symbols (circles,
squares, and triangles, respectively). The red line represents a fit to the
data. For comparison, a Breit-Wigner parametrization is also shown as the
black line \cite{Red82}.

\begin{figure}[t]
\begin{center}
{\includegraphics[width=8cm]{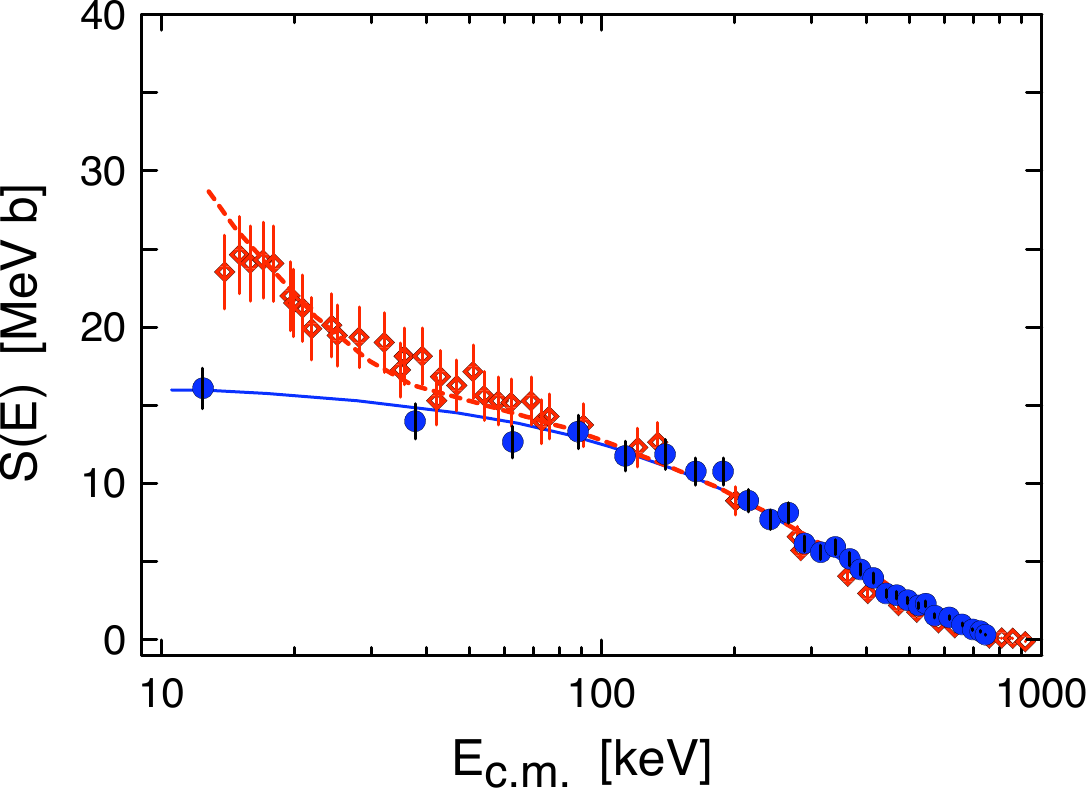}}
{\includegraphics[width=8.0cm, height=5.7cm]{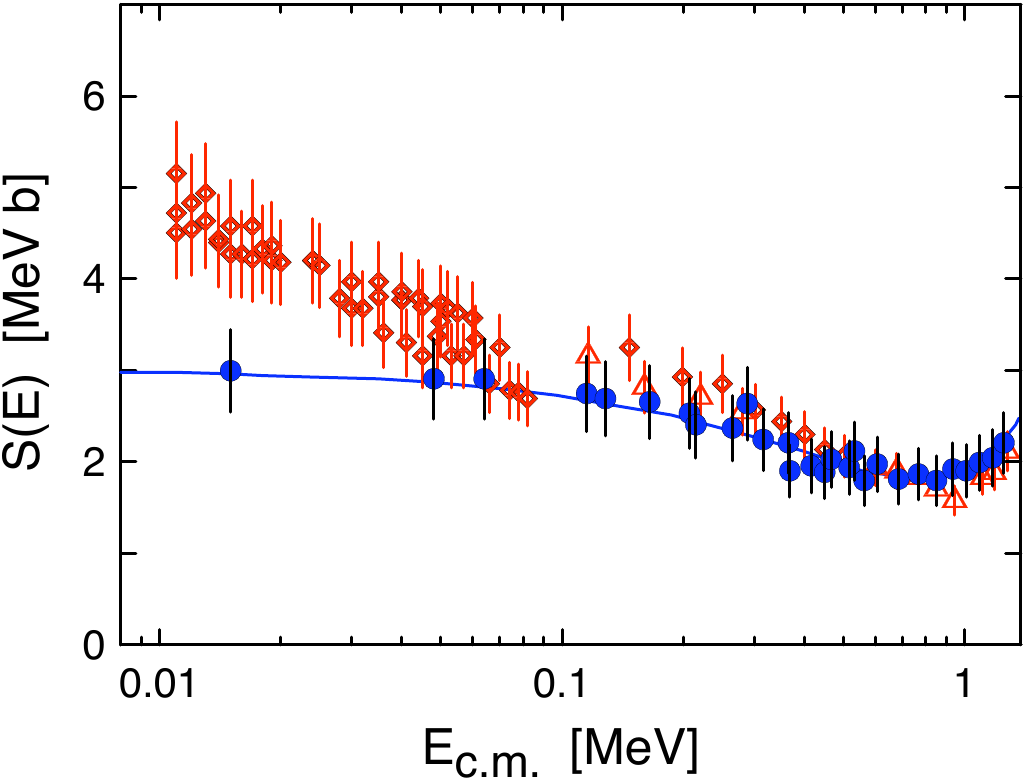}}
\end{center}
\caption{
Left: Bare $S_b(E)$ factor data (filled circles) for the reaction
$^6$Li(d,$\alpha$)$\alpha$ from the trojan horse method \cite{latuada} and
screened $S(E)$ factor data (open diamonds) from direct experiments
\cite{Eng92}. The dashed curve is a polynomial fit to $S(E)$ and the solid
curve includes the effects of electron screening with $U_e = 340$ eV. Right:
$S(E)$ for $^6$Li$(p,\alpha)^3$He obtained with the trojan
reaction $^6$Li$(d,n)^3$He (solid circles) \cite{tumino03} compared to direct
data (open triangles \cite{Elw73} and open circles \cite{Eng92}). The line
shows the result of a second-order polynomial fit to the trojan horse reaction
data \cite{tumino03}. }\label{trojan3} 
\end{figure}  

Figure \ref{trojan3} shows in the left panel the bare $S_b(E)$ factor data
(filled circles) for the reaction $^6$Li$(d,\alpha)\alpha$ obtained with the
trojan horse method \cite{latuada}. It also shows the screened $S(E)$ factor
data (open diamonds) from direct experiments \cite{Eng92}. The dashed curve is a
polynomial fit to $S(E)$ and the solid curve includes the effects of electron
screening with $U_e = 340$~eV. On the right panel one sees  $S(E)$ for
$^6$Li$(p,\alpha)^3$He obtained with the trojan horse reaction $^6$Li$(d,n)^3$He
(solid circles) \cite{tumino03} compared to direct data (open triangles
\cite{Elw73} and open circles \cite{Eng92}). The line shows the result of a
second-order polynomial fit to the trojan horse reaction data
\cite{tumino03}. It is apparent that an independent measurement of the screening
potential can be obtained in experiments with the trojan horse method
\cite{LaC05}. The results agree with the direct measurements for $U_e$
\cite{LaC05}. The method also shows that bare $S$-factors can be obtained directly
with this method. 
 
\subsubsection{Asymptotic normalization coefficients}\label{sec:ANC_th}

The asymptotic normalization
coefficient (ANC) method relies on the fact that the amplitude for
the radiative capture cross section $b+x\rightarrow a+ \gamma$
is given by
\begin{equation}
{\mathcal M}_{\pi \lambda}=\left<I_{bx}^a({\bf r_{bx}})\left|{\mathcal
      O}_{\pi\lambda}({\bf r_{bx}})\right| \psi_i^{(+)}({\bf r_{bx}})\right>,
\end{equation}
as was described in previous sections. The
 overlap integral 
\begin{equation}
I_{bx}^a({\bf r_{bx}})=<\phi_a({\boldsymbol \xi}_b, \ {\boldsymbol \xi}_x,\ {\bf %
r_{bx}}) |\phi_x({\boldsymbol \xi}_x)\phi_b({\boldsymbol \xi}_b)>
\end{equation}
corresponds to the
integration over the internal coordinates ${\boldsymbol \xi}_b$, and ${\boldsymbol \xi}_x$, of
$b$ and $x$, respectively. For low energies, the overlap integral
$I_{bx}^a$ is dominated by contributions from large $r_{bx}$. Thus,
what matters for the calculation of the matrix element ${\mathcal M}_{\pi
  \lambda}$ is the 
asymptotic value which, for charged particles is according to
Eq. \eqref{whitt},  $  I_{bx}^a\sim C_{bx}^a \ W_{-\eta_{bx},
1/2}(2\kappa_{bx} r_{bx})/r_{bx}, 
$ where $C_{bx}^a$ is the ANC and $W$ is the Whittaker function. This
coefficient is the product of the spectroscopic factor and a normalization
constant, which depends on the details of the wavefunction in the interior part
of the potential. Thus, $C_{bx}^a$ is the only unknown factor needed to
calculate the direct capture cross section. 

\begin{figure}[t]
\begin{center}
{\includegraphics[width=6cm]{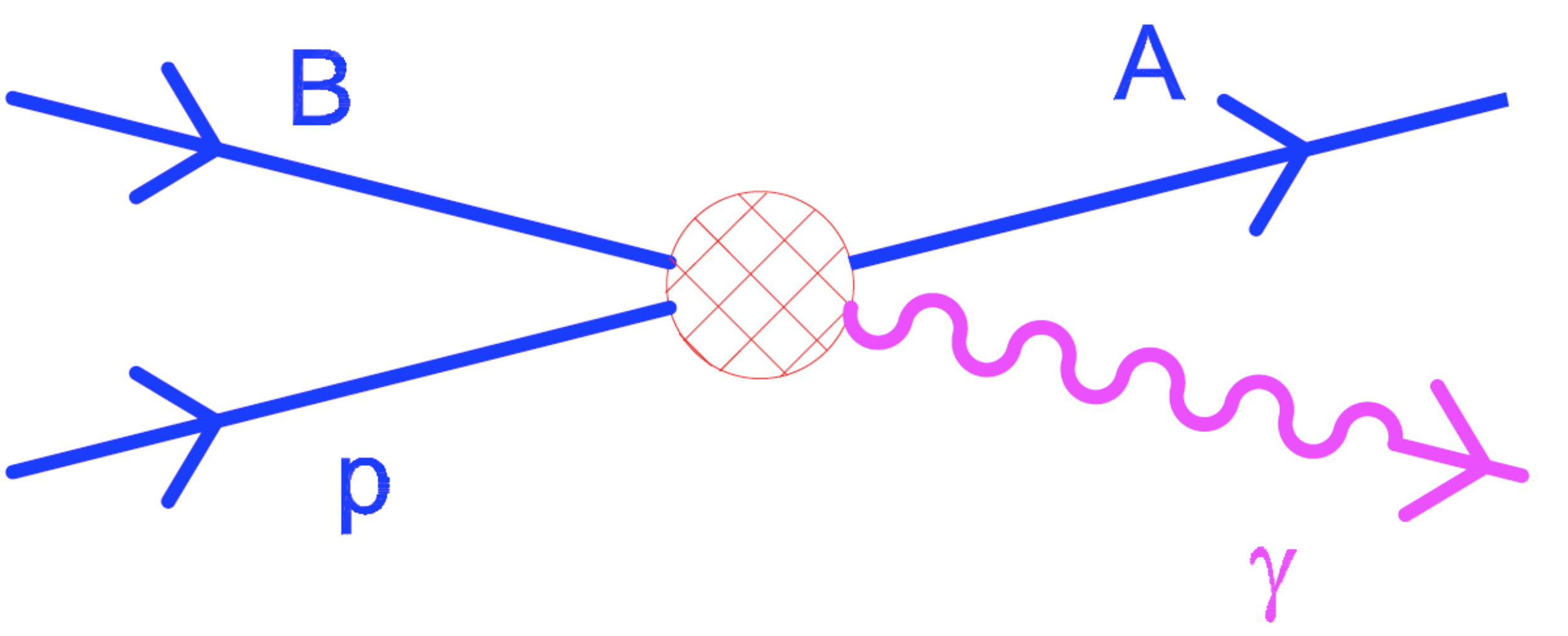}}
\hspace{1cm}
{\includegraphics[width=6cm]{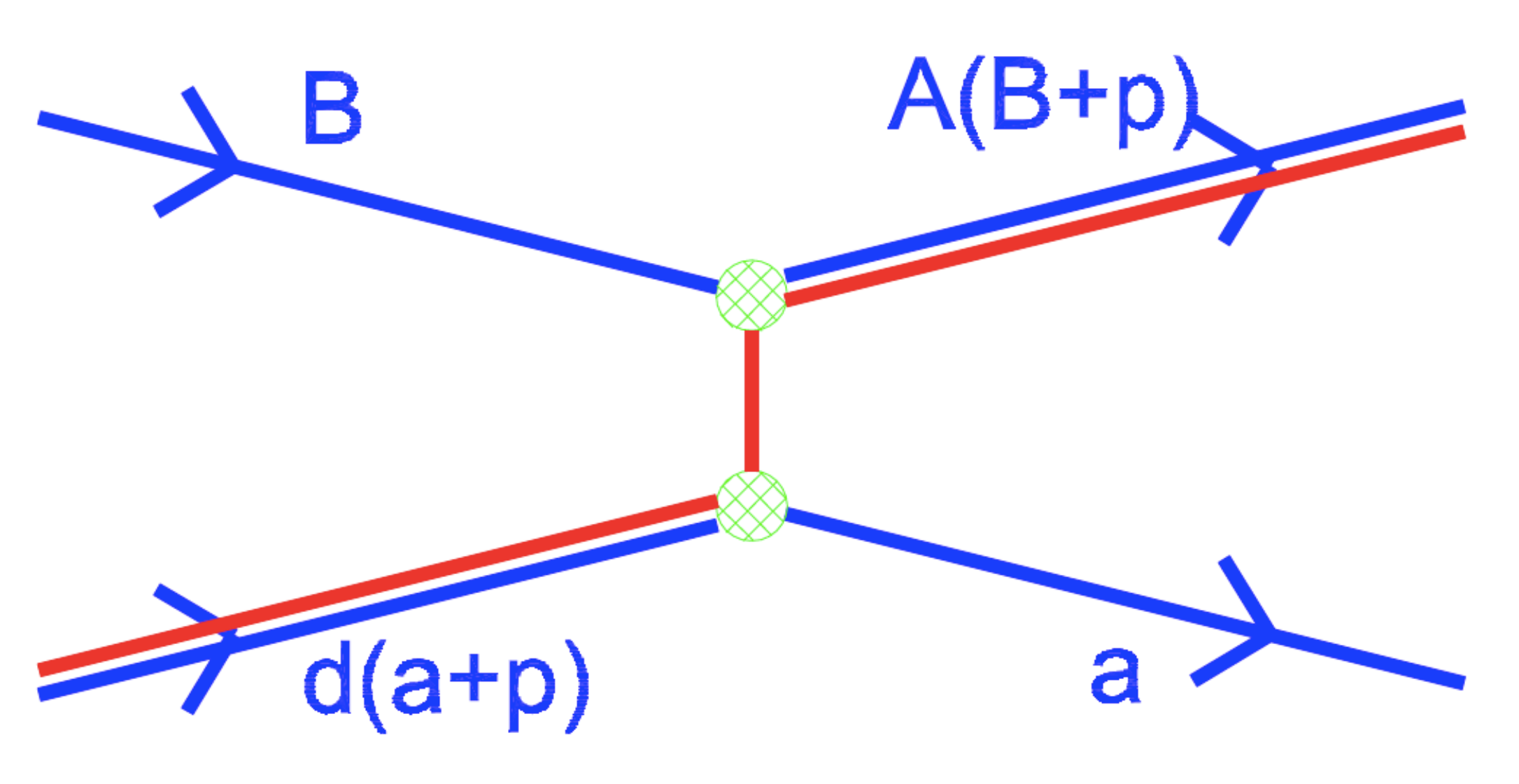}}
\end{center}
\caption{\label{anc}
Left: Schematic diagram for a radiative capture reaction. Right: The upper half
of this diagram contains part of the information of the diagram on the left. If
the collision is peripheral, only the tail of the bound-state wavefunction of
p+B is involved, allowing the extraction of its ANC by a DWBA analysis of the
experimental data. 
}
\end{figure}

The normalization
coefficients can be found from: 1) analysis of classical nuclear
reactions such as elastic scattering by extrapolation of the
experimental scattering phase shifts to the bound-state pole in the
energy plane, or 2) peripheral transfer reactions whose amplitudes
contain the same overlap function as the amplitude of the
corresponding astrophysical radiative capture cross section. 
This is shown schematically in Figure \ref{anc}, where the left
panel shows a schematic diagram for a radiative capture reaction. The upper half
of the diagram 
on the right contains part of the information of the diagram on the left. If the
collision is peripheral, 
only the tail of the bound-state wavefunction of p+B is involved, allowing the
extraction of its ANC 
by an DWBA analysis of the experimental data.
This method was proposed in Ref. \cite{MT90} and has been used with success for
many reactions of astrophysical interest, as discussed, e.g. in
Ref. \cite{Tri06} and mentioned in Section~\ref{sec:ANCexp}.  

To illustrate this technique, let us consider the proton transfer reaction $%
A(d,a)B$, where $d=a+p$, $B=A+p$. Using the asymptotic form of the
overlap integral, the DWBA cross section is given by
\begin{equation}
{d\sigma\over d\Omega} =
\sum_{J_Bj_d}\left[{(C_{Ap}^d)^2\over \beta^2_{Ap}}\right]\left[{(C_{ap}^d)^2\over \beta^2_{ap}}\right]
{\tilde \sigma}
\end{equation}
where $\tilde \sigma$ is the reduced cross section, not depending on
the nuclear structure, $\beta_{ap}$ ($\beta_{Ap}$) are the
asymptotic normalization of the shell-model bound-state proton wave
functions in nucleus $d (B)$, which are related to the corresponding
ANC's of the overlap function as $(C_{ap}^d)^2 =S^d_{ap}
\beta^2_{ap}$. Here $S^d_{ap}$ is the spectroscopic factor. Suppose
the reaction $A(d,a)B$ is peripheral. Then each of the bound-state
wavefunctions entering $\tilde \sigma$ can be approximated by its
asymptotic form and $\tilde \sigma \propto \beta_{Ap}^2
\beta_{ap}^2$. Hence 
\begin{equation}
{d\sigma\over d\Omega} =
\sum_{j_i}(C_{Ap}^d)^2(C_{ap}^d)^2 R_{Bd}, \ \ \
{\rm where} \ \ \  R_{Bd}={{\tilde
\sigma}\over \beta^2_{Ap} \beta^2_{ap}}
\end{equation} 
is independent of $\beta^2_{Ap}$
and $\beta^2_{ap}$. Thus for surface-dominated reactions, the DWBA cross
section is actually parameterized in terms of the product of
the square of the ANC's of the initial and the final nuclei $%
(C_{Ap}^d)^2(C_{ap}^d)^2$ rather than spectroscopic factors. This
effectively removes the sensitivity in the extracted parameters to
the internal structure of the nucleus.

Now consider the elastic $a+p$ scattering amplitude. It has a  pole in the
momentum plane \cite{mukhtr99} 
\begin{equation}
f_{l_{d}j_{d}}(k)= \frac{S_{l_{d}j_{d}} -1}{2\,i\,k} \stackrel{k \to 
k_{0}}{\longrightarrow}  \frac{1}{2\,i\,k_{0}}\,\frac{W_{l_{d}j_{d}}}{k 
- k_{0}},
\label{ANC1}
\end{equation}
where, for the bound state of $d$ for $k_{0}=i\,\kappa$ and 
for a resonance for $k_{0}=k_{R}$, with $k_{R}=k_{r} - i\,k_{i}$, and where
$S_{l_{d}j_{d}}$ is  the elastic $S$-matrix.
The residue in the pole $W_{l_{d}j_{d}}$ is
\begin{equation}
W_{l_{d}j_{d}}=-( - 1)^{l_d}\,ie^{i\pi \eta _d }\,{(C_{apl_d j_d 
}^d)}^2, \quad k_{0}=i\,\kappa,
\label{ANC2}
\end{equation}
\begin{equation}
W_{l_{d}j_{d}}=-( - 1)^{l_d }\,i\,{(C_{apl_d j_d (R)}^d)}^2, \quad 
k_{0}=k_{R}.
\label{residbndst2}
\end{equation}
For narrow resonances, $k_{I} << k_{r}$,
\begin{equation}
{(C_{apl_d j_d (R)}^d)}^2= 
(-1)^{l_{d}}\,\frac{\mu_{ap}}{k_{i}}e^{\pi\,\eta_{r}}\,e^{2i\,\delta_{l_{
d}j_{d}}(k_{r})}\,\Gamma_{l_{d}j_{d}}.
\label{ANC3}
\end{equation}
Here, $\eta_{r}$ is the Sommerfeld parameter for the resonance at
momentum $k_{r}$, $\,\delta_{l_{d}j_{d}}(k_{r})$ is the
potential (non-resonant) scattering phase shift taken at the
momentum $k_{r}$ \cite{mukhtr99}.

For elastic scattering close to the threshold, the bound-state pole behaves (see
Section \ref{sec:Rmatrix}) as
\begin{equation}
f_{l_{d}j_{d}}(k)\stackrel{k \to 0}{\longrightarrow}
= -\frac{1}{2\,k}\,e^{ - 2i(\phi _{l_d}  - \sigma _{l_d } 
)}\,\frac{\Gamma_{d}}{E + \varepsilon _d  + i\,\Gamma_{d}/2},
\label{ANC4}
\end{equation}
where $
\Gamma_{d}=2\,P_{l_d } (E)\,\gamma _d^2
$, and $P_{l_d } (E)$ is the Coulomb barrier penetrability. In this equation,
$\phi _{l_d}$ is the hard-sphere  
phase shift in the partial wave $l_{d}$ and
$\sigma_{l_{d}}=\sum\limits_{n=1}^{l_{d}}\,
\tan^{-1}({\eta_{d}}/{n}) $ ,$\gamma _d^2$ is the  reduced width:
\begin{equation}
\gamma _d^2  = \frac{1}{{2\mu _{ap} }}\frac{{W_{ - \eta _d ,l_d  + 
1/2}^{} (2\kappa r_c )}}{{r_c }}\,{(C_{apl_d j_d (r_c)}^d)}^2,
\label{ANC6}
\end{equation}
where   $r_{c}$ is the channel radius.

Thus, close to threshold and in the presence of a bound
state, the scattering amplitude behaves as the high-energy tail of the
resonance located at energy $E=-\varepsilon_{c}$, also 
called a ``subthreshold'' resonance. It is
not a resonance because the real resonance is located at complex
energies on the second energy sheet, while the subthreshold
resonance is a bound state located on the first energy
sheet at negative energy. At
negative energies (positive imaginary momenta), Eq.
(\ref{ANC4}) reduces to Eq. (\ref{ANC1}).
Thus, the elastic scattering amplitude
(phase shift) offers another possibility to determine
the ANC by extrapolating Eq. \eqref{ANC1} to the bound-state pole
\cite{blokh93}.

\subsection{Nucleon knockout reactions}\label{sec:knock_th}

Modern shell-model calculations are now able
to include the effects of residual interactions between pairs of
nucleons, using forces that reproduce the measured masses, charge
radii and low-lying excited states of a large number of nuclei. For
very exotic nuclei, the small additional stability that comes with
the filling of a particular orbital can have profound effects upon
their existence as bound systems, their lifetime and structure.
Thus, the verification of the ordering, spacing and the occupancy of
orbitals is essential in assessing how exotic nuclei evolve in the
presence of large neutron or proton excess and to what extent theories have
predictive power. Such spectroscopy of the single-particle structure in
short-lived nuclei typically uses direct nuclear reactions.

Heavy-ion induced single-nucleon knockout reactions -- performed at intermediate
energies and in inverse kinematics --  have become a specific and
quantitative tool for studying the location and occupancy of single-particle
states and correlation effects in the nuclear many-body system, as discussed in
Refs. \cite{Gregers,Tostevin1999,han03,Gade2008a}. In a peripheral, sudden
collision of the 
fast-moving projectile with mass $A$ with a light target (typically Be or C), a
single nucleon is removed from the projectile, producing projectile-like
residues with mass $A-1$ in the exit channel~\cite{han03}. Typically, the final
state of the target and that of the struck nucleon are not observed, but instead
the energy of the final state of the residue can be identified by measuring the
decay $\gamma$-rays emitted in flight by the excited projectile-like residues
(see Section \ref{sec:ANCexp}).

\begin{figure}[t]
\begin{center}
{\includegraphics[width=10cm]{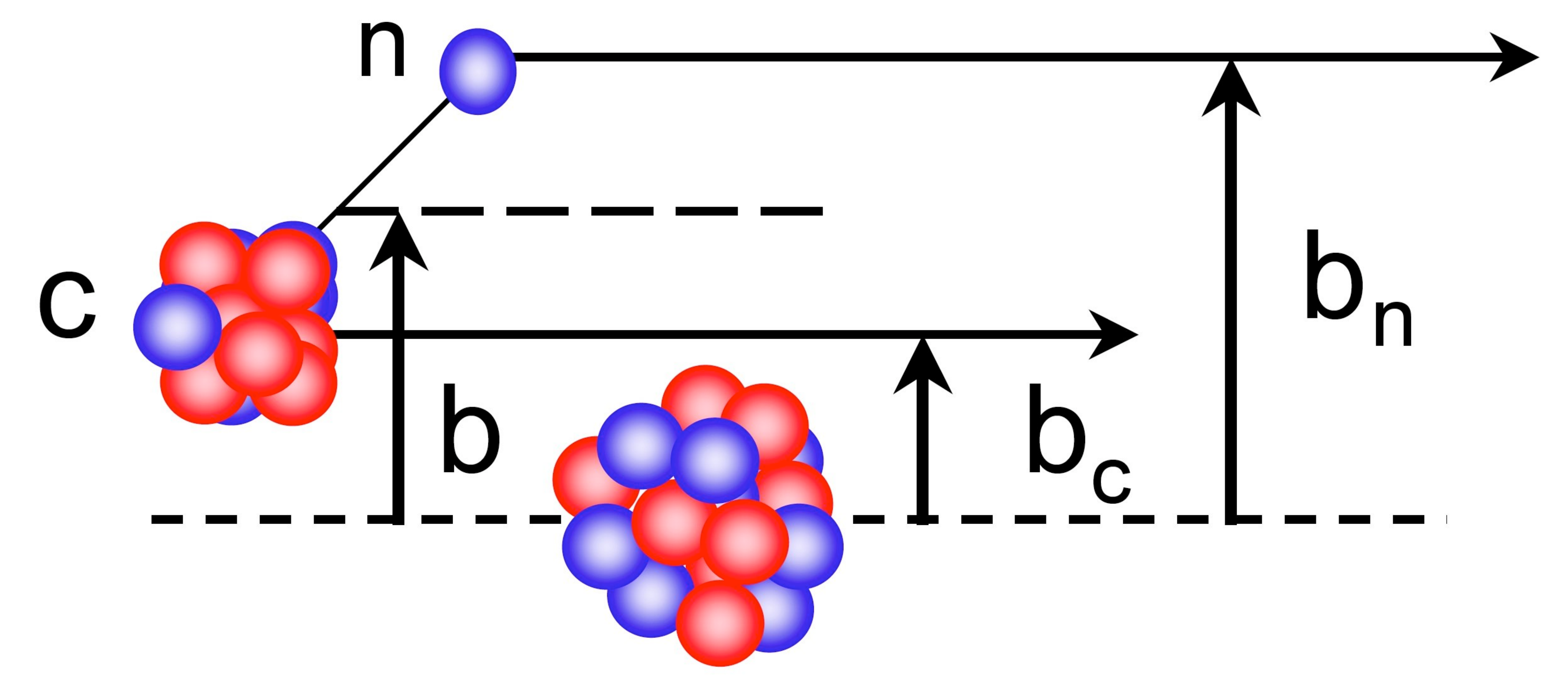}}
\end{center}
\caption{\label{knockout}
Coordinates used to describe knockout reactions in the text. (Adapted from J.A. Tostevin).
}
\end{figure}

The momentum distributions of the projectile-like residues in one-nucleon
knockout are a measure of the spatial extent of the wavefunction of the struck
nucleon, while the cross section for the nucleon removal scales with the
occupation amplitude, or probability (spectroscopic
factor), for the given single-particle configuration in the projectile ground
state. The longitudinal momentum distributions are given by (see, e.g.,
Refs. \cite{BH04,BG06}) 
\begin{align}
\frac{d\sigma_{\mathrm{str}}}{dk_{z}}  &  = (C^2S) \frac{1}{\left(  2\pi\right)  ^{2}%
}\frac{1}{2l+1}\sum_{m}\int_{0}^{\infty}d^{2}b_{n}\ \left[  1-\left\vert
S_{n}\left(  b_{n}\right)  \right\vert ^{2}\right]  \ \ \int_{0}^{\infty}%
d^{2}\rho\ \left\vert S_{c}\left(  b_{c}\right)  \right\vert ^{2}\nonumber\\
&  \times\left\vert \int_{-\infty}^{\infty}dz\ \exp\left[  -ik_{z}z\right]
\psi_{lm}\left(  \mathbf{r}\right)  \right\vert ^{2}\ ,\label{strL}%
\end{align}
where $k_{z}$ represents the longitudinal component of $\mathbf{k}_{c}$ (final momentum of the
core of the projectile nucleus)
and $(C^2S)$ is the spectroscopic factor, and $\psi_{lm}\left(  \mathbf{r}\right)$
is the wavefunction of the core plus (valence) nucleon system $(c+n)$ in a state
with single-particle angular momentum $l,m$.  In this equation,
$\mathbf{r\equiv}\left(  \mathbf{\mbox{\boldmath$\rho$}}%
,z,\phi\right)  =\mathbf{r}_{n}-\mathbf{r}_{c}$, so that
\begin{equation}
b_{c}    =\left\vert \mathbf{\mbox{\boldmath$\rho$}}-\mathbf{b}
_{n}\right\vert =\sqrt{\rho^{2}+b_{n}^{2}-2\rho\ b_{n}\cos\phi  }
  =\sqrt{r^{2}\sin^{2}\theta+b_{n}^{2}-2r\sin\theta\ b_{n}\cos
\phi  }.
\end{equation}

$S(b)$ are the $S$-matrices for core-target and nucleon-target scattering
obtained from the nuclear ground-state 
densities and the nucleon-nucleon cross sections by the relation \cite{BD04}
\begin{equation}
S(b)=\exp\left[  i\phi(b)\right]  ,\ \ \ \ \ \ \text{with}\ \ \ \ \ \ \phi
_{N}(b)=\frac{1}{k_{NN}}\int_{0}^{\infty}dq\ q\ \rho_{p}\left(  q\right)
\rho_{t}\left(  q\right)  f_{NN}\left(  q\right)  J_{0}\left(  qb\right)
\ ,\label{eikphase}%
\end{equation}
where $\rho_{p,t}\left(  q\right)  $ is the Fourier transform of the nuclear
density of the projectile (nucleon or core) and the target nucleus, and
$f_{NN}\left(  q\right)  $ is the 
high-energy nucleon-nucleon scattering amplitude at forward angles, which can
be parametrized by \cite{ray79}
\begin{equation}
f_{NN}\left(  q\right)  =\frac{k_{NN}}{4\pi}\sigma_{NN}\left(  i+\alpha
_{NN}\right)  \exp\left(  -\beta_{NN}q^{2}\right),\label{fnn}%
\end{equation}
where $k_{NN}$ is the nucleon-nucleon relative momentum, $\sigma_{NN}$ is the
nucleon-nucleon total cross section, and $\alpha_{NN}$ and $\beta_{NN}$ are
parameters fitted to (free) nucleon-nucleon scattering data. 

The first term in the integral in Eq. \eqref{strL}, $1-|S_n|^2$, represents the
probability for the knockout of the nucleon from its location at $b_n$, whereas
the second integral carries the term $|S_c|^2$ which is the probability of
core survival at impact parameter $b_c$. At high energies,  
the $S$-matrices do not depend on the longitudinal direction. That is why the
bound-state wavefunction is probed by the longitudinal Fourier transform in the
last integral of Eq. \eqref{strL}.  These results arise naturally by using
eikonal scattering waves. 
For the transverse momentum distributions, the same formalism yields
\begin{align}
\frac{d\sigma_{\mathrm{str}}}{d^{2}k_{\bot}}  &  =(C^2S)\frac{1}{2\pi}\frac{1}%
{2l+1}\ \int_{0}^{\infty}d^{2}b_{n}\ \left[  1-\left\vert S_{n}\left(
b_{n}\right)  \right\vert ^{2}\right] \nonumber\\
&  \times\sum_{m,\ p}\ \int_{-\infty}^{\infty}dz\ \left\vert \int d^{2}%
\rho\ \exp\left(  -i\mathbf{k}_{c}^{\perp}\mathbf{.\mbox{\boldmath$\rho$}}%
\right)  S_{c}\left(  b_{c}\right)  \psi_{lm}\left(  \mathbf{r}\right)
\right\vert ^{2}.\label{strT}%
\end{align}

Extensions of the nucleon knockout formalism include the treatment of
final-state interactions were discussed in Ref. \cite{BH04} where one has shown
that the inclusion of the Coulomb final-state interactions are of
relevance. They can be done by just adding the Coulomb phase
$\phi=\phi_N+\phi_C$ in the eikonal phase described above \cite{BH04}. Other
higher-order effects have been included \cite{Baz09} and a theory for
two-nucleon knockout  \cite{Tostevin2006,Sim09,Simpson2009,Bazin2003b} has been
developed. Knockout reactions represent a particular case for which higher
projectiles energies allow a simpler theoretical treatment of the reaction
mechanism, due to the simplicity of the eikonal scattering waves and the
assumption of a single-step process.       

Many new experimental approaches based on nucleon knockout reactions have been
developed and shown to reduce the uncertainties in astrophysical
(rapid proton capture) rp-process calculations by the provision of nuclear
structure data. One utilizes neutron removal from a rare-isotope beam
to populate nuclear states of interest in the knockout residue. In the first case
studied with astrophysical background, e.g., in
Ref. \cite{Clement2004,Schatz2005}, excited states in \nuc{33}{Ar} were measured
with uncertainties of several keV. The 2-orders-of-magnitude improvement in the
uncertainty of the level energies resulted in a 3-orders-of-magnitude
improvement in the uncertainty of the calculated $^{32}$Cl(p,$\gamma$)$^{33}$Ar
rate that is critical to the modeling of the rp process (see Section
\ref{sec:ANCexp} for experimental details). This approach has the potential to
measure key properties of almost all interesting nuclei on the rp-process path. 

\subsection{Quasifree (p,pN) reactions}
Quasifree $(p,pN)$ reactions ($N=$~proton or neutron) represent one of the most 
direct ways to measure single-particle properties in nuclei. They have been
used and extensively studied for over 4 decades (see \cite{JM73,Kit85} for reviews). 
In quasifree $(p,pN)$ scattering, an incident proton of intermediate energy
(200-1000~MeV) knocks out a bound nucleon. The only violent interaction of this
process occurs between the incident particle and the ejected nucleon. The
wavefunctions of incoming and outgoing nucleons are  distorted while traversing
the nucleus. From the measured energies and momenta of the emerging nucleons,
direct information on single-particle separation energy spectra and nucleon momentum
distributions can be obtained. Over the last four decades, quasifree scattering
experiments have been performed with this basic purpose. 

A popular framework used in the analysis of quasi-free scattering is the
so-called ``impulse approximation'', which yields for the quasifree cross section 
\begin{equation}
{d^3\sigma \over dT_N d\Omega_p d\Omega_N} =K {d\sigma_{pN}\over d\Omega} |F({\bf Q})|^2,
\label{int}
\end{equation}
where $K$ is a kinematic factor, $|F({\bf Q})|^2$ is the momentum distribution
of the knocked-out nucleon $N$ in the nucleus and $d\sigma_{pN}/d\Omega$ the
free $p$-$N$ cross section. In the knockout formalism, the off-shell $p$-$N$
t-matrix is required, and the factorized form that appears in Eq. \eqref{int} 
is valid only if off-shell effects are not very important. In proton-induced
knockout reactions, this is probably adequate since the energy variation of the free
nucleon-nucleon cross section is small. As expected, there is an evident
similarity between this equation and Eq. \eqref{threebody2} used in the analysis
of transfer reactions.

Improvements of the above formulation can be easily made, assuming that
(on-shell or off-shell) multiple-scattering effects are small. In the
distorted-wave impulse approximation (DWIA), the scattering matrix for the $A(p,
pN)B$ 
reaction is given by 
\begin{equation}
T_{p,pN}=\left< \chi_{{\bf K}_p^\prime}^{(-)} \chi_{{\bf K}_N^\prime}^{(-)}          
\left| \tau_{pN}\right| \chi_{{\bf K}_p}^{(+)}\psi_0\right>\label{Tmat}
\end{equation}
where $\chi_{{\bf K}_p^\prime}^{(-)}$ ($\chi_{{\bf K}_N^\prime}^{(-)}$) is the
distorted wave for an outgoing proton (knocked-out nucleon) in the presence of
the residual nucleus $B$, $\chi_{{\bf K}_p}^{(+)}$ is the distorted wave for an
incoming proton in the presence of the target nucleus $A$, and $ \phi_0$ is the
bound-state wavefunction of the knocked-out nucleon; $\phi_0$ includes the
spectroscopic amplitude. The $p$-$N$ scattering matrix is denoted by $\tau_{pN}$. 
The matrix element given by eq. \eqref{Tmat} can be written as
\begin{eqnarray}
T_{p,pN}&=&\frac{1}{2\pi}\int d^3r^\prime_{pB} d^3r^\prime_{NB}
d^3r_{pA} d^3r_{NB} 
d^3P_pd^3P_Nd^3P^\prime_pd^3P_N^\prime \tau({\bf P}_p,{\bf P}_N;{\bf P}^\prime_p, {\bf P}_N^\prime ) \nonumber \\
&\times&\tilde{\chi}_{{\bf K}_p^\prime}^{(-) *} ({\bf P}^\prime_{p}) \tilde{\chi}_{{\bf K}_N^\prime}^{(-) *}({\bf P}^\prime_{N}) \tilde{\chi}_{{\bf K}_p}^{(+)} ({\bf P}_{p})\psi_0({\bf r}_{NB})
\delta\left({\bf P}^\prime_p +{\bf P}^\prime_N  - {\bf P}_{pA}-{\bf P}_N \right),\nonumber \\
\label{tmat2}
\end{eqnarray}
where $\tilde{\chi}_{\bf K}({\bf P})$ are the Fourier transforms of $\chi_{\bf K}({\bf r})$, which
are normalized so that $\int d^3r \chi_{\bf k}({\bf r}) \chi_{\bf k^\prime}({\bf r}) = (2\pi)^3 \delta({\bf k}-{\bf k^\prime})$.

The in-medium effects due to the propagation of the incoming proton and the
outgoing particles in the nuclear interior can be included in the distorted
waves $\chi_i({\bf r})$. The effective interaction $\tau_{pN}$ must also
include medium and energy dependence effects. Spin dependence of the $p$-$N$
t-matrix and of the scattering and bound states also needs to be included.
Thus, a deviation from the simple formulation based on Eq. \eqref{Tmat} is
expected. By using inverse kinematics, quasifree scattering will  become a major
tool for the investigation of the properties of rare isotopes with
relevance for nuclear astrophysics \cite{AumannLemmon}.

\subsection{Charge-exchange reactions}\label{sec:cex_th}

Charge exchange induced in $(p,n)$ reactions is often used
to obtain values of Gamow-Teller matrix elements, $B(GT)$, which
cannot be extracted from $\beta$-decay experiments. This approach
relies on the similarity in spin-isospin space of charge-exchange
reactions and $\beta$-decay operators. As a result of this
similarity, the cross section $\sigma(p,n)$ at small momentum
transfer $q$ is closely proportional to $B(GT)$ for strong
transitions \cite{Taddeucci1987}, i.e.,
\begin{equation}
{d\sigma\over dq}(q=0)=KN_D|J_{\sigma\tau}|^2 B(\alpha) , \label{tadeucci}
\end{equation}
where $K$ is a kinematical factor, $N_D$ is a distortion factor
(accounting for initial and final state interactions),
$J_{\sigma\tau}$ is the Fourier transform of the effective
nucleon-nucleon interaction, and $B(\alpha=F,GT)$ is the reduced
transition probability for non-spin-flip, $$B(F)= {1\over 2J_i+1}|
\langle f ||\sum_k  \tau_k^{(\pm)} || i \rangle |^2,$$ and spin-flip,
$$B(GT)= {1\over 2J_i+1}| \langle f ||\sum_k \sigma_k \tau_k^{(\pm)} ||
i \rangle |^2,$$ transitions. The Fourier transform of the effective interaction
yields  largest values for the $|J_{\sigma\tau}|^2$ component at energies 
around 100-200 MeV. This indicates that the energy range of 100-200~MeV/nucleon
will be appropriate to extract $GT$ ($F$) matrix elements from studies
charge-exchange experiments in inverse kinematics involving rare-isotope beams.
 The condition $q\sim 0$ is met for very small scattering angles, such that $\theta 
 \ll 1/kR$, where $R$ is the nuclear radius and $k$ is the projectile wavenumber.

Eq. \eqref{tadeucci} is easily understood in the plane-wave
Born-approximation. Then the 
charge-exchange matrix element is given by \cite{Ber93} 
\begin{equation}
{\mathcal M}_{exch}({\bf q})=\left<\Psi_a^{(f)} ({\bf r}_a)\Psi^{(f)}_b({\bf
    r}_b) \left| e^{-i{\bf q}\cdot {\bf r}_a } 
v_{exch}({\bf q})e^{i{\bf q}\cdot {\bf r}_b} \right| 
\Psi_a^{(i)} ({\bf r}_a)\Psi^{(i)}_b({\bf r}_b) \right>, \label{Mexch}
\end{equation}
where ${\bf q}$ is the momentum transfer, $\Psi_{a,b}^{(i,f)}$ are the intrinsic
wavefunctions of nuclei $a$ and $b$ for the initial and final states, ${\bf
  r}_{a,b}$ are the nucleon coordinates within $a$ and $b$, and $v_{exch}$ is
the part of the nucleon-nucleon interaction responsible for charge exchange,
which contains spin and isospin operators. 
For forward scattering and  low-momentum transfers,
${\bf q}\sim 0$, and the matrix element \eqref{Mexch} becomes
\begin{equation}
{\mathcal M}_{exch}({\bf q} \sim 0) \sim
v_{exch}^{(0)} ({\bf q}\sim 0) \, {\mathcal M}_a(F,GT)\,
{\mathcal M}_b(F,GT)
\, ,
\label{q1}
\end{equation}
where $v_{exch}^{(0)} $ is the spinless part of the interaction, and ${\mathcal M}_{exch}(F,GT)=
\left<\Psi_{a,b}^{(f)}\vert\vert (1 \ {\rm or} \ \sigma ) \tau \vert\vert
  \Psi_{a,b}^{(i)}\right>$ are Fermi or Gamow-Teller (GT) matrix 
elements for the nuclear transition.
The result above is also valid if, instead of plane waves, one uses the eikonal
scattering waves for the nuclei. One can thus conclude that the ability to
extract information on Fermi or Gamow-Teller transition densities 
in charge-exchange reactions in a simple way depends on the validity of the
low-momentum transfer assumption in high-energy collisions.  Recently, also for
the $(\nuc{3}{He},t)$ reaction at 420~MeV, the proportionality between cross
section and Gamow-Teller strength was shown to follow simple trends as function
of the mass number~\cite{Zegers2007}.   
\begin{figure}[t]
\begin{center}
{\includegraphics[width=5cm]{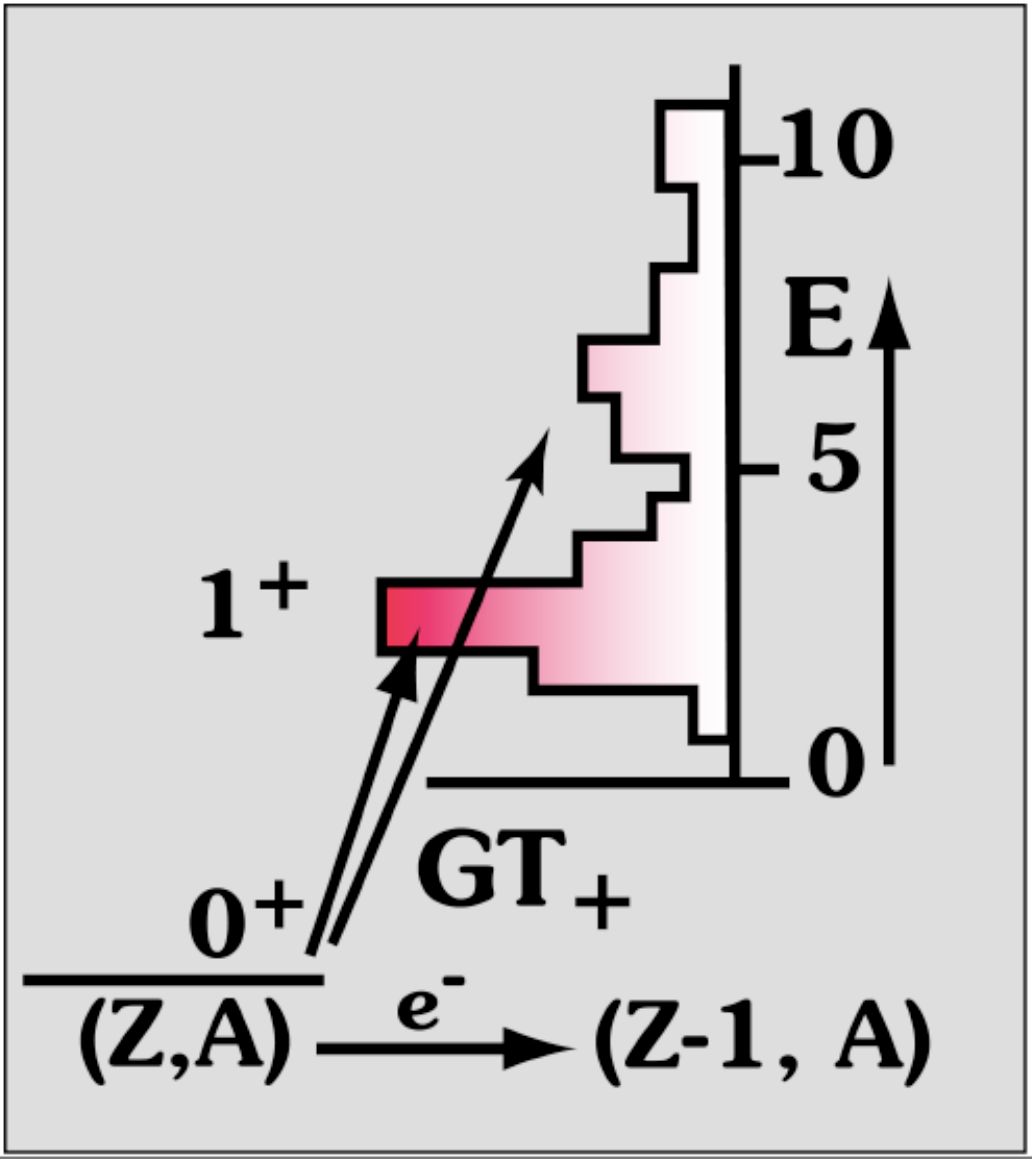}}
{\includegraphics[width=8cm]{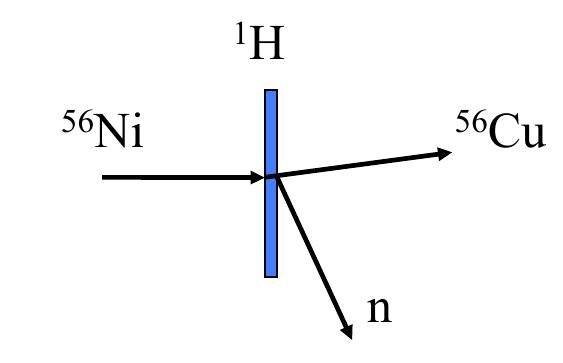}}

\end{center}
\caption{\label{electcap}
{\sl Left:} Electron captures by nuclei are dominated by Gamow-Teller (GT+)
transitions. {\sl Right:} The matrix elements for such transitions can be probed
by using rare-isotope beams interaction with protons in inverse kinematics, as
shown schematically for the reaction $^{56}$Ni$(p,n)^{56}$Cu. (Adapted from work
of D.  Frekers and R.  Zegers). 
}
\end{figure}

Another assumption in
Eq. \eqref{tadeucci}, the validity  of one-step processes, 
was proven to work rather well for $(p,n)$ reactions (with a few
exceptions). For 
heavy-ion induced charge-exchange reactions the formula might not work so
well as has been investigated in Refs. \cite{Len89,Ber93}. In Ref. \cite{Len89}
it was shown that multi-step processes involving the physical
exchange of a proton and a neutron can still play an important role
up to bombarding energies of 100 MeV/nucleon. Refs. \cite{Ber93} use
the isospin terms of the effective interaction to show that
deviations from the Taddeucci formula are common under many
circumstances. As shown in Ref. \cite{Aus94}, for important GT
transitions whose strengths are a small fraction of the sum rule, the
direct relationship between $\sigma($p,\ n$)$ and $B(GT)$ values
also can fail. Similar discrepancies have been observed
\cite{Wat85} for reactions on some odd-A nuclei including $^{13}$C,
$^{15}$N, $^{35}$Cl, and $^{39}$K and for charge exchange induced by
heavy ions \cite{St96}. It is still an open question if Taddeucci's
formula is valid in general.

Electron capture by nuclei in $pf$-shell plays a pivotal role in the
deleptonization of a massive star prior to core-collapse \cite{Bethe79}. During
the period of silicon burning, supernova collapse occurs due to a competition
of  gravity and the weak interaction, with electron captures on nuclei and
protons and $\beta$-decay playing crucial roles. Weak-interaction processes
become important when nuclei with masses $A \sim 55 - 60$ are most abundant in
the supernova core. Weak interactions change $Y_e$ and electron capture
dominates, the $Y_e$ value is successively reduced from its initial value  $\sim
0.5$. Electron capture (EC) yields more neutron-rich and heavier nuclei, as nuclei with
decreasing $Z/A$ ratios are more bound for heavier nuclei. For 
densities $\rho \le 10^{11}$ g/cm$^3$, weak-interaction processes are dominated
by Gamow-Teller and sometimes by Fermi transitions. This is shown schematically
on the left panel of figure \ref{electcap}. Fuller, Fowler and Newman
\cite{Fuller1980,Fuller1982,FFN85} made systematic estimates of EC-rates in
stellar environments. But their calculations obtain only the centroid of the
Gamow-Teller response function.  More recently, $B(GT)$-distributions have been
obtained with modern shell-model calculations
\cite{Caurier1999,Baum03,Langanke2000,LMP01}. Some marked deviations from
the previous rates 
\cite{Fuller1980,Fuller1982,FFN85} have emerged, e.g. $Y_e$ increases to about
0.445 instead of the value of 0.43 found previously
\cite{Fuller1980,Fuller1982,FFN85}.   

Most of such theoretical developments need support and validation from
experiments. A satisfactory understanding of how weak interaction rates
influence stellar evolution will rely on experiments using charge-exchange
reactions such as the traditional ($^{3}$He,t) reaction and novel
inverse-kinematics $(p,n)$ reactions on rare-isotope projectile beams (right
panel of Figure \ref{electcap}). Reactions with unstable nuclei will provide
crucial information on hitherto unknown $B(F)$ and $B(GT)$ values needed for
astrophysical purposes and the validation of the handling of weak interactions
by large-scale shell-model calculations. 

\subsection{Central collisions}

At intermediate energies of $E_{lab}\sim100-1000$~MeV/nucleon, the
nucleons and the products of their collisions can be described
individually and their propagation can be calculated by
semiclassical equations. Hadronic transport theories have been quite
successful in describing a multitude of measured particle spectra. 

The nuclear equation of state (EOS),  $e(\rho ,\delta )$, expresses the energy
per nucleon of nuclear matter as a function of the nucleon
density $\rho$ and the relative neutron excess $\delta$. It is a fundamental
quantity in theories of neutron stars and supernova explosions. The
main measured quantities, which can provide information
about the EOS are, for example, binding energies
and other data of finite nuclei. As the finite nuclei are in
states near the standard nuclear matter state with normal
nucleon density $\rho_0$ and zero neutron excess, $\delta=0$, our
knowledge about the EOS can be confirmed experimentally
only in a small region around $\rho\sim\rho_0$ and $\delta\sim 0$.
With very neutron-rich nuclei and energetic heavy-ion collisions, 
the nuclear EOS can be tested well beyond this region. 

If one assumes that the system of nucleons forms a dilute system (or gas) of
particles, such that the total volume of the gas particles is small compared to
the volume available to the gas, then $na^{2}\ll1 $,
where $n$\ is the number density of particles and $a$\ is the (interaction)
radius of a particle. Since the particles in a neutral gas do not
have long-range forces like the particles in a plasma, they are
assumed to interact only when they collide, i.e., when the
separation between two particles is not much larger than $2a$. The
term ``collision'' normally means the interaction between two such
nearby particles. A particle moves in a straight line between
collisions. The average distance travelled by a particle between
two collisions is known as the mean free path. The mean free path
depends on the cross section $\sigma \sim a^2$, and is given by
$\lambda={1}/{n\sigma}$. One consequence of the requirement that the gas be
dilute is that $\lambda\gg 
a$. If the gas is dilute, the probability of
three-body collisions is much lower than that for two-body collisions and they
can be neglected.

Assuming that these conditions are valid, a practical equation describing
nucleus-nucleus collisions  
in terms of nucleon-nucleon collisions can be deduced. A popular transport
equation is the Boltzmann-Uehling-Uhlenbeck (BUU)  equation, 
\begin{align}
  \frac{\partial f}{\partial t}+\left(  \frac{\mathbf{p}}{m}+\mathbf{\nabla
}_{\mathbf{p}}U\right)  \cdot\mathbf{\nabla}_{\mathbf{r}}f-\mathbf{\nabla
}_{\mathbf{r}}U\cdot\mathbf{\nabla}_{\mathbf{r}}f&=
  \int d^{3}p_{2}\int d\Omega\;\sigma_{NN}\left(  \Omega\right)  \left\vert
\mathbf{v}_{1}-\mathbf{v}_{2}\right\vert \nonumber\\
&  \times\left\{  f_{1}^{\prime}f_{2}^{\prime}\left[  1-f_{1}\right]  \left[
1-f_{2}\right]  -f_{1}f_{2}\left[  1-f_{1}^{\prime}\right]  \left[
1-f_{2}^{\prime}\right]  \right\}  , \label{BE}%
\end{align}
where $f$ is defined so that, if
$dN$\ is the number of particles in the volume element $d^{3}r$\ whose
momenta fall in the momentum element $d^{3}p$\ at time $t$, then the
distribution \ function{\small \ }$f\left(  \mathbf{r,p},t\right)  $\ is given
by
$
dN=f\left(  \mathbf{r,p},t\right)  d^{3}rd^{3}p \label{distbf1}%
$. The left-hand side of the above equation is due to the variations of $f$ by
means of a mean field $U$. When the right-hand side is taken as zero, the
resulting equation is known as the Vlasov equation. The right-hand side of
Eq. \eqref{BE} is the collisional term, which accounts for binary collisions
between nucleons by using the nucleon-nucleon cross section $\sigma_{NN}$. This
equation incorporates the Pauli principle through the $(1-f)$-terms to avoid
scattering into occupied states.

\begin{figure}[t]
\begin{center}
{\includegraphics[width=12cm]{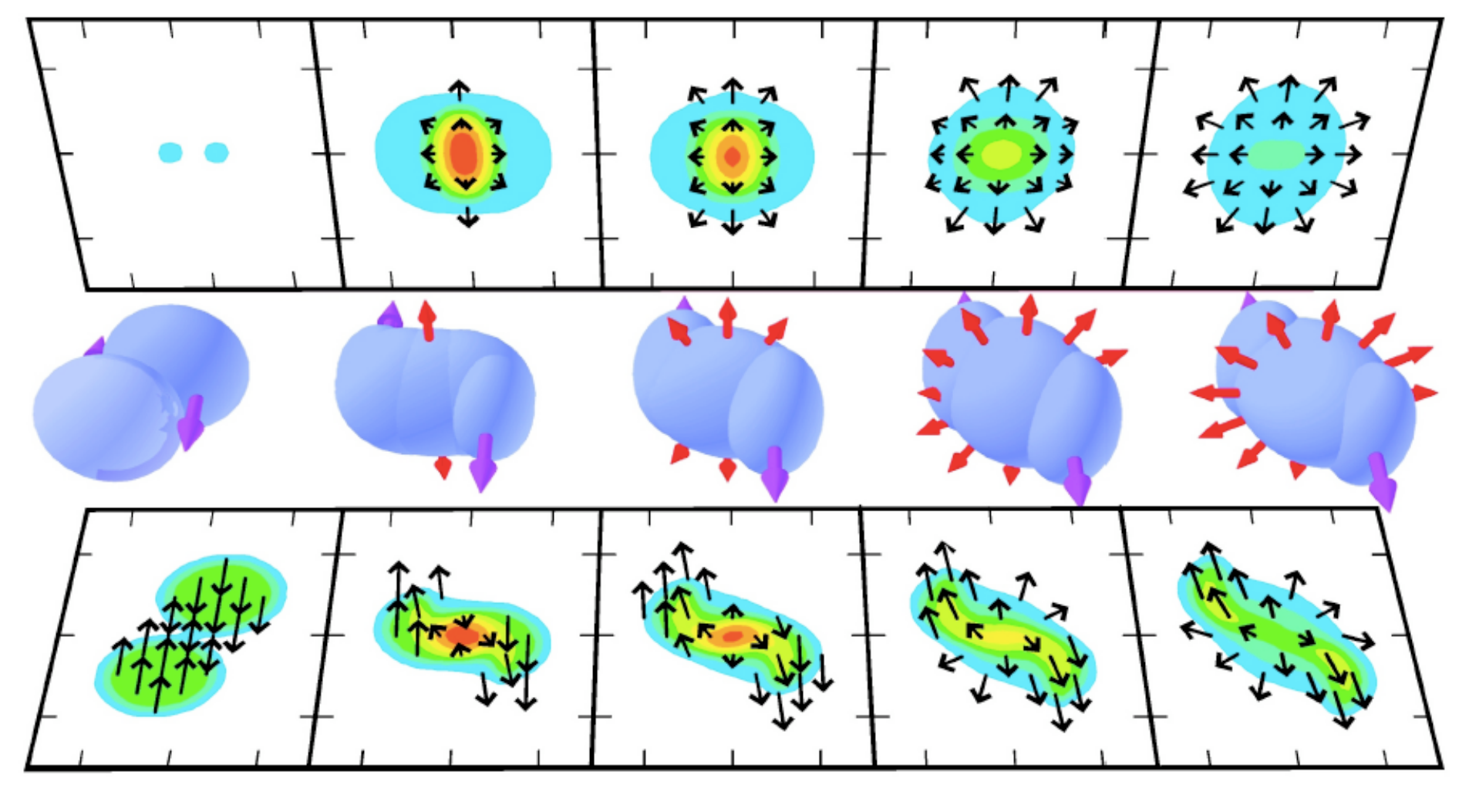}}
\end{center}
\caption{\label{central_coll}
Matter distribution at several time stages of a nearly central nucleus-nucleus
collision. (Courtesy of P. Danielewicz). 
}
\end{figure}

Eq. \eqref{BE} needs as basic ingredients the mean field $U$ and the cross
section $\sigma_{NN}$. Because these two quantities are related to each other, one
should in principle derive them in a self-consistent microscopic approach,
as the Brueckner theory, for example. However, in practice the simulations are
often done with a phenomenological mean field and free nuclear cross sections.
The most commonly used mean field is of Skyrme-type, eventually with a
momentum dependent part \cite{Gal87}.
Another important ingredient in the transport theory calculations is the
compressibility $K$ of nuclear matter, which refers to the second derivative
of the compressional energy $E$ with respect to the density:%
\begin{equation}
K=9\rho^{2}\frac{\partial^{2}}{\partial\rho^{2}}\left(  \frac{E}{A}\right)
\ . \label{Kcomp}%
\end{equation}
This is an important quantity, e.g., for nuclear astrophysics. In fact, the
mechanism of supernova explosions  is strongly dependent
on the value $K$. Supernova models might or might not lead to explosions
depending on the value of $K$. The central collisions of heavy nuclei are one of
the few probes of this quantity in the laboratory. The dependence of the
calculations on $K$ follow from the dependence of the mean field potential $U$
($U\sim E/A+$ kinetic energy terms) on the particle density $\rho$. 

Following an initial interpenetration of projectile and target densities,
the $NN$ collisions begin to thermalize matter in the overlap region making
the momentum distribution there centered at zero momentum in the c.m. system.
The density in the overlap region rises above normal and a disk of excited
and compressed matter forms at the center of the system. More and more
matter dives into the region with compressed matter that begins to expand
in transverse directions. At late stages, when the whole matter is excited,
transverse expansion dominates. A further view of the situation is illustrated in
Figure \ref{central_coll}. The measurement of the matter distribution in these
collisions and the comparison with transport theories allows one to deduce the
incompressibility of nuclear matter. With neutron-rich nuclei, one will also
able to extract the EOS dependence on the asymmetry properties of nuclear
matter. Measuring other observables, such as particle production and their
kinematic properties, also   have a valuable contribution from the use of
neutron-rich projectiles.  

\begin{figure}[t]
\begin{center}
{\includegraphics[width=10cm]{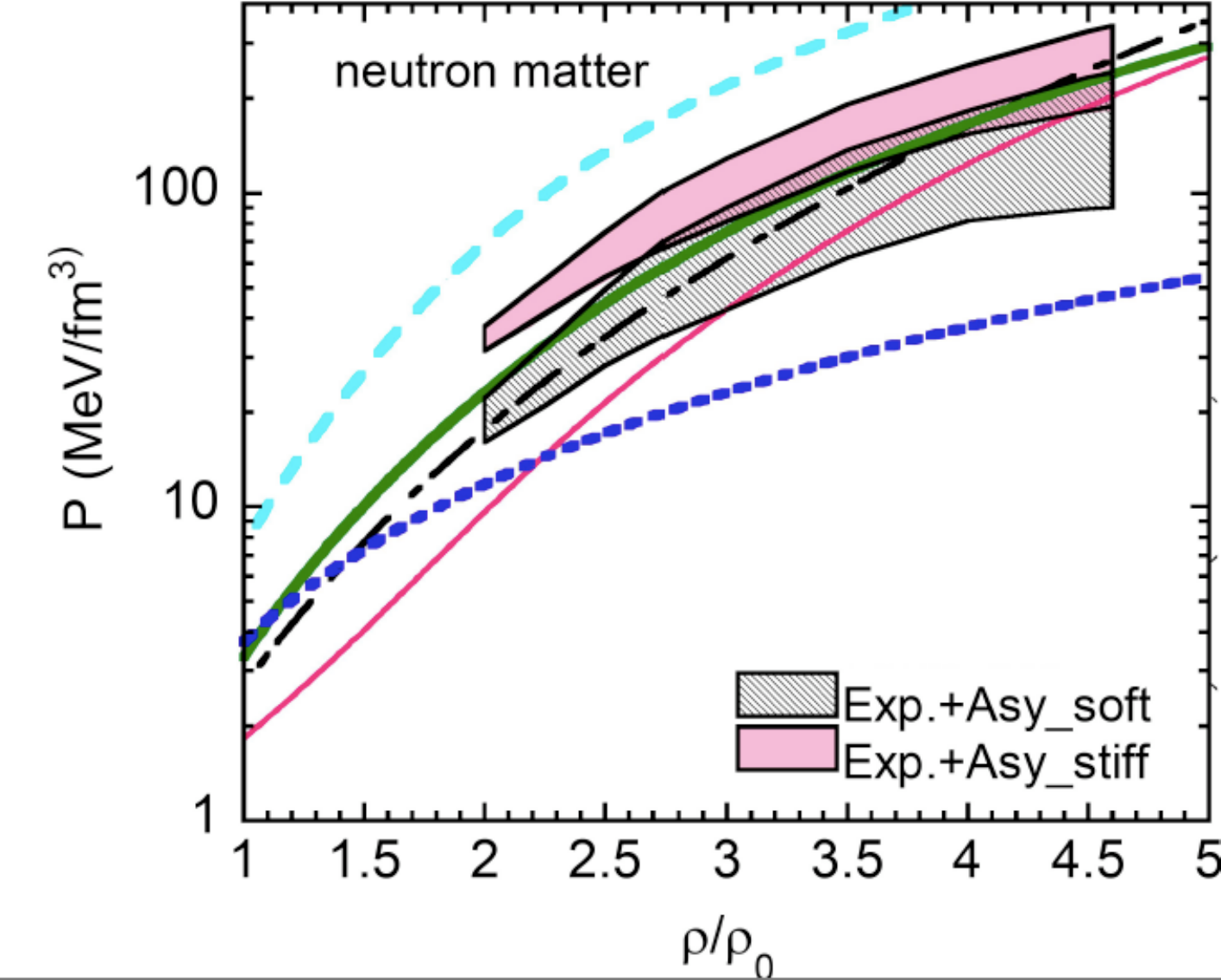}}
\end{center}
\caption{\label{central_coll2}
Pressure for neutron matter as a function of density. The different curves are
based on theoretical models. The shadow bands display the range of
possibilities for the EOS based on the analysis of experiments using a soft (lower
shadow region) and stiff (upper shadow region) EOS. (Adapted from
Ref. \cite{Lynch}).  
}
\end{figure}

Using a transport theory based on the BUU equation together with available
experimental data, 
Ref. \cite{Lynch} determined that maximum pressures achieve in central
collisions in the range of  
$P = 80$ to 130 MeV/fm$^3$ in collisions as 2 GeV/nucleon. This corresponds to 
$1.3 \times 10^{34}$ to $2.1 \times 10^{34}$ Pa). A similar analysis for
collisions at 6 GeV/nucleon yields the respective values 
of $P = 210$ to 350 MeV/fm$^3$ ($3.4 \times 10^{34}$ to $5.6 \times 10^{34}$ Pa,
respectively).  
These correspond to pressures  23 orders of magnitude larger than the maximum
pressure ever observed in the laboratory, being  
19 orders of magnitude larger than pressures within the core of the Sun. They
are, in fact, only comparable to pressures within neutron stars. The analysis
seems to be consistent with bounds for $K$ of Eq. \eqref{Kcomp} within the range
$K=170-380$ MeV \cite{Lynch}. Figure \ref{central_coll2} shows the pressure
for neutron matter as a function of the density. The different curves are based on
theoretical models. The shadow bands display the range  
of possibilities for the EOS based on the analysis of experiments using a soft
(lower shadow region) and stiff (upper shadow region) EOS (adapted from
Ref. \cite{Lynch}). Rare-isotope beams have not yet been used for experimental
work along these lines.  

\subsection{Electron scattering on radioactive beams}
Electron vs. radioactive-beam colliders have been proposed in the framework of
the physics programs of few rare-isotope beam facilities
\cite{Haik05,Sud01}. Since the reaction mechanism for electron scattering is
very well-known,  the extraction of  valuable physical quantities such as charge
distributions, transition densities, and nuclear response functions can be
obtained with good accuracy. 

In the plane-wave Born approximation (PWBA), the relation between the charge
density and the cross section is given by%
\begin{equation}
\left(  \frac{d\sigma}{d\Omega}\right)  _{\mathrm{PWBA}}=\frac{\sigma_{M}%
}{1+\left(  2E/M_{A}\right)  \sin^{2}\left(  \theta/2\right)  }\ |F_{ch}%
\left(  q\right)  |^{2}, \label{PWBA}%
\end{equation}
where $\sigma_{M}=(e^{4}/4E^{2})\cos^{2}\left(  \theta/2\right)
\sin^{-4}\left(  \theta/2\right)  $ is the Mott cross section, the term in the
denominator is a recoil correction, $E$ is the electron total energy, $M_{A}$
is the mass of the nucleus and $\theta$\ is the scattering angle.

The charge form factor $F_{ch}\left(  q\right)  $ for a spherical mass
distribution is given by%
\begin{equation}
F_{ch}\left(  q\right)  =4\pi\int_{0}^{\infty}dr\ r^{2}j_{0}\left(  qr\right)
\rho_{ch}\left(  r\right)  , \label{form}%
\end{equation}
where $q=2k\sin\left(  \theta/2\right)  $ is the momentum transfer, $\hbar k$
is the electron momentum, and $E=\sqrt{\hbar^{2}k^{2}c^{2}+m_{e}^{2}c^{4}}$.
The low-momentum expansion of Eq. \eqref{form} yields the leading terms
\begin{equation}
F_{ch}\left(  q\right)  /Z=1-\frac{q^{2}}{6}\left\langle r_{ch}^{2}%
\right\rangle +\cdots. \label{formexp}%
\end{equation}
Thus, a measurement at low-momentum transfer yields a direct assessment of the
root mean squared radius of the charge distribution, $\left\langle r_{ch}%
^{2}\right\rangle ^{1/2}$. As more details of the charge distribution
are probed, more terms of this series are needed and, for a precise description
of it, the form factor dependence for large momenta $q$ is needed as well.

The systematics of the charge-density distributions, with the inclusion
of nuclei having extreme proton-neutron asymmetry, forms a basis for
investigations  addressing both the structure of nuclei and the properties of
bulk 
nuclear matter. Information about the nuclear matter incompressibility
\cite{Gam86,Wan99}, 
the nuclear symmetry energy \cite{Vre04}, and the slope of the neutron matter
equation of state in its dependence on density \cite{Bro00} can be extracted.
Despite the importance of the nuclear equation of state (EOS), some of its
features remain fairly obscure. The electron-scattering studies will allow new
constraints on the EOS.  

Inelastic electron scattering is a powerful tool for studying
properties of excited states of nuclei, in particular their spins, parities, and the
strengths and structure of the transition operators connecting ground and
excited states (e.g. Refs. \cite{Ba62,EG88}). Electron scattering is the only 
method which can be used to determine the detailed spatial distributions of the
transition densities for a variety of single-particle and collective
transitions. These investigations provide a stringent test of nuclear many-body
wavefunctions. 

In the plane-wave Born approximation (PWBA), the cross section for inelastic
electron scattering is given by \cite{Ba62,EG88}%
\begin{align}
\frac{d\sigma}{d\Omega}  &  =\frac{8\pi e^{2}}{\left(  \hbar c\right)  ^{4}%
}\left(  \frac{p^{\prime}}{p}\right)  \sum_{L}\left\{  \frac{EE^{\prime}%
+c^{2}\mathbf{p\cdot p}^{\prime}+m^{2}c^{4}}{q^{4}}\left\vert {\mathcal M}\left(
{\bf q};CL\right)  \right\vert ^{2}\right. \nonumber\\
&  \left.  +\frac{EE^{\prime}-c^{2}\left(  \mathbf{p\cdot q}\right)  \left(
\mathbf{p}^{\prime}\cdot\mathbf{q}\right)  -m^{2}c^{4}}{c^{2}\left(
q^{2}-q_{0}^{2}\right)  ^{2}}\left[  \left\vert {\mathcal M}\left(  {\bf q};ML\right)
\right\vert ^{2}+\left\vert {\mathcal M}\left(  {\bf q};EL\right)  \right\vert
^{2}\right]  \right\}  \label{PWBA2}%
\end{align}
where $J_{i}$ $\left(  J_{f}\right)  $ $\ $is the initial (final)
angular momentum of the nucleus, $\left(  E,\mathbf{p}\right)  $ and
($E^\prime ,\mathbf{p}^{\prime}$) are the initial and final energy
and momentum of the electron, and $\left(  q_{0},\mathbf{q}\right)
=\left(\Delta E/{\hbar c},\Delta
\mathbf{p}/{\hbar}\right)  $ is the energy and
momentum transfer in the reaction. $F_{ij}\left(  q;\Pi L\right)  $
are form factors for momentum transfer $q$ and for Coulomb ($C$),
electric ($E$) and magnetic ($M$) multipolarities, $\Pi=C,E,M$,
respectively. Only for small momentum transfers and forward scattering, the
matrix elements in electron scattering are the same as those for real photons
and Coulomb excitation. The later are fixed by the energy and momentum transfer
relation $|\Delta {\bf p}| =\Delta E/c$. Electron scattering allows probing the 
momentum and energy transfer response independently. 

Due to its strong selectivity, collective and strong single-particle excitations 
can be studied particularly well in electron scattering. Electric and magnetic
giant multipole resonances are of special interest. Several of them have been
discovered and studied using electron scattering (see Ref. \cite{Bert76} and
references 
therein). Some of these giant resonances are related, e.g., to the different
charge  and matter radii of nuclei, quantities that are expected to vary strongly in
dripline nuclei.  Thus, the difference in radii of the neutron and proton
density distributions is accessible via studies of the excitation of giant
dipole resonances 
(GDR) or spin-dipole resonances. The cross section of these processes depends
strongly on the relative neutron-skin thickness \cite{Sat87}.

The reaction theory of electron scattering is  very well understood. Even in
situations for which higher order processes, such as radiative effects, are of
importance, the theory can be handled with precision.  That is why electron
scattering has the potential to become a precious tool for studies with
rare-isotope beams in the future.  
 
\begin{figure}[tb]
\begin{center}
\includegraphics[
width=4.5in]
{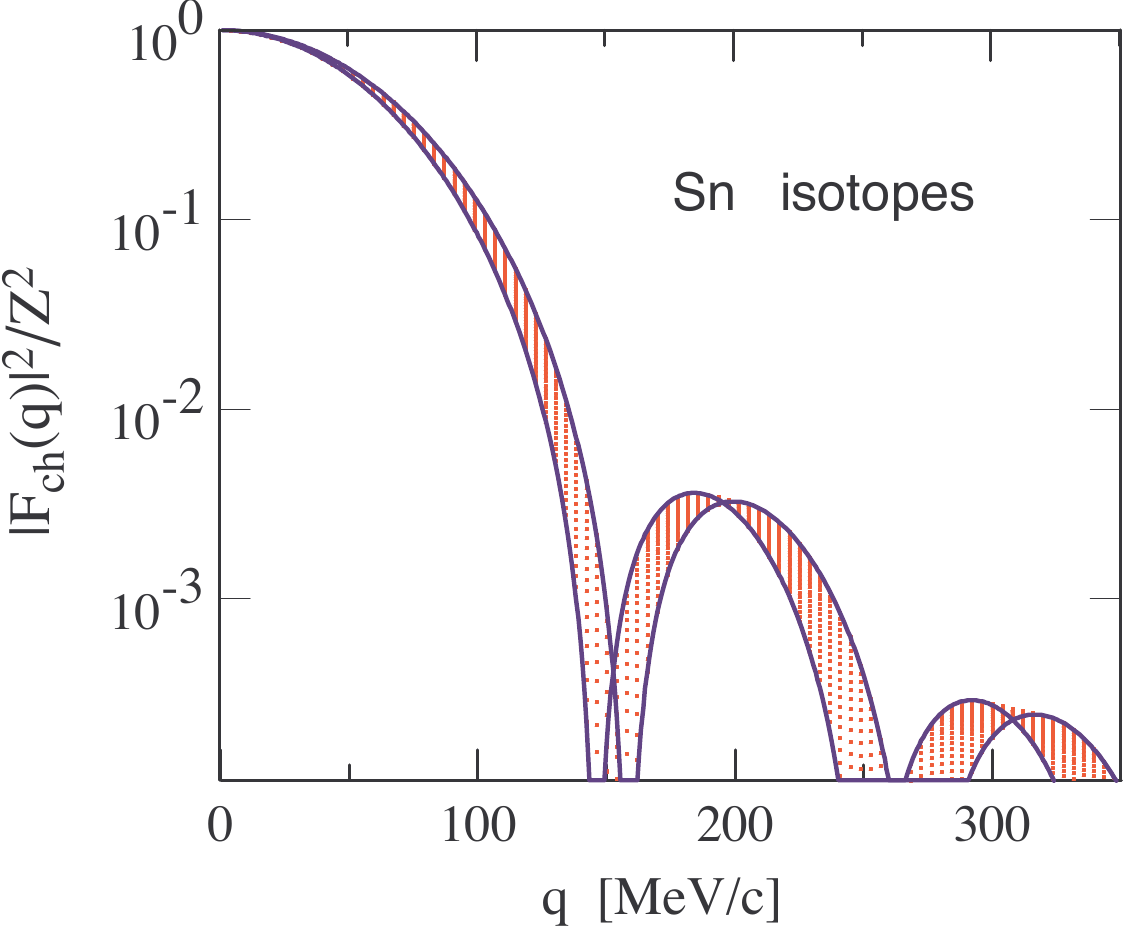}
\caption{Calculated charge form factor squared for elastic electron
scattering off tin isotopes as a function of the momentum transfer. The two
curves are for the extreme values of the asymmetry parameter
$\delta=(N-Z)/A$, that is
$\delta=0$ ($N=Z=50$), and $\delta=4/9$ ($N=90$).
\ The curves form an envelope around other curves with intermediate values
of
$\delta$. (From Ref. \cite{Ber07}).}  
\label{form_fac}
\end{center}
\end{figure}

 An example of calculated elastic electron-scattering properties for the
 scattering off exotic nuclei is given in Figure \ref{form_fac}, where the
 charge form factor squared for elastic electron scattering off tin isotopes is
 displayed as a function of the momentum transfer \cite{Ber07}. The two curves
 are for the extreme values of the asymmetry parameter $\delta=(N-Z)/A$, that is
 $\delta=0$ ($N=Z=50$), and $\delta=4/9$ ($N=90$).
The curves form an envelope around other curves with intermediate values of
$\delta$. The sensitiveness of the form factor obtained in electron
scattering on the nuclear charge distribution properties is evident and depends
essentially on the neutron skin. 
The size of the neutron skin will provide data on the volume symmetry energy which
was used as input to get the estimates presented in Figure \ref{central_coll2}. 
Recent studies of electron scattering off exotic nuclei have been published in
Refs. \cite{Roc08,Tie08,KK08,Gor98,Bertu07}. 

The electromagnetic response of light nuclei, leading to their
dissociation, has a direct connection with the nuclear physics needed in
several astrophysical sites. In fact, it has been shown \cite{Gor98} that
the existence of pygmy resonances might 
have important implications for theoretical predictions of radiative neutron
capture rates in the r-process nucleosynthesis, and consequently to the
calculated elemental abundance distribution in the universe. Recently, the
implications of inelastic electron scattering for studying the properties of
light, neutron-rich nuclei have been discussed in Ref. \cite{Bertu07}.

\section{Nuclear-astrophysics experiments with rare-isotope beams}
\subsection{Production of exotic ion beams}\label{sec:prod}
The first challenge of any experiment with short-lived ``exotic'' nuclei is
their production. Today, a broad range of rare isotopes is available for
experiments in form of ion beams. Two main production and separation mechanisms
have 
emerged as the workhorse techniques in rare-isotope beam production and are
employed in nuclear physics laboratories around the world: 
\begin{itemize}
\item Beams of short-lived nuclei are produced, separated ``in-flight'' via
  fragment separators and are directly used for experiments (in-flight
  separation).   
\item Exotic nuclei are produced, stopped, ionized, and reaccelerated (isotope
  separation on-line -- ISOL).
\end{itemize}
A third approach for the production of rare-isotope beams is the batch mode,
where a ``batch'' of radioactive material is made into a beam directly in an ion
source. This technique is limited to nuclei that have a sufficiently long
half-life and that can be produced in a chemically suitable form. This method 
was successfully used to produce exotic beams of
\nuc{7}{Be}~\cite{Gialanella1996,Gialanella2000},
\nuc{11}{C}~\cite{Joosten2000}, \nuc{14}{C}~\cite{Maier1978}, 
\nuc{18}{F}~\cite{Roberts1995,Cogneau1999}, \nuc{44}{Ti}~\cite{Sonzogni2000},
\nuc{56}{Co}~\cite{Rehm2000}, \nuc{56}{Ni}~\cite{Rehm2000} and
\nuc{76}{Kr}~\cite{Cooper2004}.  

In the following subsections we will summarize in-flight
separation and ISOL techniques and the underlying production and reaction
mechanisms.     

\subsubsection{Production of exotic nuclei -- In-flight separation}
A variety of reaction mechanisms employing various projectile-target
combinations over a broad range of collision energies are used to
produce short-lived neutron-rich or proton-rich nuclei in-flight. Peripheral
collisions at and above the Coulomb barrier create isotopes in the proximity of
the target and projectile nuclei via nucleon transfer and nucleon exchange. The highest
yield for direct one-nucleon  
transfer or exchange reactions is achieved when the momenta of the transfered or
exchanged nucleons are matched in the initial and final state (5 - 30
MeV/nucleon 
energy range)~\cite{Anyas1974}. Using stable beams and stable targets,
one-nucleon transfers 
cannot lead very far from stability. Typical production reactions are
$d(\nuc{16}{O},\nuc{17}{F})n$, $p(\nuc{7}{Li},\nuc{7}{Be})n$, and
$\nuc{3}{He}(\nuc{6}{Li},\nuc{8}{B})n$ induced by intense, low-energy beams of
\nuc{16}{O} or \nuc{6,7}{Li} impinging on light targets. The choice of inverse
kinematics yields a forward focussing of the projectile-like reaction residues, 
which can then be guided to a secondary reaction target or detector setup via
ion-optical 
transport systems like superconducting solenoids~\cite{Kolata1989,Becchetti1991}
or bending magnets~\cite{Harss2000}.  This low-energy in-flight technique is
presently being used at the {\sc TwinSol} facility at the University of Notre
Dame (ND)~\cite{Becchetti2003}, at Argonne National Laboratory
(ANL)~\cite{Harss2000}, at Texas A$\&$M University (TAMU)~\cite{Tribble2002}, at
CRIB~\cite{Yanagisawa2005} at CNS/Tokyo (Japan) and at the {\sc resolut}
facility at Florida State University (FSU).        

In the high-energy collision of a heavy-ion beam with a thin target (typically
at energies exceeding 50 MeV/nucleon), exotic nuclei are produced via projectile
fragmentation. At high energy, nucleons are removed from the projectile and the
non-interacting part of the projectile nucleus proceeds as ``spectator'' at
essentially the initial beam velocity and close to $0^{\circ}$ with narrow
angular and linear momentum distributions \cite{Greiner1975,Morrissey1989}. This
spectator or 
prefragment subsequently undergoes statistical deexcitation leading to the
observed products~\cite{Morrissey1978}. The mass, charge and
momentum distributions of the prefragments have been described with 
macroscopic as well as microscopic models~\cite{Bertsch1988}. The macroscopic
abrasion-ablation model~\cite{Bowmann1973,Gosset1977} uses  
the geometric abrasion, with the size of the prefragment calculated from the
geometry of the overlap zone in the projectile-target collision. Statistical
models can provide the proton-to-neutron ratio, assuming that both nucleon
species are abraded independently of each other. Microscopic approaches employ
intranuclear cascade models~\cite{Yariv1979,Yariv1980}, where the 
interaction of target and projectile is formulated in terms of elastic and
inelastic nucleon-nucleon scattering. The deexcitation is incorporated with
evaporation or transport models. All elements up to uranium can be produced in
projectile fragmentation.          

A fragment separator behind the production target
separates the wanted exotic nuclei from the primary beam and other reaction
residues by a combination of magnetic elements
(see~\cite{Geissel1995,Morrissey1998} for reviews). A 
selection according to the ion's magnetic rigidity separates particles with the
same momentum-to-charge ratio. A wedge-shaped degrader can remove the degeneracy as
the momentum of each species is systematically shifted by the energy loss
encountered in the
degrader material~\cite{Schmidt1987,Geissel1989}.  This in-flight separation is
chemistry-independent and a 
limitation on the lifetime of the exotic nuclei that can be produced for
experiments is solely given by the
ion's time-of-flight through the separator and to the experimental setup
(typically less than 1~$\mu$s). The available exotic beams typically have
energies exceeding 30~MeV/nucleon and event-by-event particle identification
allows 
for sensitive and efficient experiments also with cocktail beams. The drawbacks
of secondary beams produced by projectile fragmentation are the beam energies
that 
are much higher than those required by certain experiments, a poor
transversal and longitudinal beam emittance originating from the fragmentation
process and, in some cases, limited beam purity. Projectile
fragmentation as a production mechanism was pioneered at the BEVALAC in
Berkeley~\cite{Heckman1972,Symons1979} and has been the main technique for the
production of fast rare-isotope beams at GANIL (France)~\cite{GANIL}, GSI
(Germany)~\cite{GSI}, NSCL/MSU 
(US)~\cite{NSCL}, and RIBF/RIKEN (Japan)~\cite{RIKEN}.      

Although projectile fragmentation is extensively used to produce neutron-rich
nuclei, the maximum production yield is realized for neutron-deficient
species. Nuclear fission is a process known to be a source of very
neutron-rich 
nuclei~\cite{Rudstam1990}. When a fissioning nucleus is moving at energies large
compared to the fission recoil, the fission products can be collected and
separated using in-flight techniques similar to those described for projectile
fragmentation~\cite{Morrissey1998}. In a pioneering experiment at GSI, it was
shown that the fission fragments produced by a \nuc{238}{U} projectile beam
impinging upon a lead target at 750~MeV/nucleon can be separated and
identified~\cite{Bernas1994}.

\subsubsection{Isotope Separation On-Line  (ISOL) techniques}
In the ISOL method, radioactive nuclei are produced in a target, thermalized
in a catcher (often target and catcher are one and the same system), ionized in
an ion source -- if extracted from the target/catcher in atomic form -- separated and
reaccelerated~\cite{Ravn1979}. In the traditional ISOL approach, a beam of
accelerated  
stable nuclei bombards a target to produce short-lived nuclei via target
fragmentation, spallation, direct reactions, fusion or fission. The so produced
atoms are extracted from the target/catcher system -- typically with heat applied to
accelerate the diffusion process. The nuclei are transported from the
target/catcher to the ion source. After extraction from the ion source, the
low-energy ion beam is mass-separated and reaccelerated to projectile energies
required by 
the experiment. Coolers and charge breeders are used to improve the ion optical
properties of the extracted beam and to increase the reacceleration efficiency,
respectively~\cite{Ravn1979,Kugler1992,Bricault1997}. When gas catchers are
used, a fraction of the thermalizing nuclei will still be in ionic form so that
ionization in an ion source can be avoided. This scheme is referred to as Ion
Guide Isotope Separation Online (IGISOL)~\cite{Dendooven1997,Aysto2001}.  

Various projectiles over a broad range of energies impinging upon a variety of
targets are used to produce the rare isotopes of
interest. Light-ion 
induced fusion evaporation or direct reactions produce a limited number of
neutron-deficient species 
and rare isotopes close to stability at high conversion
rates. For example, at Louvain-la-Neuve, a 30~MeV proton beam on a thick
\nuc{13}{C} target ($>$1~g/cm$^2$) is used to produce a high-intensity beam of
\nuc{13}{N} with the $\nuc{13}{C}(p,n)\nuc{13}{N}$
reaction~\cite{Darquennes1990}. Heavy-ion fusion evaporation reactions 
allow to access very neutron-deficient nuclei. However, energy loss
considerations require comparably thin targets (about 1~mg/cm$^2$) and thus
result in much lower conversion rates for these reactions. 

Projectile fragmentation
reactions (e.g. heavy-ion beam on a thick carbon target~\cite{Villari2003}) and
target fragmentation  (e.g. high-energy protons on heavy
targets~\cite{Kugler1992,Bricault1997}) 
produce a variety of nuclei in the  regions close the initial projectile and
target nucleus, respectively, as well as very light nuclei. For example,
\nuc{11}{Li} is produced with this technique.  

The fission of
\nuc{235,238}{U}, \nuc{232}{Th} and long-lived actinides is used to produce
neutron-rich nuclei over a wide mass range. Low- and high-energy protons, fast
heavy ions, fast or thermal neutrons and electrons (photo fission) are employed
to induce fission. Most fission products have a large range in the target
material and thus, unlike in heavy-ion induced fusion evaporation reactions,
thick targets can be used to increase the yield of the wanted rare isotopes. In
the collision of the high-energy proton beam with a target, spallation,
fragmentation and fission can 
occur~\cite{Friedlander1963,Belyaev1980,Lynch1987,Armbruster2004}. In  spallation
reactions, a large numbers of  
protons, neutrons and $\alpha$ particles are ablated from the target nucleus,
for example \nuc{238}{U}. Spallation products are typically neutron deficient
while fission gives access to neutron-rich nuclei. 

Due to the available high flux of protons or light ions, ISOL can provide very
intense 
low-energy beams of rare isotopes with very high beam quality. However, the
effectiveness of the thick-target ISOL technique strongly depends upon the
chemistry of the wanted element, with large differences observed for different
elements~\cite{Koester2001}. Refractory elements (e.g., vanadium, zirconium and
molybdenum), that have a very high melting point and low vapor
pressure, cannot be released from an ISOL target in atomic form. Chemical
evaporation techniques and the formation of molecular sidebands are being
discussed to overcome this limitation for some refractory
elements~\cite{Koester2007,Kronenberg2008}. As the thick-target ISOL technique
relies on the 
diffusion of the wanted atoms out of the target and their effusion into
the ion source, decay losses significantly limit the intensity for species with
short lifetimes~\cite{Kirchner1992,Bennett2002,Bennett2003}. For nuclei  
with lifetimes of milliseconds or less, decay losses are often the most limiting
factor. The gas-catcher based IGISOL scheme is characterized by shorter release
times and is applicable to refractory elements and other species difficult to
ionize.       

ISOL techniques are for example used at HRIBF/ORNL (US)~\cite{HRIBF}, 
ISAC/TRIUMF (Canada) \cite{ISAC}, ISOLDE/CERN (Switzerland)~\cite{ISOLDE},
SPIRAL/GANIL (France)~\cite{SPIRAL}, and Louvain-la-Neuve (Belgium)~\cite{LLN}.
The IGISOL scheme is used at Jyv\"askyl\"a (Finland)~\cite{Aysto2001},
Louvain-la-Neuve (Belgium)~\cite{VanDuppen2000}, ANL (US)~\cite{Savard2003} and was 
recently developed for rare isotopes produced by projectile fragmentation at
NSCL/MSU (US)~\cite{Weissman2005}.   

\begin{figure}[tb]
\begin{center}
\includegraphics[
width=6.3in]
{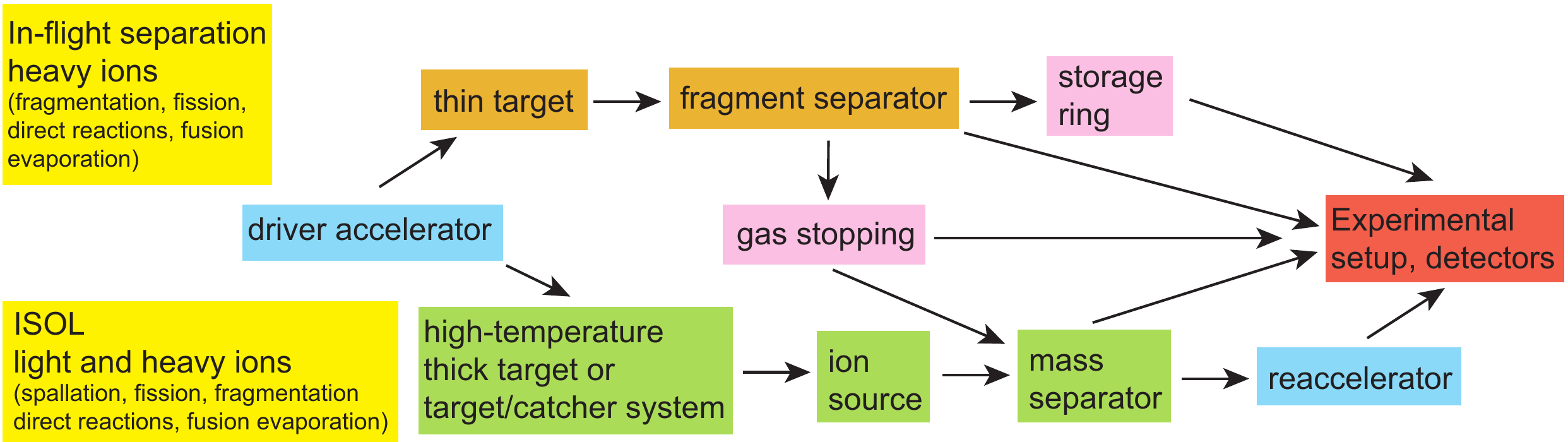}
\caption{Schematic view of the production of radioactive nuclear
beams with in-flight separation and the ISOL technique. Also shown is the hybrid 
approach envisioned for the facility for rare isotope beams (FRIB) in the US,
employing radioactive beams produced by projectile fragmentation or 
fission, thermalized in a gas stopper, for example, and reaccelerated. }  
\label{segchart1}
\end{center}
\end{figure}

\subsubsection{Outlook}

The Facility for Rare-Isotope Beams (FRIB)~\cite{FRIB} planned in the US will
provide both, fast exotic beams produced {\it in-flight} via projectile
fragmentation or 
fission and reaccelerated rare-isotope beams at various energies. In an
IGISOL-based scheme, short-lived isotopes produced in a chemistry-independent
way by projectile fragmentation or projectile fission will be separated in a
fragment separator, thermalized in a gas catcher and quickly extracted (order of
milliseconds) with the possibility of reacceleration to the beam energies
required for experiments. The gas-catcher based thermalization of fast exotic
beams produced by projectile fragmentation is presently operational at NSCL/MSU and
a reacceleration capability is presently being implemented~\cite{ReA3}. At ANL,
the CAlifornium Rare Isotope Breeder Upgrade (CARIBU) project~\cite{Savard2008}
will produce neutron-rich isotopes from a  \nuc{252}{Cf} source for
reacceleration through the ATLAS facility and provide new opportunities for
experiments with reaccelerated Cf fission fragments.    

In Europe, design studies for the next-generation European ISOL facility EURISOL
are ongoing~\cite{EURISOL}. The aim is to increase the variety of exotic nuclei
produced and to increase the production yields by orders of magnitude compared to present-generation
ISOL facilities. EURISOL would complement the Facility of Antiproton
and Ion Research (FAIR)~\cite{FAIR} in Germany, where high-energy rare-isotope
beams (GeV/nucleon) will be 
produced by fragmentation and fission, separated in-flight with the possibility
of passive deceleration or cooling in storage ring structures. The SPIRAL2 and
ISAC-II(I) upgrades are underway at GANIL~\cite{SPIRAL2} and TRIUMF~\cite{TRIUMF},
respectively. A major upgrade of the RIKEN facility (RI Beam Factory - RIBF) in
Japan became operational recently~\cite{RIBF}. 

\subsection{Experimental techniques and applications}

This section describes experimental approaches and detection
systems developed or adapted to probe properties of exotic nuclei that are
important for nuclear astrophysics. Many observables ranging from nuclear masses
and ground-state half-lives to specific reaction cross sections are nuclear
physics input crucial for the understanding of a large variety of astrophysical
processes and scenarios. We tried to focus here on the measurements and
developments that were not covered in the review by Smith and Rehm from 2001~\cite{SmithRehm2001}.

\subsubsection{Targets}

Hydrogen- and helium-induced reactions are of great importance in many
astrophysical scenarios. A variety of hydrogen and helium (enriched) target 
schemes exist. Thin plastic targets, for example, polyethylene  -- (CH$_2$)$_{\rm
  n}$ -- 
foils with thicknesses between tens of $\mu$g/cm$^2$ and several mg/cm$^2$,
have been used for proton-induced reactions. As helium does not form 
compounds, solid targets containing helium must be prepared by
implantation. Helium-implanted aluminum targets have been developed at
Louvain-la-Neuve~\cite{Vanderbist2002}. The carbon in plastic targets and the
heavier elements serving as carriers in implanted targets increase the energy
loss of the beam passing through the target and induce intense background
reactions, typically dominated by elastic scattering. Aside from physical
radiation damage at high beam intensities, hydrogen depletion in plastic targets
poses problems as well.  

Gas cells have been developed to mitigate some of these 
problems. While gas cells with windows are easy to handle, even thin windows
can degrade the beam energy and produce unwanted background reactions. A He-gas
filled chamber is used at Louvain-la-Neuve~\cite{BradfieldSmith1999}, while
cryogenic helium and hydrogen gas cells have been built for experiments at
ANL~\cite{Sonzogni2000} and GANIL~\cite{Mittig2001}, for example. When
excitation functions are of interest, the so-called thick-target inverse
kinematics (TTIK) technique~\cite{Artemov1990} can be employed. For this
approach, the beam is slowed down and stopped in a thick target with the light
reaction residues escaping and being detected. With this, a reaction can be
probed over a wide energy range with one incoming beam energy. The TTIK
technique for He-induced reactions on low-energy \nuc{14}{O} beams has been used
at CRIB in Tokyo~\cite{Notani2004} and at Texas A$\&$M University~\cite{Fu2007}
with thick He gas targets; a thick polyethylene target was used at HRIBF at Oak
Ridge National Laboratory to study resonant reactions with \nuc{17}{F} and
\nuc{18}{F} beams~\cite{GalindoUribarri2000,Bardayan2005}.   

Window-less, differentially-pumped gas   
targets eliminate background reactions and increased energy loss arising from
the presence of windows  but require several stages of pumping
at high pumping speeds to drop the pressure by orders of magnitude over
short distances. At present, window-less gas targets are used in conjunction
with the DRAGON setup~\cite{Hutcheon2003} and at
HRIBF/ORNL~\cite{Fitzgerald2005,ORNLGAS}. 

\subsubsection{Direct measurement of $(p,\gamma)$ and $(\alpha,\gamma)$
radiative capture reactions}\label{sec:capture_exp}  

Charged-particle induced radiative capture reactions such as $(p,\gamma)$ and
$(\alpha,\gamma)$ occur in 
many stellar environments, for example, in novae and X-ray bursts. In explosive
environments -- due to the high temperatures and short reaction times -- capture
reactions involving short-lived nuclei play an important role for energy
generation and nucleosynthesis. 

Resonant capture reactions on stable and long-lived radioactive targets have
been investigated with intense proton and $\alpha$
beams~\cite{Rolfs1990,Rowland2002}. The most recent examples for direct
measurements of radiative capture reactions are
\nuc{23}{Na}($p,\gamma$)\nuc{24}{Mg} and \nuc{17}{O}($p,\gamma$)\nuc{18}{F}
studied at the Laboratory for Experimental Nuclear Astrophysics (LENA) at
TUNL~\cite{Rowland2004,Fox2004,Fox2005} and the 
\nuc{3}{He}($\alpha$,$\gamma$)\nuc{7}{Be}~\cite{Gyrky2007,Bemmerer2006b,Confortola2007} 
and \nuc{14}{N}($p,\gamma$)\nuc{15}{O}
reactions~\cite{Formicola2004,Bemmerer2006a,Lemut2006,Marta2008} at the
Laboratory Underground Nuclear Astrophysics (LUNA)
facility~\cite{Greife1994,Formicola2003} located in the  
Gran Sasso Laboratory. 

For the resonant capture on short-lived nuclei, the experiments
have to be performed in inverse kinematics. Recoil separators are typically used 
to detect and identify the recoiling reaction products and to reject the direct
beam. The separators that are presently 
used for inverse-kinematics capture reactions with stable and rare-isotope beams
include DRAGON~\cite{Hutcheon2003} at TRIUMF (Canada), ARES~\cite{Couder2003} at
Louvain-la-Neuve (Belgium), the Daresbury Recoil 
Separator (DRS)~\cite{Fitzgerald2005} at HRIBF/ORNL (US), the FMA~\cite{Rehm1998}
at ANL (US) and, for stable beams only, ERNA~\cite{Schuermann2004} at Bochum
(Germany). The STrong Gradient Electro-magnetic Online Recoil separator for
capture Gamma ray Experiments (St. George) is under construction at the University
of Notre Dame and will be used for inverse-kinematics $(\alpha,\gamma)$ capture
reactions induced by a He-jet target~\cite{Couder2008}. 

The pioneering measurement with a rare-isotope beam was the first direct
determination of the 
\nuc{13}{N}$(p,\gamma)$\nuc{14}{O} reaction cross section using a radioactive
\nuc{13}{N} beam produced at Louvain-la-Neuve~\cite{Decrock1991,Delbar1993}. A
\nuc{13}{N} beam with an intensity of 3 $\times$ 10$^8$ particles per second and
an energy of 8.2~MeV impinged upon a 180~$\mu$g/cm$^2$ polyethylene target. A
surface-barrier diode located at $17^{\circ}$ with respect to the beam axis
detected the scattered \nuc{13}{N} and the carbon and hydrogen target
recoils. Assuming pure Rutherford scattering on the \nuc{12}{C} in the target,
the number of incident \nuc{13}{N} was determined. The resonant radiative proton
capture was tagged via the 5.173 MeV capture $\gamma$-ray originating from the
ground-state decay of the first excited $1^-$ state of \nuc{14}{O} measured with
a large volume Ge diode. The peak-to-total in the $\gamma$-ray spectrum was
improved by vetoing cosmic rays with Cherenkov detectors positioned above the
$\gamma$-ray detector setup. The $\Gamma_{\gamma}$ width of the dominant
\nuc{14}{O} $1^-$ level was determined for the first time from a direct
measurement. Compared to the previously adopted
\nuc{13}{N}$(p,\gamma)$\nuc{14}{O} reaction rate, the results of the measurement
by Delbar {\it et al.} suggested that the reaction actually proceeds twice as
fast~\cite{Arnould1992,Delbar1993}. After studies of the non-resonant capture
cross section~\cite{Tang2004}, the $\gamma$ width of the 5.173~MeV state in
\nuc{14}{O} now poses the largest uncertainty of this reaction
rate~\cite{Blackmon2006}.    

The later study of the
\nuc{19}{Ne}$(p,\gamma)$\nuc{20}{Na} reaction at Louvain-la-Neuve used the
resolving power of the recoil separator ARES to identify \nuc{20}{Na} events
from capture through the 2.643~MeV level~\cite{Couder2004}. An upper limit of
the resonance strength of the 2.643~MeV level was determined with this approach
that bypasses low-efficiency $\gamma$-ray detection. It was
concluded that an increased transmission and efficiency of ARES would be
required to be able to determine more than an upper limit~\cite{Couder2004}.

Some of the most recent resonant capture reactions induced by radioactive and
stable beams in inverse kinematics have
been measured at the DRAGON facility at TRIUMF. DRAGON is a recoil mass
separator at ISAC dedicated for the measurement of low-energy reactions of
astrophysical interest. Radioactive and stable beams with energies between 0.15
and 1.5~MeV are delivered by ISAC. The DRAGON facility consists of a
recirculating, differentially-pumped window-less gas target surrounded by a
30-crystal BGO $\gamma$-ray detection array, an electromagnetic separator and
a heavy-ion recoil detection system~\cite{Hutcheon2003}. At the DRAGON facility,
seven resonances in \nuc{22}{Mg}  have been
characterized in the reaction \nuc{21}{Na}$(p,\gamma)$\nuc{22}{Mg} at
center-of-mass energies between $E_{cm}=200 -
1103$~keV~\cite{Bishop2003,DAuria2004}, the strength of the $E_{cm}=184$~keV
resonance in the \nuc{26g}{Al}$(p,\gamma)$\nuc{27}{Si} reaction was determined
for the first time from a direct measurement in inverse
kinematics~\cite{Ruiz2006} and the \nuc{40}{Ca}$(\alpha,\gamma)$\nuc{44}{Ti}
reaction was probed with a \nuc{40}{Ca} beam of 0.6 - 1.15~MeV/nucleon covering
the relevant temperature range for the $\alpha$-rich freeze-out during a
core-collapse supernova~\cite{Vockenhuber2007}. 

At HRIBF/ORNL, the cross section of the $\nuc{17}{F}(p,\gamma)\nuc{18}{Ne}$
reactions was measured directly for the first time~\cite{Chipps2009}. A cocktail
beam of \nuc{17}{O} and \nuc{17}{F} was produced at HRIBF and impinged upon
ORNL's differentially pumped, windowless hydrogen target. The Daresbury Recoil
Separator was tuned to transmit the \nuc{18}{Ne} recoils to the focal plane
where they were identified in the gas-filled ionization chamber. Two
normalization schemes for counting the incoming beam were used, (i) the
detection and counting of elastically scattered protons in surface-barrier
detectors and (ii) beam-current measurements facilitated by a metal plate and
two plastic scintillator paddles; the decays of \nuc{17}{F} nuclei implanted
into the metal plate that was inserted into the beam from time to time and
retracted to sit between the scintillator paddles were
measured~\cite{Chipps2009}. The strength of the $3^+$ resonance in \nuc{18}{Ne}
(about 600~keV above the proton separation energy corresponding to a 10.83 MeV
\nuc{17}{F} beam energy for the resonant-capture measurement) was determined. At
an off-resonance \nuc{17}{F} beam energy (800 keV above the proton threshold),
an upper limit for the direct capture away from the dominating resonances could
be determined~\cite{Chipps2009}. Compared to previous nova nucleosynthesis
calculations~\cite{PareteKoon2003}, the abundance of the important
galactic $\gamma$-emitter \nuc{18}{F}, the $\beta$-decay daughter of
\nuc{18}{Ne}, 
increased in the hottest zone of a 1.35 solar mass white dwarf nova by a factor of 1.6 
over the previously adopted rate. The uncertainty of the measured rate leads to
a spread of a factor of 2.4 in the final abundance of \nuc{18}{F}, while varying
the previous resonant contribution by a factor of 10
resulted in a spread of as much as 15-16 times in the final
abundance of \nuc{18}{F}~\cite{Chipps2009}. Further measurements, for example
the determination of the $\nuc{17}{F}(p,\gamma)\nuc{18}{Ne}$ direct capture
cross section and the $\nuc{18}{F}(p,\alpha)\nuc{15}{O}$ reaction cross section, are
needed to further reduce uncertainties in the production of \nuc{18}{F} in
novae~\cite{Chipps2009}. The impact on X-ray bursts is under investigation;
there are indications that the abundances of 
\nuc{17}{O} and \nuc{17}{F} in X-ray bursts can be modified by more than an order of
magnitude at a reduction in uncertainty by a factor of 20 compared to previous
network calculations~\cite{Chipps2009}.

The cross section for radiative capture has been observed to show the
characteristic resonances of the reaction on top of a background that slowly 
varies with beam energy~\cite{Rolfs1973,Rolfs1990}. This smooth background is
attributed to the direct 
capture process -- a transition for the projectile from an
initial continuum state to a final bound state via the electromagnetic
interaction -- that depends on the properties of the bound states of 
the compound nucleus~\cite{Rolfs1973}. This direct non-resonant capture can 
become important for stellar reaction rates when the nuclear level density is
low in the region of the Gamow window. Also interference effects between
resonant and non-resonant capture can occur. For capture reactions involving
stable nuclei important in hydrostatic burning, the contributions from direct
capture have been estimated from capture cross sections measured away from
resonances~\cite{Adelberger1998}. The total cross section for
direct proton capture is of order $\mu$b~\cite{Rolfs1990} in the relevant energy
range and except for the case of
\nuc{7}{Be}$(p,\gamma)$\nuc{8}{B}~\cite{Junghans2002,Junghans2003}, direct
measurements of direct capture with radioactive beams have not been performed
due to the high beam intensity requirements. \\

\subsubsection{Coulomb dissociation -- the time-reversed approach to radiative capture}\label{sec:CDexp}

The dissociation of a fast-moving projectile in the Coulomb field of a high-$Z$
nucleus -- Coulomb excitation of continuum states -- was proposed as alternative
method to determine radiative capture cross  
sections~\cite{Baur1986,Baur1996}. For the Coulomb dissociation approach, the
residual nucleus $B$ of the radiative capture reaction of interest,
$A(x,\gamma)B$, is Coulomb excited in the electromagnetic field of a high-$Z$
nucleus to an unbound state that decays into $A+x$. Coulomb dissociation is
mediated by the absorption of a virtual photon, $B(\gamma_{virt},x)A$, which is
related to photodisintegration via the virtual photon
method~\cite{Williams1934,Weizsacker1934}.    
As photodisintegration is the time reverse process of radiative capture, the
cross sections of the two processes can be related via the detailed balance
theorem (see Section~\ref{sec:CD} for the theoretical formalism of Coulomb dissociation). 

In Coulomb dissociation measurements at intermediate beam energies, the relative
energy of the residues $A$ and $x$ is determined from their invariant mass, which
is deduced from the measured velocity vectors of $A$ and $x$. The relative
energy corresponds to the center-of-mass energy of the $A(x,\gamma)B$
capture. Experimental advantages of the Coulomb breakup method are threefold,
(i) the typically large number of virtual photons leads to a big Coulomb
dissociation cross section, (ii) the photoabsorption cross section is favored
compared to the radiative capture cross section by two to three orders of
magnitude (phase space), and (iii) thick targets can be used in the regime of fast beams (higher
luminosity)~\cite{Motobayashi1998}. A challenging requirement for experiments, 
however, is the need for the measurement of angular distributions. The
multipolarities have to be disentangled since the electromagnetic multipoles
contribute with different strengths to Coulomb excitation and capture
processes. Particularly important are the non-negligible contributions and
interference of electric dipole ($E1$) and electric quadrupole ($E2$)
excitations. The applicability of the Coulomb 
dissociation approach is limited by the fact that preferentially $E1$ modes are
excited and that the detailed balance theorem cannot be used for the extraction
of the radiative capture cross section in case the capture feeds bound
excited states that decay by $\gamma$-ray emission with unknown branching
ratio~\cite{Motobayashi1998}. The main body of 
experimental work utilizing Coulomb breakup of rare isotopes 
for nuclear astrophysics has been aimed at exploring radiative proton capture
reactions.      

An extensive experimental program that utilizes Coulomb breakup reactions for the
extraction of radiative capture cross sections was initiated first at
RIKEN/Japan by Motobayashi {\it et al.}~\cite{Motobayashi1991,Motobayashi2003}. In the
pioneering experiment, the electromagnetic transition strength for the excited
$1^-$ resonance in \nuc{14}{O} was determined in the Coulomb dissociation, 
$\nuc{14}{O} \rightarrow \nuc{13}{N} + p$, induced by 87.5~MeV/nucleon
\nuc{14}{O} impinging upon a 350~mg/cm$^2$ Pb target~\cite{Motobayashi1991}. The
Coulomb dissociation cross section was determined from coincidence spectroscopy
of protons and \nuc{13}{N} in position-sensitive detection systems consisting of
telescopes of 24 position-sensitive Si detectors for $\Delta E$ measurements
backed by CsI(Tl) scintillators for $E$ measurements~\cite{emric} and
plastic-scintillator hodoscopes with $x-y$ position sensitivity. The complete
kinematics -- total energy and relative momentum vectors of the $p+\nuc{13}{N}$
system -- were determined~\cite{Motobayashi1991}. The radiative width
$\Gamma_{\gamma}$ was deduced and found in agreement with the direct measurement
performed at Louvain-la-Neuve~\cite{Decrock1991} (see Section
\ref{sec:capture_exp} for the direct measurement). 

Subsequent experiments at RIKEN were aimed at the Coulomb dissociation of 
\nuc{8}{B}~\cite{Motobayashi1994,Kikuchi1997,Kikuchi1998} to extract the
reaction rate of the crucial $\nuc{7}{Be}(p,\gamma)\nuc{8}{B}$ reaction, which
is of great importance for the neutrino production in the Sun through the
$\beta$ decay of \nuc{8}{B}. In the first experiment in 1994, the breakup of
a \nuc{8}{B} beam of 46.5 MeV/nucleon incident on a 50~mg/cm$^2$ \nuc{208}{Pb}
target, used a two-layered, $x-y$ position-sensitive plastic-scintillator
hodoscope with an active area of 
1 $\times$ 0.96~m$^2$ for the detection of the coincident $\nuc{7}{Be} + p$
breakup products. The hodoscope with a $\Delta E$ layer of 5~mm thick plastic
(10 segments) and the $E$ plane of 60~mm thick plastic (16 segments) was
positioned 5~m downstream of the Pb target. The energy of the breakup fragments
was deduced from the time-of-flight (TOF) over the 5~m flight path, the angle
from the location of the hits in the segmented hodoscope. The particle
identification was performed with the $\Delta E - E$ method and from TOF-E
information. A helium-filled bag was inserted between the target and the
hodoscope to reduce the background due to breakup in air~\cite{Motobayashi1994}.
In a second experiment~\cite{Kikuchi1997,Kikuchi1998}, the hodoscope was
modified for higher 
count-rate capability and moved closer to expand the angular range covered from
6$^{\circ}$ to 10$^{\circ}$. Also, the NaI(Tl) scintillator $\gamma$-ray detection
array DALI was positioned around the target to tag the fraction of the breakup
leading to the 429~keV bound excited state in \nuc{7}{Be}. The larger angular
range covered by the hodoscope allowed for angular-distribution measurements and
an assessment of the $E2$ contribution to the breakup
process~\cite{Kikuchi1997,Kikuchi1998}. Similarly, the reactions
$\nuc{8}{B}(p,\gamma)\nuc{9}{C}$, $\nuc{11}{C}(p,\gamma)\nuc{12}{N}$ and
$\nuc{12}{N}(p,\gamma)\nuc{13}{O}$, important for the hot $pp$ mode nuclear buring
in hydrogen-rich, very massive objects \cite{Wiescher1989}, were studied
via Coulomb breakup at RIKEN to extract the reaction rates relevant to explosive
hydrogen burning~\cite{Motobayashi2003}. Subsequent measurements on $sd$-shell
nuclei aimed at the 
study of the breakup of \nuc{23}{Al}~\cite{Gomi2005} and
\nuc{27}{P}~\cite{Togano2005} into $\nuc{22}{Mg} + p$ and $\nuc{26}{Si} + p$,
respectively, continued to use $\gamma$-ray spectroscopy with DALI and employed
a position-sensitive Si detector telescope in front of the plastic hodoscope for
the identification of the projectile-like breakup residues. The experimental
results in the $sd$ shell are relevant to the reaction path in 
Ne novae~\cite{Gehrz1985}, where at high temperature and density many nuclear reactions
involving rare isotopes contribute in the hot CNO cycle and the NeNa- and
MgAl-cycles~\cite{Champagne1992}. A specific signature is the nucleosynthesis of
long-lived galactic $\gamma$ emitters such as \nuc{22}{Na} and \nuc{26}{Al}, and
nucleosynthesis up to the silicon and calcium range~\cite{Starrfield2000}. Assuming
temperature and density conditions given by nova models~\cite{Iliadis2002}, the radiative
width obtained in the RIKEN study indicates that the main reaction path
favors the $\beta$ decay rather than the proton capture on \nuc{22}{Mg}~\cite{Gomi2005}. For
$\nuc{26}{Si} + p$, the preliminary result of the $\gamma$-decay width of the
first excited state in \nuc{27}{P} is ten times smaller than the value
estimated based on a shell-model calculation in \cite{Caggiano2001}. This
indicates that the $\nuc{26}{Si}(p,\gamma)\nuc{27}{P}$ reaction does not
contribute significantly to the amount of \nuc{26}{Al} produced in
novae~\cite{Togano2005}.  

At GANIL, the Coulomb dissociation cross sections of $\nuc{14}{O} \rightarrow
\nuc{13}{N} + p$~\cite{Kiener1993} and $\nuc{12}{N} \rightarrow \nuc{11}{C} +
p$~\cite{Lefebvre1995} were measured. Radioactive beams of \nuc{14}{O} at
70~MeV/nucleon and \nuc{12}{N} at 65.5~MeV/nucleon were provided by the ALPHA
spectrometer and guided onto 100~mg/cm$^2$ and
120~mg/cm$^2$ \nuc{208}{Pb} targets, respectively. In both experiments, the heavy
breakup residues were characterized with the SPEG spectrometer and its
focal-plane detection system consisting of two drift chambers  
and a plastic scintillator. The protons were detected in CsI(Tl) detectors
downstream of the Pb target~\cite{Kiener1993,Lefebvre1995}.             
             
Much higher beam energies are available for Coulomb dissociation experiments at
GSI. The breakup of $\nuc{8}{B} \rightarrow \nuc{7}{Be} + p$ was measured in two
experiments at 254~MeV/nucleon \nuc{8}{B} beam
energy~\cite{Iwasa1999,Schumann2003}. Both experiments utilized the KaoS
spectrometer~\cite{Senger1993}. Beam particles were identified with the
TOF-$\Delta E$ method 
using a plastic timing detector upstream of the target and a large-area
scintillator wall close to the focal plane of KaoS. Momentum vectors of the
reaction residues were analyzed by trajectory reconstruction with KaoS using
position 
sensitive microstrip detectors and two 2d multi-wire proportional counters
(MWPC). The first experiment used a 199~mg/cm$^2$ \nuc{208}{Pb} target, the
second experiment used a Pb target of 50~mg/cm$^2$ thickness. Tracking of the angle of
the incoming beam with parallel-plate avalanche counters allowed to measure
angular distributions to disentangle contributing multipolarities in the second
experiment~\cite{Schumann2003,Schumann2006}.       
      
At MSU, inclusive and exclusive measurements of the $\nuc{8}{B} \rightarrow
\nuc{7}{Be} + p$ Coulomb dissociation were
performed~\cite{Davids2001a,Davids2001b}. Inclusive measurements 
used 44 and 81~MeV/nucleon \nuc{8}{B} beams provided by the A1200 fragment
separator at NSCL. The S800 spectrograph~\cite{Bazin2003} was used to detect the
\nuc{7}{Be} 
residues from the Coulomb dissociation of \nuc{8}{B} impinging upon
27~mg/cm$^2$ Ag and 28~mg/cm$^2$ Pb targets. The longitudinal momentum
distributions of \nuc{7}{Be} were derived from trajectory reconstruction
using the position information provided by the position-sensitive cathode
readout drift chambers of the spectrograph's focal-plane detection
system~\cite{Yurkon1999}. Particle identification was achieved with the $\Delta
E - E$ method 
employing the S800 ionization chamber for $\Delta E$ and the first focal-plane
plastic scintillator for $E$ measurements~\cite{Yurkon1999}. In comparison to
model calculations, 
a rather high $E2$ contribution to the Coulomb breakup was deduced, which has not
been confirmed by any other measurement to date. An exclusive measurement at
83~MeV/nucleon on a 47~mg/cm$^2$ Pb target in front of a 1.5~T dipole 
magnet was performed in addition. Two multi-wire drift chambers (MWDCs) each
were used to track the flight 
paths of the \nuc{7}{Be} and the proton after passage through the magnetic
field. An array of plastic scintillators was placed behind the MWDCs. Particle
identification was performed with the energy loss measured in the plastic
scintillator array and the TOF information. The invariant mass method was used
to calculate the relative energy of the breakup
residues~\cite{Davids2001a,Davids2001b}.       

In a recent comprehensive study, M. Gai~\cite{Gai2006} summarized the results
of the various \nuc{8}{B} Coulomb dissociation measurements and the extracted 
astrophysical $S_{17}(0)$ factor in comparison to the results of direct capture
measurements.  

Recently, the neutron capture reaction on \nuc{14}{C} has been explored using
Coulomb dissociation of $\nuc{15}{C} \rightarrow \nuc{14}{C} + n$ at
69~MeV/nucleon in the field of a 224~mg/cm$^2$ Pb target at
RIKEN~\cite{Nakamura2009}. The \nuc{14}{C} residues were analyzed with a
magnetic spectrometer that used a drift chamber and plastic scintillator
hodoscopes. The momentum vector of \nuc{14}{C} was determined by combining the
tracking data and time-of-flight information. Particle identification was
achieved with $\Delta E$ - TOF and tracking information. The breakup neutron was
detected by two layers of neutron hodoscope arrays with an active area of $2.14
\times 0.92$~m$^2$ positioned almost 5~m downstream of the Pb target. The neutron
momentum vector was extracted from the hit position of the neutron and its
TOF. The spectral shape and amplitude was found consistent with similar data
taken at higher beam energies at GSI~\cite{Nakamura2009}. The deduced direct
capture cross section agrees with the most recent direct capture measurement and
demonstrates that the Coulomb breakup with a neutron in the exit channel is a
useful tool to derive the radiative capture cross section. A program aimed at
Coulomb breakup of neutron-rich nuclei, not exclusively with an astrophysical
background, is also ongoing at GSI at higher beam
energies~\cite{Pramanik2003,Adrich2005,Nociforo2005,Aumann2006}.     

\subsubsection{Direct $(p,\alpha)$ and $(\alpha,p)$ measurements}

Proton-induced reactions are of great importance in many astrophysical scenarios
as the corresponding reaction cross sections are high due to the small Coulomb
barrier. The predominant reactions in nova outbursts, for example, are of type
$(p,\gamma)$ and $(p,\alpha)$. A significant fraction of the energy produced
during nova explosions stems from proton-induced reactions on pre-existing
carbon, nitrogen, and oxygen seed nuclei. The $(p,\alpha)$ reactions play
a crucial role in cycling reaction flow back to lower masses.  At sufficiently
high temperatures, breakout sequences can occur and heavier nuclei are synthesized
through reactions on neutron-deficient nuclei. In X-ray bursts, for example, a
sequence of $(\alpha,p)$ and $(p,\gamma)$ reactions ($\alpha$p process) leads to
nuclei up to mass $A=40$, eventually followed by rapid proton capture reactions
(rp process) and $\beta$ decays resulting in the formation of nuclei with $A
\approx 100$~\cite{Schatz1998}.

The study of reactions with charged particles in the exit channel 
advanced tremendously with the advent of large-area silicon strip detectors. Si
strip detectors can be single-sided -- providing segmentation in one dimension --
or double-sided for two-dimensional segmentation. Detector thicknesses between
50-1000~$\mu$m are typically available. A high degree of segmentation also
allows for higher total counting rates (without pileup) than could be achieved
with a single detector of the same total area. When operated in transmission
geometry, silicon detectors can be stacked to enable $Z$ identification of a broad
range of particles by measuring energy loss and total energy ($\Delta E-E$
method).   

The Louvain-Edinburgh Detector Array
(LEDA)~\cite{Davinson2000} was one of the pioneering charged-particle arrays for
the detection of reaction residues in low-energy nuclear astrophysics
experiments. For LEDA, the strips are curved in a circular pattern to realize an
annular geometry. The CD-like detector consists of eight sectors with 16 strips
(5~mm pitch) per sector. For such a configuration, the range of scattering
angles and the angular resolution are determined by the target-detector
distance. The removal of two sectors from the complete annular LEDA detector
allows the remaining six sectors to be arranged as a six-sided cone
(referred to as the LAMP configuration) providing very large
solid-angle coverage at poorer angular resolution~\cite{Davinson2000}. ORNL's
Silicon Detector Array (SIDAR)~\cite{Bardayan2000} consists of single-sided
silicon strip detectors of the LEDA-type (128 strips each) and a smaller CD-like
detector with  64 strips in total for the detection and identification of the
heavy 
recoil~\cite{SIDAR,Bardayan2000}. The TRIUMF UK Detector Array (TUDA) facility
at ISAC/TRIUMF  is a cylindrical scattering chamber, where large-area silicon
detectors of the LEDA-type or similar CD-shaped detectors can be
mounted~\cite{Buchmann2008}.      

For the TUDA facility at TRIUMF, a large solid angle active target for the
detection of low-energy charged particles in measurements of
astrophysical reaction rates is being developed. The TRIUMF Annular Chamber for
Tracking and Identification of Charged particles (TACTIC)~\cite{Laird2007} is a
cylindrical ionization (time projection) chamber with segmented anode strips for
the energy loss and total energy determination of light charged particles
emerging from reactions. Measurements of the drift time provide trajectory
reconstruction. Gas electron multipliers (GEM) will amplify the signals. Digital
readout and pulse shape analysis of the signals is
envisioned~\cite{Laird2007}. The chamber gas will serve as target similar to the
active target device MAYA~\cite{Demonchy2007}.

The \nuc{18}{F}$(p,\alpha)$\nuc{15}{O} reation has received much attention as it
is of importance for nova $\gamma$-ray astronomy~\cite{Gomez-Gomar1998}. The
$\gamma$-ray spectrum emitted from novae in the first few hours after expansion
is dominated by the positron annihilation from the $\beta^+$ decay of radioactive
nuclei. One of the main contributors is \nuc{18}{F} with a $\beta$-decay
half-life of 110~min. Therefore, its production and destruction -- via
\nuc{18}{F}$(p,\alpha)$\nuc{15}{O} -- is of great importance.  Direct
measurements with  \nuc{18}{F} radioactive beams have been performed at
LLN~\cite{Coszach1995,Graulich1997,Graulich2000,Sereville2009},
ANL~\cite{Rehm1996a} and ORNL~\cite{Bardayan2001,Bardayan2002,Chae2006}. Most of
the early measurements~\cite{Coszach1995,Graulich2000,Rehm1996a,Bardayan2001}
where performed at center-of-mass energies exceeding 550~keV, which dominate the
reaction rate at temperatures above 0.4~GK. Graulich {\it et al.} in
1997~\cite{Graulich1997} found evidence for a $3/2^-$ resonance at
$E_{cm}=330$~keV but with insufficient statistics to determine the resonance
strength. In 2002 Bardayan {\it et al.}~\cite{Bardayan2002} used a \nuc{18}{F} beam
provided by the Holifield facility at ORNL with an intensity of 2 $\times$
10$^6$ particles/s interacting  with a CH$_2$ target to measure the 
\nuc{18}{F}$(p,\alpha)$\nuc{15}{O} reaction. The $\alpha$-\nuc{15}{O}
coincidences were unambiguously identified with the SIDAR setup and the $3/2^-$
resonance strength at $E_{cm}=330$~keV could be determined for the first
time. This resonance dominates the reaction rate over a range of temperatures
important for ONeMg novae~\cite{Bardayan2002}. The region of lower
center-of-mass energies, which is relevant for typical nova temperatures
($\approx$ 0.25~GK), has remained unaccessible to
direct measurements. Recent work at LLN~\cite{Sereville2009} and
ORNL~\cite{Chae2006} constrains the interference of $3/2^+$ resonances located
just above the proton separation energy from new measurements at $E_{cm} > 400$~keV
and $E_{cm}=663-877$~keV, respectively, and from the wealth of previous data in
comparison to $\mathcal{R}$-matrix calculations.         

In explosive hydrogen burning up to a temperature of 0.2~GK, the burning with
carbon takes place through a series of reactions known as the hot CNO cycle. At
higher temperatures of about 0.4~GK, the \nuc{14}{O} waiting point can be
bypassed by a chain of reactions initiated by
\nuc{14}{O}$(\alpha,p)$\nuc{17}{F}. At still higher temperatures, the breakout
from the CNO cycle becomes possible. The reaction that dominates the leak rate
is \nuc{15}{O}$(\alpha,\gamma)$\nuc{19}{Ne}~\cite{Fisker2006}. However, in the
regime of high temperature, alternative breakout routes have been suggested, in
particular the \nuc{18}{Ne}$(\alpha,p)$\nuc{21}{Na}
reaction~\cite{Goerres1995}. While a direct measurement of the key rate
\nuc{15}{O}$(\alpha,\gamma)$\nuc{19}{Ne} has not been feasible so far, the
\nuc{18}{Ne}$(\alpha,p)$\nuc{21}{Na}~\cite{BradfieldSmith1999a,Groombridge2002}
and \nuc{14}{O}$(\alpha,p)$\nuc{17}{F}~\cite{Notani2004} reactions have been
probed directly in experiments induced by \nuc{18}{Ne} and \nuc{14}{O}
radioactive beams at LLN and CRIB-Tokyo, respectively. The experiments at
Louvain-la-Neuve covered the energy region of
$E_{cm}=2.04-3.01$~MeV~\cite{BradfieldSmith1999a} and
$E_{cm}=1.7-2.9$~MeV~\cite{Groombridge2002}. The measurements of the
\nuc{18}{Ne}$(\alpha,p)$\nuc{21}{Na} reaction rate  were performed with a 
scattering chamber that consisted of a gas target and a vacuum side. The
incoming \nuc{18}{Ne} beam interacted with the 500~mbar He gas target. Two
double-sided silicon strip detector telescopes in the gas volume for $\Delta
E-E$ measurements were used to detect the protons and to reconstruct their
trajectories. The normalization of the incoming \nuc{18}{Ne} beam rate was based on
the detection of elastically scattered \nuc{18}{Ne} off a gold foil in
surface-barrier
detectors on the vacuum side of the scattering
chamber~\cite{BradfieldSmith1999,BradfieldSmith1999a,Groombridge2002}.
Groombridge {\it et al.}~\cite{Groombridge2002} identified eight states in the
compound nucleus. Calculations of the enhanced stellar reaction rate using the
new resonances as input show a good agreement with theoretical predictions
by~\cite{Goerres1995}. The experimental reaction rate represents a lower limit
and causes a breakout from the CNO cycle via
\nuc{18}{Ne}$(\alpha,p)$\nuc{21}{Na} to be delayed by several hundred ms when
compared to calculations based on previous rates; further
measurements are required at lower energies to map the lower-lying resonances
above the $\alpha$ threshold~\cite{Groombridge2002}.  

At the
CRIB facility, a \nuc{14}{O} beam interacted with a He  gas target operated at a
temperature of 30~K, enhancing the density compared to room temperature by a
factor of 10~\cite{Notani2004}. The \nuc{14}{O}$(\alpha,p)$\nuc{17}{F} reaction products were
identified with the $\Delta E - E$  
methods in silicon detector telescopes. The reaction cross
section was measured with the thick-target technique over an energy range
of $E_{cm}=0.8-3.8$~MeV. This constituted the first direct measurement of the
$\nuc{14}{O}(\alpha,p)\nuc{17}{F}$ reaction. The measured cross section differs
from conclusions based on an indirect measurement and from a direct measurement
of the time-reversed reactions as the contributions of the \nuc{17}{F} excited
states could not be taken into account~\cite{Notani2004}. The result seems to
suggest an increase of 50\% for the $\nuc{14}{O}(\alpha,p)\nuc{17}{F}$ reaction
rate which might impact the ignition phase of X-ray bursts~\cite{Notani2004}.           

\subsubsection{Resonance properties from elastic and inelastic scattering experiments}

Some reaction rates of astrophysical interest are totally or partially dominated 
by the contribution of resonances. Important experimental methods to study
the properties of resonance is resonant elastic and inelastic scattering. In
particular, proton resonance scattering measurements have provided an extensive
amount of data on unbound states in proton-rich nuclear systems relevant to
reaction rates in explosive burning scenarios.   

In inverse-kinematics resonant scattering, a rare-isotope beam bombards
a proton-rich target -- typically (CH$_2)_{\rm n}$, rarely cryogenic H$_2$
targets. The spectrum of the scattered protons depends sensitively on resonances
present in the compound nucleus. When a resonant state is scanned in the
appropriate energy range, the scattered-proton spectrum shows a distinct
structure that  allows to extract resonance properties -- energy, width, spin and
parity -- in comparison to $\mathcal{R}$-matrix theory. The energy of the scattered proton
is related to the resonant energy and the shape of the spectrum allows to
determine the width and the spin of the resonant state~\cite{Angulo2004}. The
protons are typically detected in position-sensitive Si $\Delta E - E$
telescopes or with annular Si strip detectors of the LEDA type. Important for
the extraction of resonance properties from the shape of the proton recoil
spectra in comparison to theory is a complete understanding of the experimental
resolutions that contribute to the shape of the detected proton
spectrum~\cite{Angulo2004}.       

Two different experimental
approaches are used, the thick-target and the thin-target technique. In the
thin-target approach, see for
example~\cite{Bardayan1999,Bardayan2000,Blackmon2003a}, the energy region of
interest is scanned by using beams at different energy. At ORNL, resonances in
the compound nucleus \nuc{19}{Ne} were studied with a \nuc{18}{F} beam at 15
different energies between 10 and 14~MeV~\cite{Bardayan2000} irradiating a
35~$\mu$g/cm$^2$ CH$_2$ target. The scattered protons were detected in SIDAR and
\nuc{18}{F} passing through the target was identified in an ionization
chamber~\cite{Bardayan2000}. In 
the thick-target approach, the beam is slowed down and stopped in the target and
the elastic proton scattering can be performed over a range of energies that
depends on the choice of target thickness and beam energy. This technique relies
on the fact that the protons escape from the target as their energy loss is 
negligible compared to that of the 
heavy ion. This technique was used for example by Angulo {\it et al.} at
Louvain-la-Neuve~\cite{Angulo2003a}. A \nuc{18}{Ne} beam at 28~MeV impinged upon a
520~$\mu$g/cm$^2$ CH$_2$ target. The recoil protons were detected at 20 different
angles with LEDA and the first excited state of
\nuc{19}{Na} could be characterized~\cite{Angulo2003a}.  

Inverse-kinematics resonant proton scattering measurements have been performed,
for example, with the thick-target technique at the TUDA facility at
TRIUMF/ISAC, $\nuc{21}{Na} +p$~\cite{Ruiz2002,Ruiz2005} and
$\nuc{20}{Na}+p$~\cite{Murphy2006}; with the thin and thick-target approach at
HRIBF/ORNL, $\nuc{17}{F}+p$~\cite{Bardayan1999,Blackmon2003a} and
$\nuc{18}{F}+p$~\cite{Bardayan2000,Bardayan2004}; at Louvain-la-Neuve with the
thick-target technique, $\nuc{13}{N,C}+p$~\cite{Delbar1992},
$\nuc{19}{Ne}+p$~\cite{Coszach1994},
$\nuc{18}{F}+p$~\cite{Graulich2000},$\nuc{18}{Ne}+p$~\cite{Angulo2003a},
and $\nuc{7}{Be}+p$~\cite{Angulo2003}; at
CRIB in Tokyo with the thick-target technique,
$\nuc{11}{C}+p$~\cite{Teranishi2003}, $\nuc{22}{Mg}+p$~\cite{He2007}
$\nuc{13}{N}+p$~\cite{Teranishi2007,Wang2008}, and
$\nuc{7}{Be}+p$~\cite{Yamaguchi2009}; at the BEARS facility in Berkeley,
$\nuc{14}{O} +p$~\cite{Guo2005}, $\nuc{11}{C}+p$~\cite{Perajarvi2006} and
$\nuc{15}{O}+p$~\cite{Lee2007}; as well as at Texas
A\&M~\cite{Tabacaru2006,Perajarvi2006}, Notre
Dame~\cite{Rogachev2001,Rogachev2007}, ANL~\cite{Harss2002} and
Spiral/GANIL~\cite{deSantos2004}.             

Inelastic resonance proton scattering has been employed to
probe properties of states in the compound nucleus in the cases where particle
decay to excited states is energetically possible. Recently, proton emitting
states in \nuc{19}{Ne} were studied with the inelastic scattering reaction
H$(\nuc{19}{Ne},p)\nuc{19}{Ne}^*(p)\nuc{18}{F}$ at Louvain-la-Neuve. Resonance
energies and widths were assigned from the shape of the scattered proton
spectrum, while spins were assigned using proton-proton angular correlations between
recoil and decay protons~\cite{Dalouzy2009}. 

Inelastic scattering gains importance when rates are derived from
measurements of the inverse reaction. For example, when the reaction
$A(p,\alpha)B$ is 
measured, $A$ is typically in its ground state and the rate of 
$B(\alpha,p)A_{gs}$ can be deduced from the detailed 
balance. However, contributions from $B(\alpha,p)A^*$ leading to an excited
state in $A$ are not accessible from the inverse reaction with $A$ in the ground
state. Inelastic scattering $A+p \rightarrow A^*$ can then reveal important
excitations that may contribute. One example is the
$\nuc{14}{O}(\alpha,p)\nuc{17}{F}$ reaction, which was studied from
measurements of the inverse reaction
$\nuc{17}{F}(p,\alpha)\nuc{14}{O}$~\cite{Harss1999,Blackmon2001}. Proton
inelastic scattering $\nuc{17}{F} +p$ were
measured at ANL~\cite{Harss2002} and ORNL~\cite{Blackmon2003a} to characterize
resonances of importance for the $\nuc{14}{O}(\alpha,p)\nuc{17}{F}$ reaction
rate. For the measurement of inelastic scattering, thin targets are typically
used to separate elastic and inelastic channels from the spectroscopy of the
protons~\cite{Blackmon2003a} or $\gamma$-ray spectroscopy is used to tag bound
final states, see for example~\cite{Yamaguchi2009}, where NaI detectors were
used to measure $\nuc{7}{Be}(p,p')\nuc{7}{Be}^*$.   

\subsubsection{Nucleon transfer reactions, nucleon knockout and population of
excited states in fragmentation and projectile fission}\label{sec:ANCexp}

For stable nuclei, light-ion induced transfer reactions in normal kinematics
have been widely used to explore stellar reaction rates by extracting
spectroscopic information on resonances close to the threshold that dominate the
reactions of interest. The parameters of a resonance -- energy,
orbital angular momentum, partial and total width, spectroscopic factors and
decay modes, for example -- can be determined. In general, the sensitivity of
transfer reactions to the single-particle degree of freedom continues to provide
important data to benchmark nuclear structure models, in particular  the nuclear
shell model which provides important input to the modeling of many astrophysical
processes (see Section~\ref{sec:trf_th} for details on the theoretical treatment
of transfer reactions). 
 
Neutron capture cross sections are important in the r process, in which heavier
nuclei are formed from seed elements by consecutive neutron capture reactions
and $\beta$ decays. In an environment of high neutron density, tens of neutron
captures may occur until the $\beta$ decay half-lives become shorter than the
half-life against neutron capture. Consequently, the process proceeds off the
valley of stability towards neutron-rich nuclei. For most neutron capture cross
sections, statistical models can be applied. However, near closed shells -- in a
regime of low level density -- direct capture becomes
important~\cite{Rauscher1998}. For direct capture calculations, level properties
(excitation energies, spins and parities) have to be known accurately unlike for
statistical calculations where averages over resonances are considered.       

Following pioneering inverse-kinematics $(d,p)$ one-neutron transfer experiments
induced by stable Xe beams impinging upon deuterated Ti
targets performed at GSI~\cite{Kraus1991}, a program of low-energy $(d,p)$
transfer experiments in mass regions relevant to the r process was started at
HRIBF at ORNL. The experimental study of $d(\nuc{124}{Sn},p)$ in inverse
kinematics at energies close to the Coulomb barrier proved that $Q$-value spectra
and angular distributions can be extracted for low-energy inverse-kinematics
transfer reactions with heavy beams~\cite{Jones2004}. This experiment used a
deuterated polyethylene target (CD$_2$) of 100~$\mu $g/cm$^2$ thickness that was
angled to achieve an effective thickness of 200~$\mu $g/cm$^2$ and to enable
proton detection under $\theta_{lab}=90^\circ$. Two silicon telescopes were
positioned covering angles of $\theta_{lab}=70-102^\circ$ and
$\theta_{lab}=85-110^\circ$, respectively. The silicon detector array SIDAR  was
mounted in half-lampshade configuration covering backward angles
of $\theta_{lab}=130-160^\circ$. The states were determined from the $Q$-value
spectrum and the angular distributions were used to determine the
orbital angular momentum ($\ell$ value) of the transferred neutron. An absolute
normalization of the cross sections was performed using the elastically
scattered deuterons~\cite{Jones2004}. The results were found in agreement with
normal-kinematics \nuc{124}{Sn}$(d,p)$ and demonstrated that this important mass
region for the r process can be accessed with low-energy, inverse-kinematics
transfer reactions. 

The first $(d,p)$ transfer on a neutron-rich r process nucleus,
\nuc{2}{H}(\nuc{82}{Ge},$p$)\nuc{83}{Ge} at 330~MeV on a 430~$\mu$g/cm$^2$
(CD$_2)_{\rm n}$ target, was subsequently performed at
HRIBF~\cite{Thomas2005}. Protons were detected with SIDAR, subtending   
$\theta_{lab}=105-150^\circ$. A segmented ion chamber was employed to determine
the $Z$ of the projectile-like reaction residues. The measured reaction
$Q$-value provided an indirect measurement of the mass of \nuc{83}{Ge}. The
extracted neutron separation energy $S_n=3.69$~MeV was found low suggesting that
the \nuc{82}{Ge}$(n,\gamma)$\nuc{83}{Ge} reaction rate has a significant direct
neutron capture component. The excitation energy of the $1/2^+$ excited state
followed the falling trend as semi-magic \nuc{79}{Ni} is
approached~\cite{Thomas2005}. \nuc{84}{Se} was the main contaminant in the beam
and the $(d,p)$ neutron transfer to \nuc{85}{Se} was studied at the same
time~\cite{Thomas2007}. Direct-semidirect contributions to the neutron capture
cross sections were computed~\cite{Thomas2007}. Recently, neutron capture
extracted from $(d,p)$ transfer reactions was benchmarked for
$\nuc{48}{Ca}(d,p)\nuc{49}{Ca}$ at different deuteron
energies~\cite{Mukhamedzhanov2008a}.    
           
One of the key reactions in nova $\gamma$-ray astronomy,
\nuc{18}{F}$(p,\alpha)$\nuc{15}{O}, has been explored indirectly by exploiting
\nuc{2}{H}(\nuc{18}{F},p)\nuc{19}{F} neutron transfer reactions and
the \nuc{19}{Ne}--\nuc{19}{F} mirror
symmetry~\cite{Sereville2003,Kozub2005,Kozub2006}. Spectroscopic information on
excited states just above the proton separation energy in the compound nucleus
\nuc{19}{Ne} is needed for the determination of the
\nuc{18}{F}$(p,\alpha)$\nuc{15}{O} reaction rate at nova temperatures. Some of the
levels in $\nuc{18}{F}+p$ cannot be reached in resonance scattering experiments since the
resonances are well below the Coulomb barrier. With the availability of
a \nuc{18}{F} rare-isotope beam, neutron spectroscopic
factors of the analog levels in the mirror nucleus of \nuc{19}{Ne}, \nuc{19}{F}, can be studied
with $(d,p)$ reactions in the corresponding energy region and -- invoking mirror
symmetry -- serve as input to calculate the proton width in \nuc{19}{Ne}. At
Louvain-la-Neuve, a 14~MeV \nuc{18}{F} beam impinging upon a 100~$\mu$g/cm$^2$
CD$_2$ target was used to induce the $(d,p)$ neutron transfer
reaction~\cite{Sereville2003}. The relevant levels are above the $\alpha$
separation energy leading to $\nuc{19}{F}^* \rightarrow \nuc{15}{N}
+\alpha$. The measurement employed the LEDA detector for the detection of
\nuc{15}{N} (downstream) and the LAMP detector (backward) for proton
detection. Spectroscopic factors deduced in comparison to DWBA calculations were
used to put new limits to the contribution of low-energy resonances to the
\nuc{18}{F}$(p,\alpha)$\nuc{15}{O} reaction rate~\cite{Sereville2003}. A similar
experiment was performed at HRIBF at ORNL at a much higher energy with a
108.49~MeV beam of \nuc{18}{F} on a 160(10)~$\mu$g/cm$^2$ CD$_2$
target~\cite{Kozub2005,Kozub2006}. The protons emerging from the
$\nuc{2}{H}(\nuc{18}{F},p)\nuc{19}{F}$ transfer reaction were detected with the
silicon strip detector array SIDAR at $\theta_{lab}=118-157^{\circ}$. Another
silicon strip detector at the focal plane of the Daresbury Recoil Separator was
used to tag particle-stable $A=19$ reaction residues and 
$\alpha$-decaying final states in coincidence with the protons
detected in SIDAR. Neutron spectroscopic factors for eight~\cite{Kozub2005} and
13~\cite{Kozub2006} analog levels of astrophysical relevance in
the mirror \nuc{19}{Ne} were determined. The results implied significantly reduced
\nuc{18}{F}$(p,\gamma)$\nuc{19}{Ne} and \nuc{18}{F}$(p,\alpha)$\nuc{15}{O} rates
compared to what was reported previously~\cite{Kozub2005}. 

The $(d,n)$ transfer reaction has been
used at the RESOLUT facility at Florida State University as surrogate for proton capture. The
reaction $\nuc{25}{Al}(d,n)\nuc{26}{Si}^*\rightarrow p+\nuc{25}{Al}$ was induced
by a 91.5~MeV beam of \nuc{25}{Al} impinging upon a 1.66~mg/cm$^2$ thick CD$_2$
target~\cite{Peplowski2009}. The decay protons were 
detected under forward angles with the $\Delta E - E$ method using a hybrid ion
chamber and a double-sided silicon strip detector backed by a second,
unsegmented silicon detector. The lowest $\ell=0$ proton resonance in
\nuc{26}{Si} could be identified. This shifts the main experimental uncertainty
in the synthesis of \nuc{26}{Al} to the destruction of \nuc{26m}{Al} through the
$(p,\gamma)$ reaction, which is planned to be studied at RESOLUT in the
future~\cite{Peplowski2009}.

The use of heavy-ion induced, peripheral, proton-adding transfer reactions of type
(\nuc{14}{N},\nuc{13}{C}) \cite{Azhari2001,Tang2003,Tang2004}, 
(\nuc{10}{B},\nuc{9}{Be})~\cite{Mukhamedzhanov1997,Azhari1999,Azhari2001} and
(\nuc{13}{C},\nuc{12}{C})~\cite{Trache2003} in inverse kinematics as well as
$(\nuc{3}{He},d)$ in normal
kinematics~\cite{Gagliardi1999,Mukhamedzhanov2003,Mukhamedzhanov2006} 
has been pioneered at Texas A \& M  to indirectly assess the non-resonant
direct-capture contribution in $(p,\gamma)$ reactions. Asymptotic normalization
coefficients (ANCs)~\cite{Xu1994,Gagliardi1999,Azhari2001} are deduced from the
measured angular distributions in transfer reactions at energies above the
Coulomb barrier where the cross sections are orders of magnitude higher than for
direct capture processes at astrophysical energies (see Section
\ref{sec:ANC_th}). Radioactive beams of 
\nuc{7}{Be} (84~MeV), \nuc{11}{C} (10~MeV/u), and \nuc{13}{N} (11.8~MeV/u) were
produced at Texas A \& M and separated with the MARS spectrometer.
$\nuc{10}{B}(\nuc{7}{Be},\nuc{8}{B})\nuc{9}{Be}$,
\nuc{14}{N}(\nuc{7}{Be},\nuc{8}{B})\nuc{13}{C}, 
\nuc{14}{N}(\nuc{13}{N},\nuc{14}{O})\nuc{13}{C},  
and \nuc{14}{N}(\nuc{11}{C},\nuc{12}{N})\nuc{13}{C} transfer reactions
induced by \nuc{10}{B} and melamine $(C_3 N_6 H_6)$ targets, respectively,
were used to determine the ANCs for $\nuc{8}{B} \rightarrow \nuc{7}{Be} + p$,
$\nuc{14}{O} \rightarrow \nuc{13}{N} + p$, and $\nuc{12}{N} \rightarrow
\nuc{11}{C} + p$ to deduce the non-resonant capture rates for
the $\nuc{7}{Be}(p,\gamma)\nuc{8}{B}$, $\nuc{13}{N}(p,\gamma)\nuc{14}{O}$ and
$\nuc{11}{C}(p,\gamma)\nuc{12}{N}$ reactions. In these experiments, the
elastically 
scattered projectile beam and transfer products are observed simultaneously in
two detector telescopes consisting of position-sensitive 16-strip Si $\Delta
E$ detectors backed by Si $E$ detectors. Particle identification is then done
via the $\Delta E - E$ method. The reaction telescopes were cooled down to
-6$^\circ$C to reduce thermal noise. A slightly different approach was taken at
HRIBF at ORNL to study the $\nuc{17}{F}(p,\gamma)\nuc{18}{Ne}$ direct capture
cross section~\cite{Blackmon2003}.  A beam of 170 MeV \nuc{17}{F} impinged upon
a melamine target inducing the \nuc{14}{N}(\nuc{17}{F},\nuc{18}{Ne})\nuc{13}{C}
reaction. Recoiling \nuc{18}{Ne} nuclei were detected via energy loss in
position-sensitive Si strip detectors covering center-of-mass angles from 2 to
$9^\circ$. To be able to distinguish the transfer to individual excited states,
$\gamma$ rays emitted by \nuc{18}{Ne} were detected in CLARION in coincidence
and allowed to tag the final state. Preliminary results are reported~\cite{Blackmon2003}.  
 
The ANC technique was
also applied to a sub-Coulomb $\alpha$-transfer reaction
$\nuc{13}{C}(\nuc{6}{Li},d)$ performed at FSU with a \nuc{13}{C} beam at 60~MeV to
study the astrophysical reaction rate of
$\nuc{13}{C}(\alpha,n)\nuc{16}{O}$ which -- at low energies -- is dominated by
a $1/2^+$ sub-threshold resonance in \nuc{17}{O}~\cite{Johnson2006}. The application of the
ANC method also to 
one-nucleon removal and breakup reactions induced by weakly-bound projectile
nuclei and light as well as heavy target nuclei was
proposed~\cite{Trache2001,Trache2004}.

\begin{figure}[t]
\begin{center}
{\includegraphics[width=8.5cm]{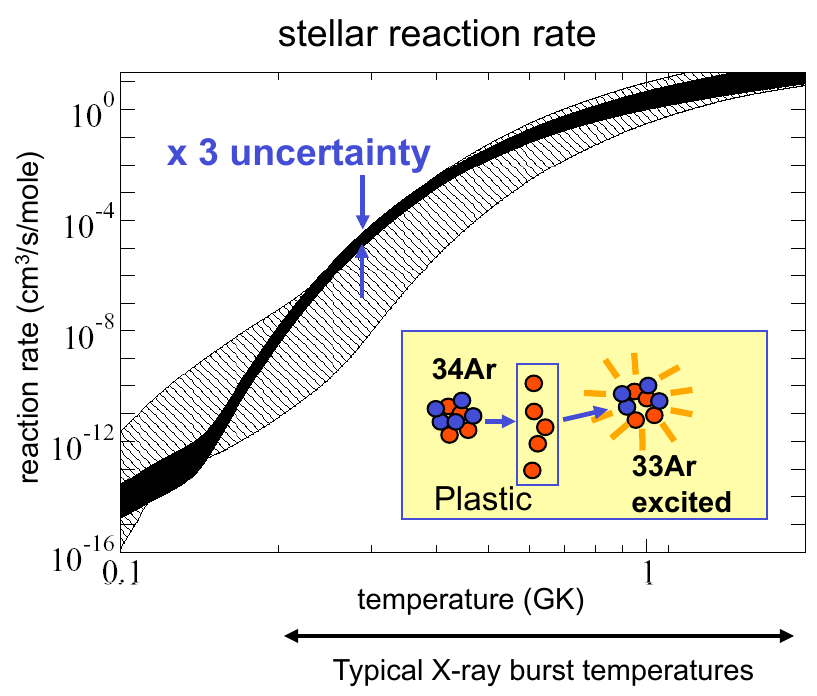}}
\end{center}
\caption{\label{Ar33}
 Total astrophysical $\nuc{32}{Cl}(p,\gamma)\nuc{33}{Ar}$ ground-state capture
rate as function of T$_9$~\cite{Schatz2005}. The black region shows the capture rate restricted by
the one-neutron knockout measurement of Clement {\it et al.}~\cite{Clement2004}
relative to the previously possible range of rates (gray band). The total
estimated uncertainty of the rate ranges from a factor of 3 to a factor of
6. (Figure adapted from Hendrik Schatz and Ref. \cite{Schatz2005}).   
}
\end{figure}

A different experimental approach that utilizes the high luminosity of fast-beam
experiments at fragmentation facilities was developed at NSCL. The one-neutron
removal from a projectile beam of \nuc{34}{Ar} upon collision with a CH$_2$
target at 
84~MeV/nucleon was used to populate excited states in
\nuc{33}{Ar}~\cite{Clement2004}. In-beam $\gamma$-ray spectroscopy using the
segmented germanium array SeGA~\cite{Mueller2001} provided precision information
on the excitation energy of states in the proximity of the proton separation
energy and for states above $S_p$ that have a $\gamma$ branch. High-resolution
$\gamma$-ray 
spectroscopy allowed to determine the excitation energy with uncertainties of
several keV. The 2-orders of magnitude improvement in the uncertainty of the
level energy translated into an 3-orders of magnitude improvement for the rate
of $\nuc{32}{Cl}(p,\gamma)\nuc{33}{Ar}$~\cite{Clement2004} (see
Figure~\ref{Ar33}). Similar measurements
have been performed to precisely determine relevant level energies in
\nuc{30}{S}~\cite{Galaviz2008}. Pioneered at NSCL, \nuc{9}{Be}-induced one-nucleon removal
reactions from fast exotic beams have proven to be a crucial tool to study the
detailed evolution of nuclear shell structure in exotic
nuclei~\cite{han03,Gade2008}. Their reach to the most exotic nuclei and the
sensitivity to the single-particle degree of freedom continue to probe the
predictability achieved by large-scale shell model calculations, which provide
crucial input for many astrophysical processes (see also Section \ref{sec:knock_th}). 

A less selective method is the population of isomeric excited states in
projectile fragmentation or fission and the spectroscopy of the level structure
fed by the long-lived state(s). At GSI, $\gamma$-ray decays of excited states in the
$N=82$ r-process waiting-point nucleus \nuc{130}{Cd} were observed unambiguously
for the first 
time~\cite{Jungclaus2007}. An $8^+$ isomeric state in \nuc{130}{Cd} with a
two-quasiparticle structure was populated in the fragmentation of a\nuc{136}{Xe}
beam and in the fission of \nuc{238}{U}. In the two parts of the
experiment, heavy, neutron-rich beams were produced by fragmentation of a 
\nuc{136}{Xe} beam at 750~MeV/nucleon impinging upon a 4~g/cm$^2$ \nuc{9}{Be} target and by
a 650~MeV/nucleon \nuc{238}{U} beam fissioning upon collision with a 1~g/cm$^2$ thick
\nuc{9}{Be} 
target. The produced \nuc{130}{Cd} nuclei were separated with the GSI fragment
separator (FRS) and identified via the measured energy loss, time of flight, magnetic
rigidity, and various position measurements. The nuclei were implanted into a
passive stopper surrounded by 15 large-volume Ge cluster
detectors~\cite{Eberth1996}.  The new $2^+_1$ energy of 1395~keV for
\nuc{130}{Cd} is not consistent with the previous tentative assignment of the
$2^+$ state at 957~keV~\cite{Kautzsch2000}. The evidence for the existence of an
$8^+$ isomer in \nuc{130}{Cd} is in line with expectations based on the valence
analog \nuc{98}{Cd}. These new results on the level scheme of \nuc{130}{Cd}
provided no evidence for a reduction of  the $N=82$ shell gap in the vicinity of
\nuc{132}{Sn} contrary to what was implied by the results of a previous
experiment~\cite{Dillmann2003}. In a similar measurement at GSI, excited states
in the magic nucleus \nuc{204}{Pt} (neutron number $N=126$) were populated in
the fragmentation of \nuc{208}{Pb} at 1~GeV/nucleon beam
energy~\cite{Steer2008}. Medium spin isomers with half-lives between 150~ns and
55~$\mu$s and the level structures fed by their decays were identified. The data
suggested a revision of the two-body interactions for $N=126$ and $Z < 82$,
which is important for the evolution of nuclear structure toward the waiting
points of the r process in this region~\cite{Steer2008}.        

\subsubsection{Weak-interaction strength}\label{sec:weak}

Supernovae are a major source of nucleosynthesis. For both types of supernovae
-- core-collapse and thermonuclear -- weak interaction rates play are crucial
role~\cite{Fuller1980,Fuller1982,Langanke2003}. In particular electron capture
rates for many nuclei in the $fp$-shell and beyond are among the important
nuclear physics ingredients needed for supernova models~\cite{Hix2003}. The
weak-interaction rates are largely provided by nuclear structure calculations,
as for example by large-scale shell-model 
calculations~\cite{Caurier1999,Langanke2000}. Important experimental benchmarks
necessary for the reliable modeling of weak-interaction rates by nuclear theory
are provided by measured Gamow-Teller (GT) strength distributions (see Section
\ref{sec:cex_th} for details on the theoretical formalism). 

GT transition strength
($B(GT)$ values) can be obtained from Gamow-Teller $\beta$ decays within
excitation-energy limitations given by decay $Q$-value window. Charge-exchange 
reactions, however, can map the GT strength distributions over a wider range of
excitation energies. A variety of charge-changing reactions can be employed, for
example, 
$\nuc{A}{Z}(p,n)\nuc{A}{(Z+1)}$~\cite{Ikeda1963,Taddeucci1987},
$\nuc{A}{Z}(n,p)\nuc{A}{(Z-1)}$~\cite{Jackson1988}, 
$\nuc{A}{Z}(d,\nuc{2}{He})\nuc{A}{(Z-1)}$~\cite{Frekers2004},
$\nuc{A}{Z}(\nuc{3}{He},t)\nuc{A}{(Z+1)}$~\cite{Fujiwara2000,Fujita2002,Zegers2008,Zegers2008a},
and 
$\nuc{A}{Z}(t,\nuc{3}{He})\nuc{A}{(Z-1)}$ \cite{Galonsky1978,Zegers2006,Howard2008,Hitt2009}. In
charge-exchange reactions at beam 
energies exceeding 100~MeV/nucleon, the cross section at low momentum transfer
(small angles) is proportional to the $B(GT)$ transition
strength~\cite{Goodman1980,Osterfeld1992,Alford1998,Taddeucci1987,Zegers2007}.
GT transitions have been probed extensively with the charge-exchange reactions 
induced by stable projectiles on stable targets (see references
above). Charge-exchange reactions on short-lived exotic nuclei in inverse 
kinematics are proposed to be developed into spectroscopic tools at
FAIR (Germany)~\cite{Aumann2007} and at NSCL (US)~\cite{Gelbke2009}. At FAIR,
the utilization of charge-exchange reactions is proposed by the EXL~\cite{EXL}
and by the R$^3$B~\cite{R3B} collaborations. At R$^3$B, $(p,n)$ charge-exchange
reactions will be induced by rare-isotope beams at relativistic energies impinging upon a liquid
hydrogen target with the emerging slow neutrons measured by plastic scintillators
surrounding the target~\cite{R3B}. The EXL collaboration envisions the study of
charge exchange via $(p,n)$,
$(\nuc{3}{He},t)$, and $(d,\nuc{2}{He})$ reactions with unmatched luminosity
induced by rare isotopes circulating in a storage ring passing through a gas jet
target~\cite{EXL}. Two avenues are taken at NSCL, the first program utilizes the 
$\nuc{A}{Z}(\nuc{7}{Li},\nuc{7}{Be})\nuc{A}{(Z-1)}$ induced charge-exchange
reaction where the spin transfer is tagged with the 430 keV $\gamma$-ray decay
$(1/2^- \rightarrow 3/2^-)$ in the target-like reaction residue \nuc{7}{Be}~\cite{LiBe}. For
a second program, the Low-Energy Neutron Detector Array (LENDA)~\cite{LENDA} is presently
constructed for slow-neutron detection in $(p,n)$ charge-exchange reactions
induced by rare-isotope beams bombarding a hydrogen target. In all NSCL
experiments, the projectile-like fragments will be tracked and identified with the S800
spectrograph~\cite{Gelbke2009}.

\subsubsection{Beta-decay half-lives and $\beta$-delayed neutron emission probabilities}

Isotopes of elements beyond iron are almost exclusively produced in
neutron-capture processes. The two main neutron-capture processes for
astrophysical nucleosynthesis are the slow s process and the rapid r process
(see~\cite{Wallerstein1997} for a review). The attributes slow and rapid refer
to the timescale between  
subsequent neutron captures relative to the competing time scale for $\beta$
decay. While the s process stays close to the valley of stability, the r process
leads to very neutron-rich nuclei. To determine the path of the r process, among
others, $\beta$-decay properties are crucial. Beta decay
competes with the neutron capture and the drives the path toward stable nuclei
and fission determines the heaviest nuclei produced. On the proton-rich side of
the nuclear chart, $\beta$-decay half-lives are an important input for
nucleosynthesis models that describe the reaction flow in the rp process and the
final abundances. While much data has been available on the proton-rich side of
the nuclear chart, some information in key regions, as for example around the
nuclei \nuc{92,94}{Mo} and \nuc{96,98}{Ru}, which are found in the solar system in
unexpectedly large abundances, could only be gathered recently with technical 
improvements crucial for experiments on proton-rich nuclei.   

Two approaches have been taken to measure the $\beta$-decay half-lives and
$\beta$-delayed neutron emission probabilities. At the
Jyv\"askyl\"a (IGISOL) and the CERN/ISOLDE facilities, neutron-rich isotopes of
interest are produced by fission, mass-separated and implanted into a collection
tape, which moves at pre-set times to suppress the $\beta$ and $\gamma$ activity
from longer-lived isobaric contaminants. Time spectra are recorded for
$\gamma$-ray, neutron and $\beta$ events where the half-lives and production
rates can be derived from the growth and decay curves of the collected
activities during pulsed beam mode. At fragmentation facilities, tracking and
event-by-event particle identification become possible in the regime of fast
beams~\cite{Mantica2005}.  The fast-moving beam particles are, for example,
implanted into a 
double-sided silicon strip detector (DSSD) with subsequent $\beta$ decays
detected on an event-by-event basis and correlated with the position of a
specific implantation. All events are typically time-stamped and the decay curve
is constructed from the time difference between implant and correlated
decay. This method is applicable at rates of less than 0.1 particles per second.

In early experiments at Jyv\"askyl\"a,
the $\beta$-decay half-lives and $\beta$-delayed neutron emission branching
ratios of neutron-rich Y, Nb, and Tc isotopes
produced in the proton-induced fission of \nuc{238}{U} were
determined~\cite{Mehren1996,Wang1999}.  The mass-separated 
nuclei of interest were implanted onto a collection tape and their decays
were measured inside the Mainz $4\pi$ neutron long counter used for the detection of $\beta$-delayed
neutrons. The $\beta$-particle detection and $\gamma$-ray detection was performed
with a thin plastic scintillator and a planar germanium detector,
respectively, positioned inside the Mainz counter. The neutron long counter
consisted of 42 \nuc{3}{He} ionization 
chamber tubes, arranged in two concentric rings embedded in a polyethylene
matrix surrounding the implantation point~\cite{Mehren1996,Wang1999}. At
CERN/ISOLDE, the $\beta$-decay properties of \nuc{130,131,132}{Cd} were
studied~\cite{Hannawald2000,Dillmann2003}. In 2000, the heavier
\nuc{131,132}{Cd} isotopes were produced from proton-induced uranium fission and
purified with resonance-ionization laser ion source (RILIS) and subsequent
separation of the extracted nuclei with the General Purpose Separator
(GPS). Beta-decay half-lives and $\beta$-delayed neutron decay branching
ratios $P_n$ were determined using the $\beta n$ spectroscopy with the Mainz
neutron long counter and a $\Delta E$ $\beta$ plastic
scintillator~\cite{Hannawald2000}. Superior
separation was achieved for the study of \nuc{130}{Cd} and its decay daughter
\nuc{130}{In}, in the vicinity of the $Z=50$, $N=82$ r-process waiting
point. \nuc{130}{Cd} was produced by fission induced with fast reaction neutrons
-- generated from a 
1~GeV proton beam impinging upon a Ta or W rod -- interacting with the uranium
carbide/graphite production target. Highest chemical selectivity was achieved
with RILIS and separation using the High-Resolution Separator
(HRS). $\beta\gamma$ and $\beta\gamma\gamma$ spectroscopy around the
moving-tape-collector was performed with $\Delta E -E$ $\beta$ telescope and
four large-volume high-purity germanium detectors in close
geometry~\cite{Dillmann2003}.

On the proton-rich side of the nuclear chart and important for ONe novae,
\nuc{23}{Al} was studied at Texas A$\&$M~\cite{Iacob2006}. \nuc{23}{Al} was
produced via the $p(\nuc{24}{Mg},\nuc{23}{Al})2n$ reaction, separated by the
MARS separator and collected with a moving-tape system to a 
detector setup for $\beta$ and $\beta \gamma$ spectroscopy consisting of a thick
plastic scintillator for $\beta$ detection and a high-purity germanium detector
for $\gamma$-ray detection~\cite{Iacob2006}. Beta branching ratios and log
$ft$ values for transitions to levels in \nuc{23}{Mg} were extracted. From this,
the ground-state spin and parity $5/2^+$ of \nuc{23}{Al} was determined
unambiguously. This excludes the large increases in the radiative proton
capture cross section for the reaction $\nuc{22}{Mg}(p,\gamma)$ at astrophysical
energies, which were implied by earlier claims that the spin and parity of the
\nuc{23}{Al} ground state are $1/2^+$~\cite{Iacob2006}.  

Rare-isotope beams produced by projectile fragmentation are used at NSCL
to measure the $\beta$-decay properties of exotic nuclei of importance to
nuclear astrophysics. In-flight tracking and event-by-event particle
identification become possible in the regime of fast beams. The fast-moving
beam particles are implanted into a double-sided silicon strip detector (DSSD)
with 40 horizontal and vertical strips resulting in a pixelation of
1600, the center piece of NSCL's Beta Counting Station
(BCS)~\cite{Prisciandaro2003}. Subsequent $\beta$ decays are detected on an
event-by-event basis and correlated with the position of a specific
implantation. Downstream of the DSSD, six single-sided Si strip detectors
(SSSDs) and two Si PIN detectors comprise the $\beta$ calorimeter part of the
BCS. All events in the BCS carry an absolute time stamp and the time
difference between implant and decay serves to map the $\beta$ decay
curve. Coincident $\gamma$-ray spectroscopy and $\beta$-delayed neutron
spectroscopy can be performed with the Segmented Germanium Array
(SeGA)~\cite{Mueller2001} and the Neutron Emission
Ratio Observer (NERO)~\cite{Lorusso2008} -- a neutron long counter with
\nuc{3}{He} and BF$_3$ gas counters. As the $\gamma$-rays are emitted by 
nuclei at rest, the segmentation of SeGA is not used. NERO consists of 60
proportional gas-counter tubes -- 16 filled with 
\nuc{3}{He} and 44 filled with B$_3$F -- embedded in a polyethylene
moderator. The counters are arranged parallel to the beam direction in three
concentric rings around the vacuum beam line that accommodates the BCS. A
DSSD-based $\beta$ counting system similar to the BCS is being
developed for experiments at GSI with the RISING setup~\cite{Kumar2009}. A
challenge in implant and decay setups is the large difference in the energy
response to fragments that are stopped and implanted in the DSSD (several GeV)
and the energy signals of charged-particle decays (hundreds of keV). The
$\beta$-decay detection system for RISING will use logarithmic preamplifiers to
cover the required dynamic range~\cite{Kumar2009} while NSCL's BCS is using
preamplifiers with 
dual gain capability, low gain for the fragment implantation and high gain for
the detection of decays~\cite{Prisciandaro2003}.     

The half-life of the doubly-magic nucleus \nuc{78}{Ni} is important input for
r-process model predictions of the nucleosynthesis around
$A=80$~\cite{Hosmer2005}. At NSCL, \nuc{78}{Ni} was produced via projectile
fragmentation from a beam of \nuc{86}{Kr} at 140~MeV/nucleon impinging upon a
beryllium target of 376~mg/cm$^2$ thickness. Each nucleus in the secondary beam
was individually identified in flight by measuring energy loss and time of
flight (TOF). The ions were continuously implanted into the DSSD. In the regime of
very low beam rate for this exotic nucleus, the typical total implantation rate for
the entire detector was under 0.1 per second. The DSSD registered the time and
position of the decays following the implantation. This allowed the correlation
of 7 implanted nuclei with subsequent decay events. The measured half-life
indicated a rather short time scale for the buildup of heavy elements beyond $N=50$
compared to some earlier predictions and thus unraveled an acceleration of 
the r process in this region~\cite{Hosmer2005}. 

For a similar experiment near the
$N=82$ neutron shell closure, neutron-rich Tc, Ru, Rh, Pd, and Ag isotopes were
produced at the NSCL by fragmentation of a 121.8 MeV/nucleon \nuc{136}{Xe} primary beam
impinging onto a 206 mg/cm$^2$ Be target~\cite{Montes2006}. The secondary beam
was then transmitted to the experimental vault, where a second plastic
scintillator provided the stop signal for the TOF measurement (relative to a
timing signal from a scintillator located at the intermediate image of the A1900
fragment separator) before the beam was
implanted into the NSCL's BCS for the study of the subsequent $\beta$ decay. The
average implantation rate in the DSSD was 0.4 Hz. Implantation and decay 
events were time stamped and correlated via their pixel locations. Coincident
$\beta$-delayed neutrons were detected with NERO. $\beta$-delayed neutron
emission branchings (or upper limits) for the neutron-rich nuclei
\nuc{116-120}{Rh}, \nuc{120-122}{Pd}, and \nuc{124}{Ag} were determined as well
as $\beta$-decay half-lives for neutron rich \nuc{114-115}{Tc},
\nuc{114-118}{Ru}, \nuc{116-121}{Rh}, and \nuc{119-124}{Pd} near the proposed
the r-process path~\cite{Montes2006}. A little lower in $Z$, the $\beta$-decay
properties of Y, Zr and Mo nuclei around $A=110$ were
measured~\cite{Pereira2009}. Fast beams of
neutron-rich Y, Zr, Nb, Mo, and Tc isotopes were produced at  
NSCL by fragmentation of a 120~MeV/nucleon \nuc{136}{Xe} beam on a thick Be
target. The species of interest were separated and identified with the A1900
fragment separator using the $B\rho -\Delta E - B\rho$ method and implanted in
the BCS which was surrounded by NERO. New
half-lives for \nuc{105}{Y}, \nuc{106,107}{Zr}, and \nuc{111}{Mo}, along with
new $P_n$ values for \nuc{104}{Y} and \nuc{109,110}{Mo} and $P_n$ upper limits
for \nuc{103-107}{Zr} and \nuc{108,111}{Mo} were determined. Analysis of the
measured $T_{1/2}$ and $P_n$ values in the framework of QRPA calculations
brought new insights in terms of deformation and shape coexistence, compatible
with the hypothesis of a quenched $N=82$ shell gap invoked by some to explain the
r-process abundances around $A=110$. An experimental complication at high $Z$ is
the presence of charge states in the secondary
beam~\cite{Pereira2009}. Measurements of the ion's total kinetic energy with the
PIN detectors and the DSSD of the BCS are used to separate the fully stripped
ions from charge states; isomer tagging with germanium detectors of the SeGA
array is typically performed to confirm the particle identification. 

The $N=Z$ nuclei \nuc{84}{Mo} and \nuc{100}{Sn} and nuclei in the vicinity of
$N=Z=50$, \nuc{96}{Cd} and \nuc{98}{In}, that lie along the rp-process path became
accessible for $\beta$-decay measurements at NSCL only after implementation of
the Radio-Frequency Fragment Separator (RFFS). With the energies available at NSCL,
secondary beams optimized on neutron-deficient species are of low purity due to
extended low-momentum tails of higher-rigidity fragments which overlap with the
fragment momentum distribution of interest. The RFFS applies a transverse RF
electric field which deflects nuclear species in the secondary beam based on
their phase difference with the primary beam. Secondary beams containing
\nuc{96}{Cd}, \nuc{98}{In}, \nuc{100}{Sn}~\cite{Bazin2008} and
\nuc{84}{Mo}~\cite{Stoker2009} were produced by
projectile fragmentation of \nuc{112}{Sn} and \nuc{124}{Xe} primary beams,
respectively, and purified by the RFFS. The BCS was used in both cases to
measure the $\beta$-decay half-lives. The new result for \nuc{84}{Mo} resolved a
previously reported deviation from theoretical predictions~\cite{Stoker2009} and
the measurements 
around \nuc{96}{Cd} revealed that the rp process in X-ray bursts is not the main
source of the unexpectedly large amount of \nuc{96}{Ru} in the solar
system~\cite{Bazin2008}.  

More technical details and applications to nuclear structure research can be
found in the recent review articles by Mantica~\cite{Mantica2005} and Rubio and
Gelletly~\cite{Rubio2009}.  

\subsubsection{Mass measurements}

Atomic masses of rare isotopes across the nuclear chart are among the key
input data for large-scale reaction network calculations that quantify the
nucleosynthesis, for example, via the rp process, $\nu$p process and the
r process (see for example~\cite{Schatz2006a}). Experimental methods for the
determination of atomic masses 
basically fall into two categories. Approaches that measure the $Q$ values in
decays or reactions make use of Einstein's mass-energy equivalence; mass
measurements that are based on the deflection of ions in electromagnetic fields
determine the mass-to-charge ratio. The highest precision in mass spectrometry
today is obtained by frequency measurements. The revolution or cyclotron
frequencies of ions in a magnetic field are measured to deduce the
mass-to-charge ratio in Penning trap experiments, in time-of-flight measurements
involving cyclotrons and in storage rings. The different experimental methods
are optimized for different beam energy regimes and are thus linked to the
production and separation schemes of the rare isotopes.

Penning traps use the three-dimensional confinement of ions with static magnetic
and electric fields. Mass measurements in a trap are based on the determination
of the cyclotron frequency of the stored ions in a magnetic field. The ion
motion in the combined magnetic and electric fields is the superposition of
three (ideally) independent harmonic motions with eigenfrequencies $\omega_z$
(axial motion), $\omega_{+}$ (modified cyclotron motion) and $\omega_{-}$
(magneton motion). Their frequencies are:
\begin{equation}
\omega_z=\sqrt{\frac{qV_0}{mr^2}},~~~~~~\omega_{\pm}=\frac{\omega_c}{2} \pm
\sqrt{\frac{\omega_c^2}{4}-\frac{\omega_z^2}{2}}, ~~~~~~\omega_c=\omega_{+} +
\omega_{-}   
\end{equation}         
with the cyclotron frequency $\omega_c=qB/m$ and the characteristic trap
dimension $r$. The sum $\omega_{-} + \omega_{+}=qB/m$ is directly proportional
to the charge-to-mass ratio. Inside the trap, the ions are excited by an
azimuthal quadrupolar RF field with frequencies around $\omega_c$. The ions are
released from the trap and their kinetic energy is determined from a
time-of-flight measurement. In case of a resonant excitation $\omega_c$, the
radial kinetic energy of the released ion will be maximal and its time of flight
minimal. Unknown masses are then determined relative to calibration measurements
of nuclei with precisely known mass. Relative mass uncertainties of order
$10^{-8}$ are routinely achieved. The Penning trap setups used for mass
spectrometry of exotic nuclei are ISOLTRAP~\cite{Bollen1996,Blaum2005} at
CERN/ISOLDE (Switzerland), CPT~\cite{Clark2003,Savard2006} at 
ANL (US), SHIPTRAP~\cite{Dilling2000,Block2005,Block2006} at GSI (Germany),
JYFLTRAP~\cite{Kolhinen2004,Jokinen2006} at Jyv\"askyl\"a (Finland),
LEBIT~\cite{Schwarz2003,Bollen2004,Ringle2006} 
at NSCL (US) and TITAN~\cite{Dilling2006} at TRIUMF (Canada).

At GSI, rare-isotopes are produced in-flight, selected in the Fragment Separator
(FRS) and subsequently injected into the Experimental Storage Ring
(ESR)~\cite{Franzke2008}. For two ion 
species circulating in the ring the relative difference in the mass-to-charge 
ratio $m/q$ of the revolving ion species is expressed to first order as: 
\begin{equation}
\frac{\Delta f}{f}=-\frac{1}{\gamma^2_t} \frac{\Delta
(m/q)}{m/q}+\left(1-\frac{\gamma^2}{\gamma_t^2}\right)\frac{\Delta \nu}{\nu},
\label{eq:ESR}
\end{equation}
where $f$, $m/q$,$v$ and $\gamma$ are the mean values of frequency,
mass-to-charge ratio, velocity and Lorentz factor, respectively, and $\Delta
f=f_1-f_2$, $\Delta (m/q)=(m_1/q_1)-(m_2/q_2)$ and $\Delta v=v_1 - v_2$ are the
corresponding differences for the two ion species. The transition point
$\gamma_t$ characterizes the point where the revolution frequency becomes
independent of the energy for each species with fixed $m/q$. This quantity can
be varied within certain limits by adjustments of the ring's ion optics. The ESR
can be operated in two different optics modes to deduce mass-to-charge
ratios. For longer-lived nuclei ($T_{1/2}$ of the order of seconds), the
velocity spread $\Delta v /v$ of the ions can be reduced by electron
cooling (Schottky method). Then the  term of eq.~(\ref{eq:ESR}) that depends on
the velocity 
spread becomes negligible and the $m/q$ can be directly determined from the
frequency of revolution. The frequency is determined via the
Schottky noise pickup technique. Many species can be stored in the ring
simultaneously, with ions of precisely known mass serving as
calibrants. Relative mass uncertainties are typically of order
$10^{-7}$~\cite{Radon1997,Franzke2008}. For 
short-lived nuclei, the ring can be operated in isochronus mode. Here, the
Lorentz factor is chosen to match $\gamma_t$, where the revolution time becomes
independent of the velocity and the mass-to-charge ratio can be
deduced. Typically, thin-foil timing detectors are used to measure the
time of flight. Since this approach does not require cooling, short-lived ions
($T_{1/2}$ of order 10~$\mu s$) can be studied~\cite{Hausmann2000,Franzke2008}.
Future measurements at the ESR will probably also address the masses of
neutron-rich nuclei, in particular those involved in the r process of stellar
nucleosynthesis, as these can be produced efficiently by uranium projectile
fission with subsequent separation in-flight by the FRS at GSI.

Another frequency-based mass measurement concept is implemented at ISOLDE with
the radiofrequency transmission mass spectrometer
MISTRAL~\cite{Coc1988,Lunney2006}, where in a homogeneous magnetic field a
controlled and coherent 
manipulation of the ion trajectory is performed to determine the mass via the
cyclotron frequency.   

Time-of-flight measurements in combination with magnetic spectrometers or
cyclotrons are used in various experimental schemes to determine the
mass-to-charge ratio of short-lived ions. An ion's motion in a magnetic field
can be characterized by the magnetic rigidity $B \rho$ which is connected to the
mass-to-charge ratio via the ion's velocity $v$ or cyclotron frequency
$\omega_c$:   
\begin{equation}
B\rho = \frac{\gamma m v}{q} =\frac{vB}{\omega_c},
\end{equation}
where $\gamma$ is the Lorentz factor. The magnetic rigidity can be measured with
high-resolution magnetic spectrometers, the velocity can be determined from the acceleration
potential at low energy~\cite{Barillari2003} (ISOLDE) or directly from
time-of-flight 
measurements at higher beam energies~\cite{Bianchi1989,Matos2008} at GANIL and
NSCL, for example. For a given
detector time resolution, the resolution of the time-of-flight measurement is
limited by the total flight time. An increased flight time can be realized with
the use of cyclotrons~\cite{Auger1994}. For two ions with masses $m$ and $m +
\delta m$, accelerated simultaneously in a cyclotron, the time difference
$\delta T$, after $n_T$ turns is given to first order by $\delta T/T= \delta m
/m$. Relative to a well-known mass $m$, the mass $m + \delta m$ can be derived
from the flight times of the ions.

$Q$-values of radioactive decays can provide accurate mass differences between
parent and daughter nuclei. For a long time, $\beta$ decay was one of the main
sources of mass determinations ($\beta^+$ decay and electron capture (EC) for
proton-rich nuclei and $\beta^-$ decay on the neutron-rich side of the nuclear
chart). If the decay proceeds to excited states in the daughter nucleus, the
mass difference has to be adjusted for the excitation energy, meaning that in
addition to the determination of the $\beta$ end-point energy, spectroscopy of
the daughter has to be performed in coincidence (see~\cite{Mittig1997} for a
review). $Q_{\alpha}$ or $Q_p$ measurements are less complicated owing to the
fact that the $\alpha$ particle or proton do not share their energy with a
neutrino, thus making the energy difference a more direct observable for
ground-state to ground-state decays~\cite{Lunney2003}.   
Mass differences may also be derived from the energy balance in
reactions. In a two-body reaction, $A(B,C)D$, the mass excesses $\Delta$ are
related via the amount of released energy $Q$:  
\begin{equation}
Q=\Delta_A + \Delta_B -\Delta_C - \Delta_D
\end{equation}
If three of the masses are well known, the determination of the $Q$-value from a
measurement of the reaction kinematics will allow the mass of the remaining
reaction partner to be extracted. This methods also works for extracting the mass
of unbound systems when the decay particle is detected (invariant mass
method)~\cite{Penionzhkevich2001}. For both approaches based on decay or
reaction $Q$ values, the mass differences that enter the energy balance must be
linked to known masses.

Measured masses are evaluated in the 2003 Atomic Mass
Evaluation (2003AME)~\cite{Wapstra2003,Audi2003}. Below, we summarize some of
the more recent mass measurements of exotic nuclei with relevance to nuclear
astrophysics.

The rapid neutron-capture process (r process) is responsible
for the synthesis of roughly half of the elements
beyond iron. However, its site and reaction path is not known with
certainty. The abundance patterns of various r-process models show a significant
dependence on the underlying nuclear
physics
(see~\cite{Cowan1991,Kratz1993,Truran2002,Qian2003,Arnould2003,Wanajo2004} for
reviews). High precision mass measurements of \nuc{71m,72-81}{Zn} were performed
with ISOLTRAP at CERN~\cite{Baruah2008}.  From \nuc{81}{Zn}
and \nuc{80}{Zn}, the neutron separation energy and neutron-capture $Q$-value of
\nuc{80}{Zn} were determined for the first time. Depending on the stellar
environment, the r-process path either includes the slow $\beta$ decay of
\nuc{80}{Zn} making it a major waiting point or it proceeds rapidly to
\nuc{81}{Zn} and beyond via neutron capture~\cite{Baruah2008}.  The new 
results improved the mass-related uncertainties for r-process calculations
compared to the previously used 2003AME. In the temperature regime below 1.5~GK,
where the $(n,\gamma) \rightleftharpoons (\gamma,n)$ equilibrium begins to
break down, the previously dominating mass uncertainties to the reaction flow in
this region have become negligible with the new results. High-precision Penning
trap mass measurements of 
\nuc{132,134}{Sn} carried out with the ISOLTRAP setup revealed a 0.5~MeV
deviation of 
the binding energy of \nuc{134}{Sn} compared to the previously accepted
value~\cite{Dworschak2008}. The new deduced value for the $N=82$ shell gap in
\nuc{132}{Sn} was found to be larger than the $N=28$ shell gap of
the stable doubly magic nucleus \nuc{48}{Ca}. The $N=82$ shell gap is thought to
influence the fission cycling in the r process~\cite{MartinezPinedo2007} since, in
the presence of a shell gap, the r process slows down and more neutrons are
created by photodisintegration inducing more fission~\cite{Dworschak2008}. 

X-ray bursts are initiated when the temperature
and the density in the accreted layer on a neutron star
become high enough to allow a breakout from the hot
CNO cycle. The nuclei start to capture protons and proceed along a capture
chain via the rp process. The capture rate is favorable compared to
the $\beta$-decay rate of the nuclei involved until a nucleus
with a small $Q$ value for the proton-capture reaction is encountered. There, an
equilibrium between the $(p,\gamma)$ proton capture 
and  $(\gamma,p)$ photodisintegration 
develops and the rp process is delayed until the
subsequent $\beta$-decay of this so-called ``waiting-point''
nucleus~\cite{Wallace1981,Schatz1998}. The effective 
lifetime of a waiting-point nucleus depends strongly  
on its proton separation energy which can be directly determined from mass
differences. The effective lifetime of a nucleus in an X-ray burst is
exponentially dependent on the proton separation energy and thus accurate mass
values -- preferably with less than 10~keV uncertainty -- are  desired for X-ray
burst model calculations~\cite{Brown2006,Schatz2006}. The selfconjugate nuclei
\nuc{64}{Ge}, \nuc{68}{Se} and \nuc{72}{Kr} are believed to be important, long-lived
waiting point nuclei~\cite{Schatz1998}. 

Mass measurements of \nuc{68}{Se},
\nuc{68}{As} and \nuc{68}{Ge} with the Canadian Penning Trap (CPT) confirmed 
\nuc{68}{Se} to be a waiting point nucleus, causing a considerable delay in the
rp process~\cite{Clark2004}. Penning trap mass spectrometry of \nuc{63,64}{Ga},
\nuc{64,65,66}{Ge}, \nuc{66,67,68}{As} and \nuc{69}{Se} was performed at the
LEBIT facility at  NSCL~\cite{Schury2007}. By using theoretical Coulomb-shift
energies and measured masses of \nuc{65,66}{Ge}, \nuc{67}{As}, and \nuc{69}{Se},
mass values for \nuc{65}{As}, \nuc{66,67}{Se}, and \nuc{69}{Br} were predicted
with an uncertainty of about 100~keV~\cite{Schury2007}. The results together
with the measured masses were used to calculate improved effective lifetimes of
the rp-process waiting-point $N=Z$ nuclei \nuc{64}{Ge} and \nuc{68}{Se}. It was
found that  \nuc{64}{Ge} is less of a waiting point while \nuc{68}{Se} poses a
larger delay in the rp process than previously
thought~\cite{Schury2007}. Recently, the masses of \nuc{68}{Se}, \nuc{70}{Se},
\nuc{70m}{Br} and \nuc{71}{Br} were measured with experimental uncertainties of
0.15-15 keV with the LEBIT facility at NSCL~\cite{Savory2009}. The new and
improved data were used in conjunction with Coulomb displacement energies as
input for rp-process calculations. An increase of the effective lifetime of the
waiting point nucleus \nuc{68}{Se} was found and more precise information on the
luminosity of type I X-ray bursts and final element abundances after the burst
were 
obtained~\cite{Savory2009}. At ISOLTRAP, the mass of the selfconjugate nucleus
\nuc{72}{Kr} was measured for the first time with Penning trap mass spectrometry
yielding a precision of 8~keV~\cite{Rodriguez2004}. The masses of
\nuc{73,74}{Kr} were determined with an order of magnitude improvement in
accuracy compared to previous results. \nuc{72}{Kr} was found to be an important
waiting point in the rp process during X-ray bursts, delaying energy generation
with at least 80\% of its $\beta^+$-decay half-life~\cite{Rodriguez2004}. 

Of interest is the possible existence of an endpoint to the rp process. Reaction
network calculations suggest that the rp process in X-ray bursts terminates at
tellurium and is limited by the SnSbTe cycle~\cite{Schatz2001}. The nuclei
\nuc{104-108}{Sn}, \nuc{106-110}{Sb}, \nuc{108,109}{Te}, and 
\nuc{111}{I}, in the vicinity of the expected endpoint of
the astrophysical rp process, were produced in fusion-evaporation reactions with
a \nuc{58}{Ni} beam irradiating a natural Ni target at the IGISOL facility in
Jyv\"askyl\"a~\cite{Elomaa2009}. Their mass values were precisely measured at
the  
JYFLTRAP Penning trap facility. For \nuc{106,108,110}{Sb}, these were
the first direct mass measurements. One-proton
separation energies were derived and the low value for \nuc{106}{Sb} excludes a
strong SnSbTe cycle starting at \nuc{105}{Sn}~\cite{Elomaa2009}. Masses of 34
neutron-deficient nuclei along the rp-process path 
were measured  with SHIPTRAP at GSI with an uncertainty of 10
keV~\cite{Block2006}.  The nuclei relevant for the rp-process path above $A=90$
were produced in fusion-evaporation reactions induced by \nuc{40}{Ca},
\nuc{50}{Cr} and \nuc{58}{Ni} beams impinging upon a \nuc{58}{Ni}
target. Preliminary results were reported for \nuc{101,102,103,104}{Cd},
\nuc{102,103,104,105}{In} and \nuc{105,106}{Sn}, and \nuc{99,101,103}{Ag},
\nuc{112,111,110,109}{Te}, \nuc{113,112,111}{I}, \nuc{113}{Xe},
\nuc{107,109,111}{Sb}. Except for \nuc{105}{In} and several tellurium isotopes,
the measured masses were found to agree well with the 2003AME. The rp process
proceeds along the Sn isotopic chain until the proton separation energy of Sb
isotopes becomes large enough to delay the competing photodisintegration and
hence, causes the rp process to proceed up to tellurium where SnSbTe cycles
prevent the production of heavier elements. With the masses measured at SHIPTRAP,
more accurate proton separation energies can be deduced in order to clarify
which SnSbTe cycle dominates~\cite{Block2006}. The $Q$ values for the proton
emission from \nuc{104}{Sb} and \nuc{105}{Sb} were also recently deduced
indirectly from $\alpha$-decay measurements carried out at HRIBF
(ORNL)~\cite{Mazzocchi2007}. 

The $\nu$p process occurs in explosive environments when
proton-rich matter is ejected in the presence of strong
neutrino fluxes~\cite{Frohlich2003}. This includes the inner ejecta of
core-collapse supernova~\cite{Buras2006}, for example. Neutron-deficient nuclei
from yttrium to palladium of relevance to the $\nu$p 
process were produced in fusion-evaporation reactions and studied at SHIPTRAP
(GSI) and JYFLTRAP (Jyv\"askyl\"a)~\cite{Weber2008}. Precise masses of 21 nuclei
were determined with Penning trap mass spectrometry, the masses of \nuc{88}{Tc},
\nuc{90-92}{Ru}, \nuc{92-94}{Rh}, and \nuc{94,95}{Pd} for the first time. The
data was used for $\nu$p-process nucleosynthesis calculations to probe their
impact on  reaction flow and final abundances. In particular the reaction flow
around \nuc{88}{Tc} was found to be significantly modified as the new
measurements give a proton separation energy that differs from the previously
used  2003AME by 1~MeV. However, the final abundances were basically
unchanged~\cite{Weber2008}.

Direct mass measurements of the short-lived nuclei \nuc{44}{V}, \nuc{48}{Mn},
\nuc{41}{Ti} and \nuc{45}{Cr} were performed with the isochronous mass
spectrometry at the experimental storage ring (ESR) at
GSI~\cite{Stadlmann2004}. An accuracy of about 100-500 keV was achieved. The
results largely confirmed the previously employed theoretical mass
predictions and thus lead only to 
small changes in the rp process in astrophysical X-ray burst
models~\cite{Stadlmann2004}.

The mass and electron capture $Q$ value of \nuc{26}{Si} was measured
with JYFLTRAP Penning trap facility and a ten times improved precision was
achieved~\cite{Eronen2009}. This leads to an improvement of the reaction $Q$
value for the $\nuc{25}{Al}(p,\gamma)\nuc{26}{Si}$  proton capture. The new
result changed the stellar production rate of \nuc{26}{Si} at nova ignition
temperatures by about 10\% and makes the limiting factor for higher precision of
the reaction $Q$ value the mass of \nuc{25}{Al}~\cite{Eronen2009}. 

\section{Outlook}
Tremendous progress has been made over the past years in experimental nuclear
astrophysics as well as in the developments of the multi-faceted theoretical
framework that is crucial to link the study of nuclear structure and reactions
to astrophysical processes in stars and explosive scenarios. The future for
nuclear physics and nuclear astrophysics looks bright with 
next-generation rare-isotope facilities on the horizon like GSI/FAIR in Germany and
FRIB in the United States and major upgrades on the way at GANIL in France and
TRIUMF in Canada.  

Short-term, the RIBF facility in JAPAN, which became operational recently, and
significant capability upgrades at NSCL, TRIUMF, ANL, HRIBF/ORNL, and CERN/ISOLDE
(see Section \ref{sec:prod}) will continue to provide a multitude of exciting opportunities to
advance the experimental research in nuclear astrophysics at all fronts. For
example, the nuclear equation of state (EOS) -- crucial for an  
understanding of neutron stars and supernova explosions -- can be probed with 
fast-beam heavy-ion collisions induced by projectiles with extreme $N/Z$
ratios utilizing the present NSCL fast-beam capability as well as the different
energies available at GSI and RIBF. Pioneering programs at NSCL aimed at
charge-exchange reactions on exotic nuclei will measure electron-capture rates
for exotic nuclei and generate crucial input for the nucleosynthesis in
supernovae. The CARIBU at the ATLAS facility at ANL will provide low-energy
and reaccelerated beams of neutron-rich \nuc{252}{Cf} fission products and
enable experiments in the proximity of r-process path in several mass
regions~\cite{Savard2008}. The ReA3 reaccelerator addition underway at NSCL will
deliver for the first time reaccelerated beams of rare isotopes produced by
projectile  fragmentation and fission with final energies ranging from 0.3 to
3~MeV/nucleon for ions with a charge-to-mass ratio of 0.25, and 0.3 to 6
MeV/nucleon for ions with a ratio of charge to $A$ of 0.5, which are of great
interest for direct and indirect nuclear astrophysics
measurements across the chart~\cite{ReA3,Gelbke2009}.  

The important scientific questions to be addressed both experimentally and
theoretically in nuclear physics of exotic nuclei with relevance for
astrophysics comprise: (a) How do loosely-bound systems survive and what are
the general laws of their formation and destruction? (b) Are new types of
radioactivity possible? (c) Are new types of nuclear symmetry and spatial
arrangement possible? (d) What are the limits of nuclear existence? (e) How
do the properties of nuclear matter change as a function of density,
temperature and proton-to-neutron ratio? (f) How do thermal and quantum
phase transitions occur in small systems? (g) What determines the shape and
symmetry properties of an exotic nucleus? (h) How does quantum tunneling of
composite particles occur in the process of reactions and decay? (i) What
are the manifestations of fundamental forces and symmetries in unusual
conditions? (j) How were the elements heavier than iron formed in stellar
explosions? (k) How do rare isotopes shape stellar explosions? (l) What is
the role of rare isotopes in neutron stars?

These questions provide extreme challenges for experiments and theory. On
the experimental side, producing the beams of radioactive nuclei needed to
address the scientific questions has been an enormous challenge. Pioneering
experiments have established the techniques and present-generation facilities
have produced first exciting science results, but the field is still at the
beginning of an era of discovery and exploration that will be fully underway
once the range of next-generation facilities will become operational.  The
theoretical challenges relate to wide variations in nuclear composition and
rearrangements of the bound and continuum structure, sometimes involving
near-degeneracy of the bound and continuum states \cite{unedf}. The extraction
of reliable information from experiments requires a solid understanding of the
reaction process, in addition to the structure of the nucleus. In
astrophysics, new observations, for example the expected onset of data
on stellar abundances, will require rare-isotope science for their
interpretation.         

\section*{Acknowledgement} 

This work was supported in part by the U.S. DOE
grants DE-FG02-08ER41533 and DE-FC02-07ER41457 (UNEDF, SciDAC-2), by the
Research Corporation, and by the US National Science Foundation under
grant PHY-0606007. A.G. acknowledges discussions with Hendrik Schatz.

\bibliographystyle{elsart-num}

\end{document}